\documentstyle[12pt]{article}
\def\be{\begin{equation}}
\def\ee{\end{equation}}
\def\bea{\begin{eqnarray}}
\def\eea{\end{eqnarray}}
\def\l#1{\label{#1}}
\def\r#1{(\ref{#1})}
\def\c#1{\cite{#1}}
\begin{document}

\title{
LARGE-N THEORY FROM THE AXIOMATIC POINT OF VIEW
}
\author{
  O.Yu. Shvedov
\\
{\small {\em Sub-Dept. of Quantum Statistics and Field Theory,}}\\
{\small{\em Department of Physics, Moscow State University }}\\
{\small{\em Vorobievy gory, Moscow 119899, Russia}} }

\maketitle

hep-th/0009035

\footnotetext{e-mail: olshv@ms2.inr.ac.ru,
shvedov@qs.phys.msu.su}

\begin{abstract}

The state space and observables for the leading order of the large-$N$
theory are constructed. The obtained model ("theory of infinite number
of fields")   is   shown   to  obey  Wightman-type  axioms  (including
invariance under boost transformations) and to  be  nontrivial  (there
are scattering processes,  bound states,  unstable particles etc). The
considered class  of  exactly  solvable  relativistic  quantum  models
involves good  examples  of  theories  containing such difficulties as
volume divergences associated  with  the  Haag  theorem,  Stueckelberg
divergences and infinite renormalization of the wave function.

\end{abstract}

PACS numbers: 11.30.Cp, 11.10.Cd, 11.15.Pg, 11.10.Gh.
\newpage

\section{Introduction}

Large-N expansion   is   widely   used   in   quantum   field   theory
\c{Wilson,CJP,GN}. This   approximation   allows    us    to    obtain
non-perturbative results  and  investigate  the behavior of the Green
functions, the effective action,  dynamical and  spontaneous  symmetry
breaking.

The traditional  approaches to the 1/N-expansion enable us to evaluate
different quantities mentioned above.  However,  some problems of  the
large-N theories   remain  to  be  understood.  What  are  states  and
observables in the theory  of  infinite  number  of  fields?  Can  one
determine such a theory as a large-N limit?

From the axiomatic field theory point  of  view  \c{W56,SW,BLOT},  the
relativistic quantum field theory is constructed if:

(i) the Hilbert state space $\cal H$ is specified;

(ii) the  operators $U_g:  {\cal H} \to {\cal H}$ corresponding to the
Poincare transformations  $g$  are  specified;  the   group   property
$U_{g_1}U_{g_2} = U_{g_1g_2}$ is satisfied;

(iii) the field operators are constructed.

The introduced objects should obey certain (Wightman-type) axioms.

It happens that the axiomatic formulation of the large-N  QFT  can  be
obtained  within  the  third-quantized  approach developed recently in
\c{MS}.  It is interesting that the large-N limit of QFT may be viewed
as  a theory of a variable number of fields.  This is analogous to the
statistical physics:  the system  of  a  large  but  fixed  number  of
particles  can  be  considered as a set of quasiparticles which can be
created and annihilated.  Analogously, the large-N field system can be
treated  from  the  "quasifield" point of view:  there is an amplitude
that there are no fields,  that there is one field,  two  fields  etc.
Thus,  the  large-N  limit  of  QFT is {\it not} a field theory in the
usual  treatment  since  one  cannot  define  usual  field  operators,
However,   the   property  of  the  relativistic  invariance  remains.
Moreover,  we will introduce the analog of notion of  field  which  is
very useful for constructing boost transformations.

The models of infinite number of fields constructed in this paper seem
to  be  remarkable  from  the  point of view of the constructive field
theory \c{Hepp,GJ}.  The old  problem  of  QFT  is  to  construct  the
nontrivial  model  of field theory obeying all Wightman axioms.  Such
examples were constructed  for  the  cases  of  2-  and  3-dimensional
space-time.  The  models  presented  here  are  considered  in  higher
dimensions.

The models considered in this paper are good examples of theories that
contain such  difficulties  as  Stueckelberg  divergences \c{Stu}
and  volume divergences  associated with the Haag theorem.
There was a hypothesis \c{Hepp} that the models
with the Stueckelberg divergences cannot be constructed with the  help
of the Hamiltonian methods. However, we show this hypothesis to be 
not correct.

This paper is organized as follows.

As an example, we consider the
$\lambda   (\varphi^a\varphi^a)^2$   model   in
(d+1)-dimensional space-time with the following Lagrangian:
$$
{\cal L}   =   \frac{1}{2}   \partial_{\mu}  \varphi^a  \partial_{\mu}
\varphi^a -  \frac{m^2}{2}  \varphi^a\varphi^a  -   \frac{\lambda}{4N}
(\varphi^a\varphi^a)^2,
$$
we sum over repeated indices $a=1,...,N$, $\mu=0,...,d$. In section II
the $N=\infty$-limit of the model is  heuristically  constructed.  The
Hamiltonian,  momentum,  angular  momentum  and  boost  generator  are
presented. It is heuristically shown that  they  formally  obey  usual
commutation   relations   of   the   Poincare  algebra.  However,  the
divergences  shows  us  that  the   obtained   expressions   are   not
mathematically well-defined.

Section III is devoted  to  the  problem  of  renormalization  of  the
Hamiltonian. The momentum and angular momentum are also
investigated in section III.  The spectral and vacuum axioms are  
checked.

It is not easy to construct operators of boost transformation (Lorentz
rotation) and check the group properties.  It is convenient  first  to
introduce the  composed  field being an analog of the large-N operator
$\sum_{a=1}^N \varphi^a(x_1)...\varphi^a(x_k)$.     Such     operators
(multifields)  being analogs of fields of ordinary QFT are constructed
in section IV. They are shown to be operator distributions. The cyclic
property of the vacuum state is checked. The invariance of multifields
under spatial rotations and space-time translations is checked.

Section V  deals  with  construction  of   the   operator   of   boost
transformation. This  allows us to construct the representation of the
Poincare group and check the relativistic invariance of the theory.
The results of section IV are essentially used.

Section VI contains concluding remarks.

\section{What is a theory of infinite number of fields?}

This section  deals  with  investigation  of  the  theory of N fields,
$\varphi^1$, ...,  $\varphi^N$  as  $N\to\infty$.  Such  models   were
considered in  context  of calculations of physical quantities such as
Green functions.  It   seems   to   be   useful   to   formulate   the
$N=\infty$-theory: to    determine    the    state   space,   Poincare
transformations, field operators etc.

\subsection{Multifield canonical operator}

In  the
functional Schrodinger representation states of the  $N$-field  system
at time      $t$      are     specified     by     the     functionals
$\Psi_N^t[\varphi^1(\cdot),...,\varphi^N(\cdot)]$ depending on the field
configurations $\varphi^1({\bf x})$,...,  $\varphi^N({\bf x})$,  ${\bf
x}=(x_1,...,x_d)$. The inner  product  is  formally  written  via  the
functional integral
$$
(\Psi_N,\Psi_N) = \int D\varphi^1 ...D\varphi^N |
\Psi_N[\varphi^1(\cdot),...,\varphi^N(\cdot)]|^2.
$$
The evolution equation has the form
\be
i\frac{d}{dt}\Psi_N^t = {\cal H}_N\Psi_N^t
\l{ba1}
\ee
with the  Hamiltonian  presented  as a sum of the free Hamiltonian and
interaction
\be
{\cal H}_N = \int d{\bf x} \left[
- \frac{1}{2} \frac{\delta^2}{\delta \varphi^a({\bf x})
\delta \varphi^a({\bf x})} + \frac{1}{2} (\nabla \varphi^a)({\bf x})
(\nabla \varphi^a)({\bf x}) + \frac{m^2}{2} \varphi^a({\bf x})
\varphi^a({\bf x})   +   \frac{\lambda}{4N}   (   \varphi^a({\bf   x})
\varphi^a({\bf x}))^2 \right]
\l{b0}
\ee
If one   considers  states  of  a  few  number  of  particles  in
comparison with $N$,  one can suppose that almost all fields  are  in
the vacuum  state.  This treatment leads us to the following structure
of the  wave  functional  $\Psi^t_N$.  If  all  fields  $\varphi^1$,...,
$\varphi^N$ are  in  the  same  (vacuum)  state,  the  $N$-field state
$\Psi_N$ is
\be
\Psi_N[\varphi^1(\cdot),...,\varphi^N(\cdot)] =
c \Phi_0[\varphi^1(\cdot)] ... \Phi_0[\varphi^N(\cdot)].
\l{b1}
\ee
If $(N-1)$  fields are in the state $\Phi_0$,  while 1 field is in the
state
$f_1$, the $N$-field state can be written as
\be
\Psi_N[\varphi^1(\cdot),...,\varphi^N(\cdot)] =
\frac{1}{\sqrt{N}}                                        \sum_{a=1}^N
\Phi_0[\varphi^1(\cdot)]...\Phi_0[\varphi^{a-1}(\cdot)]
f_1[\varphi^a(\cdot)]\Phi_0[\varphi^{a+1}(\cdot)]
... \Phi_0[\varphi^N(\cdot)]
\l{b2}
\ee
Without loss of generality,  one can  suppose  that  $(\Phi_0,f_1)=0$.
Otherwise, one  could  decompose  the functional $f_1$ into two parts,
one of them being proportional to $\Phi_0$,  another being  orthogonal
to $\Phi_0$.  The  case  $f_1=const  \Phi_0$  does  not  lead to a new
functional since expressions \r{b1} and \r{b2} coincide then.

Analogously, the state corresponding to $(N-k)$ fields in  the  vacuum
state and  $k$  fields  in  the  state  $f_k(\varphi^1,...,\varphi^k)$
being symmetric with respect to transpositions  of  $\varphi^1$,  ...,
$\varphi^k$ and satisfying the orthogonality condition
\be
\int D\varphi_1  \Phi_0^*[\varphi^1(\cdot)]  f_k[\varphi^1(\cdot),...,
\varphi^k(\cdot)] = 0
\l{b2*}
\ee
has the form
\be
\Psi_N[\varphi^1(\cdot),...,\varphi^N(\cdot)] =
\frac{1}{\sqrt{N^kk!}} \sum_{1\le a_1 \ne ... \ne a_k \le N}
f_k[\varphi^{a_1}(\cdot),..., \varphi^{a_k}(\cdot)]        \prod_{a\ne
a_1...a_k} \Phi_0[\varphi^a(\cdot)].
\l{b3}
\ee
Finally, one can consider the  superposition  of  states  \r{b3}  with
rapidly decreasing at $k\to\infty$ set of norms $||f_k||$. This is the
most general  form  of
a  state "with a few number of particles",  provided that
one takes into account symmetric functionals $\Psi$ only.
Nonsymmetric functionals are investigated in appendix B.

We see that symmetric
states in the theory of a large number of fields occur  to
be specified by infinite sets
\bea
f= \left(
\matrix{
f_0 \nonumber
\\
f_1[\varphi^1(\cdot)]
\nonumber \\
...
\nonumber \\
f_k[\varphi^1(\cdot),...,\varphi^k(\cdot)]
\nonumber \\
...
}
\right)
\l{b3*}
\eea
where $f_k$ are symmetric functionals satisfying eq.\r{b2*}.
One can  say that $f_k$ is a probability amplitude that $k$ fields are
in the non-vacuum state.  We see that the theory of a large number  of
fields is  equivalent  to  the  theory of a variable number of fields.
This observation is  analogous  to  the  quasiparticle  conception  in
statistical physics.

The mapping $K_N: f\mapsto \Psi_N$ of the form
\be
(K_Nf)[\varphi^1(\cdot),...,\varphi^N(\cdot)] = \sum_{k=0}^{N}
\frac{1}{\sqrt{N^kk!}} \sum_{1\le a_1 \ne ... \ne a_k \le N}
f_k[\varphi^{a_1}(\cdot),..., \varphi^{a_k}(\cdot)]        \prod_{a\ne
a_1...a_k} \Phi_0[\varphi^a(\cdot)]
\qquad \qquad \qquad
\qquad
\l{b4}
\ee
will be  called  as a multifield canonical operator analogously to the
multiparticle canonical operator used in statistical physics \c{MS}.
The orthogonality condition \r{b2*} implies that
\be
||K_Nf||^2 = \sum_{k=0}^N \frac{N!}{N^k(N-k)!}
||f_k||^2 \to_{N\to\infty} \sum_{k=0}^{\infty} ||f_k||^2.
\l{b5}
\ee
We see  that  sets  \r{b3*}  may  be identified with
states of the system of
$N=\infty$ fields,  while  the relation \r{b5} can be considered as an
argument that the norm of a state should be chosen as
$$
||f||^2 = \sum_{k=0}^{\infty} ||f_k||^2.
$$
Thus, decomposition \r{b4} gives us a relationship
between the theory of a
large number of fields and the theory of a variable number of fields.

\subsection{Representation of operators}

Let us  write  operators  of physical quantities in the representation
\r{b3*}. It will be convenient to present them via the third-quantized
creation and annihilation operators which can be introduced as follows
\c{MS}. The creation  operator  $A^+[\varphi(\cdot)]$  increases  the
number of       fields,       i.e.       transforms       the      set
$f=(0,0,...,0,f_{k-1},0,...)$                                     into
$(0,...,0,(A^+[\varphi(\cdot)]f)_k,0,...)$, The functional
$(A^+[\varphi(\cdot)]f)_k[\varphi^1(\cdot),...,\varphi^k(\cdot)]$
being the  $k$-th  component  of  the  set $(A^+[\varphi(\cdot)]f)$ is
expressed via the $(k-1)$-th component of $f$:
\be
(A^+[\varphi(\cdot)]f)_k[\varphi^1(\cdot),...,\varphi^k(\cdot)]
= \frac{1}{\sqrt{k}} \sum_{a=1}^k
\delta
(\varphi(\cdot)-\varphi^a(\cdot))f_{k-1}[\varphi^1(\cdot),...,
\varphi^{a-1}(\cdot), \varphi^{a+1}(\cdot), ...,\varphi^k(\cdot)].
\l{b6}
\ee
The annihilation operator $A^-[\varphi(\cdot)]$ is
\be
(A^-[\varphi(\cdot)]f)_{k-1}[\varphi^1(\cdot),                    ...,
\varphi^{k-1}(\cdot)] =      \sqrt{k}       f_k       [\varphi(\cdot),
\varphi^1(\cdot), ..., \varphi^{k-1}(\cdot)].
\l{b7}
\ee
The condition \r{b2*} is not invariant under  transformations  \r{b6},
\r{b7}. Consider the modified creation and annihilation operators:
\bea
\tilde{A}^+ [\varphi(\cdot)]       =       A^+[\varphi(\cdot)]       -
\Phi_0^*[\varphi(\cdot)] \int D\phi \Phi_0[\phi(\cdot)]A^+[\phi(\cdot)]
\nonumber \\
\tilde{A}^- [\varphi(\cdot)]       =       A^-[\varphi(\cdot)]       -
\Phi_0[\varphi(\cdot)] \int D\phi \Phi^*_0[\phi(\cdot)]A^-[\phi(\cdot)]
\nonumber
\eea
To write   operators  in  the  representation  \r{b3*},  consider  the
orthonormal basis $(\Phi_0,\Phi_1,...)$ in the space of functionals,
$$
\int D\varphi   \Phi_i^*[\varphi(\cdot)]   \Phi_j[\varphi(\cdot)]    =
\delta_{ij}
$$
which contains  the  vacuum functional $\Phi_0$ entering to expression
\r{b4}. Investigate the following ``elementary'' operators
\be
{\cal O}_N^{ij} \Psi [\varphi^1(\cdot),...,\varphi^N(\cdot)] =
\sum_{a=1}^N \Phi_i[\varphi^a(\cdot)] \int D\phi \Phi_j^*[\phi(\cdot)]
\Psi[\varphi^1(\cdot),...,\varphi^{a-1}(\cdot),
\phi(\cdot),\varphi^{a+1}(\cdot),...,\varphi^N(\cdot)].
\l{b8}
\ee
Apply them to the expression \r{b4}.  It is necessary to distinguish 4
cases.

(i) $i=j=0$.

Due to the condition \r{b2*}, one has
$$
({\cal O}_N^{00}K_Nf)[\varphi^1(\cdot),...,\varphi^N(\cdot)] =
\sum_{k=0}^{N} (N-k)
\frac{1}{\sqrt{N^kk!}} \sum_{1\le a_1 \ne ... \ne a_k \le N}
f_k[\varphi^{a_1}(\cdot),..., \varphi^{a_k}(\cdot)]        \prod_{a\ne
a_1...a_k} \Phi_0[\varphi^a(\cdot)].
$$
This means that the operator ${\cal O}^{00}_N$
acts  in  the  space  \r{b3*}  as
$N-\hat{n}$, i.e.
$$
{\cal O}_N^{00} K_N f = K_N (N-\hat{n})f,
$$
where $\hat{n} = \int D\varphi \tilde{A}^+[\varphi(\cdot)]
\tilde{A}^-[\varphi(\cdot)]$ is the operator of number of fields,
$$
(\hat{n} f)_k = kf_k.
$$

(ii) $i=0, j\ne0$.

It follows from the symmetry condition that
\bea
({\cal O}^{0j}_N K_Nf)[\varphi^1(\cdot),...,\varphi^N(\cdot)] =
\sum_{k=0}^{N}
\frac{1}{\sqrt{N^kk!}}
\sum_{p=1}^k \frac{1}{\sqrt{k}}
\sum_{1\le a_1 \ne ... \ne a_k \le N}
\Phi_0[\varphi^{a_p}(\cdot)]
\nonumber \\
\times
(\int D\phi \tilde{A}^-[\phi(\cdot)] \Phi_j^*[\phi(\cdot)]
f)_{k-1}[\varphi^{a_1}(\cdot),...,
\varphi^{a_{p-1}}(\cdot), \varphi^{a_{p+1}}(\cdot), ...
\varphi^{a_k}(\cdot)]
\prod_{a\ne
a_1...a_k} \Phi_0[\varphi^a(\cdot)].
\eea
After redefining      $a_1=b_1$,       ...,       $a_{p-1}=b_{p-1}$,
$a_{p+1}=b_p$,..., $a_k = b_{k-1}$,  $a_k=b$ we obtain that the symbol
$\sum_k$ can be substituted  by  $k$,  while  $\sum_j$  transforms  to
$(N-k+1)$. Thus, one obtains the following commutation rule:
$$
{\cal O}^{0j}_N K_N f = K_N (N-\hat{n}) \frac{1}{\sqrt{N}}
\int D\phi \tilde{A}^-[\phi(\cdot)] \Phi_j^*[\phi(\cdot)]
f.
$$

(iii) $i\ne 0$, $j=0$.

Due to eq.\r{b2*}, we have
\bea
({\cal O}_N^{i0} K_Nf)
[\varphi^1(\cdot),...,\varphi^N(\cdot)] = \sum_{k=0}^{N}
\frac{1}{\sqrt{N^kk!}} \sum_{1\le a_1 \ne ... \ne a_k \le N}
f_k[\varphi^{i_1}(\cdot),..., \varphi^{i_k}(\cdot)]
\nonumber \\
\times
\sum_{a\ne a_1...a_k} \Phi_i [\varphi^a(\cdot)]
\prod_{b\ne
a,a_1...a_k} \Phi_0[\varphi^b(\cdot)].
\eea
After symmetrization the commutation rule takes the form
$$
{\cal O}^{i0}_N K_N f = K_N \sqrt{N}
\int D\phi \tilde{A}^+[\phi(\cdot)] \Phi_i[\phi(\cdot)]
f.
$$

(iv) $i\ne 0$, $j\ne 0$.

Analogously, we find that
$$
{\cal O}_N^{ij} K_N f = K_N
\int D\varphi \tilde{A}^+[\varphi(\cdot)] \Phi_i[\varphi(\cdot)]
\int D\phi \tilde{A}^+[\phi(\cdot)] \Phi_j^*[\phi(\cdot)]
$$
Any operator  can  be  represented  via  elementary  operators \r{b8}.
Consider an example.

The operator $\sum_{a=1}^N \varphi^a({\bf  x})\varphi^a({\bf  x})$  is
expressed as
$$
\sum_{a=1}^N \varphi^a({\bf  x})\varphi^a({\bf  x})
= \sum_{ij=0}^{\infty}  \int  D\phi  \Phi_i^*[\phi(\cdot)]   \phi({\bf
x})\phi({\bf x}) \Phi_j[\phi(\cdot)] {\cal O}_N^{ij}.
$$
Therefore, the following commutation rule takes place:
$$
\lambda \sum_{a=1}^N \varphi^a({\bf  x})\varphi^a({\bf  x}) K_N
= K_N \tilde{\cal Q}_N({\bf x}),
$$
where the operator $\tilde{\cal Q}_N({\bf x})$
consists of the constant term of order
$O(N)$, the  linear  in  creation-annihilation operators term of order
$O(\sqrt{N})$ and the regular as $N\to\infty$ term which is  quadratic
in creation and annihilation operators:
\bea
\tilde{\cal Q}_N({\bf x}) =
\lambda (N-\hat{n}) (\Phi_0, \phi({\bf x}) \phi({\bf x}) \Phi_0) +
\lambda \sqrt{N} \int  D\phi  \tilde{A}^+[\phi(\cdot)]
\phi({\bf x})\phi({\bf
x}) \Phi_0[\phi(\cdot)]
\nonumber
\\ +
\frac{\lambda}{\sqrt{N}} (N-\hat{n})
\int  D\phi  \tilde{A}^-[\phi(\cdot)]  \phi({\bf x})\phi({\bf
x}) \Phi^*_0[\phi(\cdot)]
+ \lambda \int  D\phi  \tilde{A}^+[\phi(\cdot)]  \phi({\bf x})\phi({\bf
x}) \tilde{A}^-[\phi(\cdot)].
\eea

\subsection{Evolution equation at $N=\infty$}

Analogously to  the  previous subsection,  the operators $\sum_{a=1}^n
(\nabla \varphi^a)^2({\bf x})$  and  $\sum_{a=1}^N  (-  \frac{\delta^2}
{\delta\varphi^a({\bf x})\delta  \varphi^a({\bf  x})})$  can  be  also
written in the representation \r{b3*}.  Since the  Hamiltonian  \r{b0}
contains the  considered  operator  expressions  only,  it can be also
commuted with the multifield canonical operator,
$$
{\cal H}_N K_N = K_N \tilde{\cal H}_N.
$$
The transformed Hamiltonian $\tilde{\cal H}_N$ is
\bea
\tilde{\cal H}_N =  \int d{\bf x} \left[
(N-\hat{n}) (\Phi_0, {\cal E}_0({\bf x}) \Phi_0) +
\sqrt{N}
\int D\phi  \tilde{A}^+[\phi(\cdot)]
{\cal E}_0({\bf x}) \Phi_0[\phi(\cdot)] \right.
\nonumber \\
\left.
+ \frac{1}{\sqrt{N}} (N-\hat{n})
\int D\phi \Phi^*_0[\phi(\cdot)]
{\cal E}_0({\bf x})
\tilde{A}^-[\phi(\cdot)]
+
\int D\phi \tilde{A}^+[\phi(\cdot)]
{\cal E}_0({\bf x})
\tilde{A}^-[\phi(\cdot)]
+ \frac{1}{4N\lambda} \tilde{\cal Q}_N^2({\bf x}),
\right]
\l{b9}
\eea
where
$$
{\cal E}_0  ({\bf x}) = - \frac{1}{2} \frac{\delta^2}{\delta \phi({\bf
x}) \delta \phi({\bf  x})}  +  \frac{1}{2}  (\nabla   \phi)^2({\bf   x})   +
\frac{m^2}{2} \phi^2({\bf x}).
$$
Expression \r{b9}  contains  the  terms of order $O(N)$,  $O(N^{1/2})$,
$O(1)$ and the terms damping as $N\to\infty$:
\be
\tilde{\cal H}_N = N\tilde{\cal H}^0 + N^{1/2} \tilde{\cal
H}^1 + \tilde{\cal H}^2 + O(N^{-1/2}).
\l{b9*}
\ee
The operator  $\tilde{\cal H}^0$
is  a  multiplication  by  the divergent c-number
quantity
$$
\tilde{\cal H}^0 = \int d{\bf x} (\Phi_0, {\cal E}_0({\bf x}) \Phi_0) +
\frac{\lambda}{4} \int   d{\bf   x}  (\Phi_0,  \phi({\bf  x})\phi({\bf
x})\Phi_0)^2.
$$
As usual in QFT, the vacuum energy is set to zero by adding a constant
to the  Hamiltonian,  so  that  the  Hamiltonian  is  defined  up to a
constant, and the quantity $\tilde{\cal H}^0$ can be neglected.

The operator  $\tilde{\cal H}^1$
is  a  linear   combination   of   creation   and
annihilation operators:
$$
\tilde{\cal H}^1 = \int D\phi \tilde{A}^+[\phi(\cdot)] Z[\phi(\cdot)]
+ \int D\phi \tilde{A}^-[\phi(\cdot)] Z^*[\phi(\cdot)]
$$
with
$$
Z[\phi(\cdot)] = \int d{\bf x}
\left[
- \frac{1}{2}\frac{\delta^2}{\delta \phi({\bf x})\delta \phi({\bf x})}
+ \frac{1}{2} (\nabla \phi ({\bf x}))^2 + \frac{m^2 + \lambda (\Phi_0,
\phi^2({\bf x}) \Phi_0)}{2} \phi^2({\bf x})
\right] \Phi_0[\phi(\cdot)].
$$
The operator $\tilde{\cal H}^1$ vanishes if and only if
\be
Z= const \Phi_0.
\l{b10}
\ee
We choose the functional $\Phi_0$ to be a vacuum state functional  for
the field of the mass $\mu$,
\be
\Phi_0[\phi(\cdot)] = const exp
[- \frac{1}{2}\int d{\bf  x}  \phi({\bf  x})  \sqrt{-\Delta  +  \mu^2}
\phi({\bf x})],
\l{b10*}
\ee
so that eq.\r{b10} will take the form
\be
\mu^2 = m^2 + \lambda (\Phi_0,\phi^2({\bf x})\Phi_0).
\l{b11}
\ee
This is a well-known equation in the $1/N$-expansion theory (see,  for
example, \c{CJP}).

The remaining  nonvanishing  as  $N\to\infty$  part of the Hamiltonian
is quadratic in creation and annihilation operators,
\bea
\tilde{H} \equiv
\tilde{\cal H}^2 = \int D\phi \tilde{A}^+[\phi(\cdot)]
:\int d{\bf x}
{\cal E}({\bf x})
: \tilde{A}^-[\phi(\cdot)]
+ \frac{\lambda}{4}
\int d{\bf x}
Q_0^2({\bf x}).
\l{b12}
\eea
where $:\hat{O}: = \hat{O} - (\Phi_0, \hat{O} \Phi_0)$,
$$
{\cal E}({\bf x}) =
\left[
- \frac{1}{2}\frac{\delta^2}{\delta \phi({\bf x})\delta \phi({\bf x})}
+ \frac{1}{2} (\nabla \phi ({\bf x}))^2 + \frac{\mu^2
}{2} \phi^2({\bf x})
\right]
$$
$$
Q_0({\bf x}) =
\int D\phi
(\tilde{A}^+[\phi(\cdot)] + \tilde{A}^-[\phi(\cdot)])
\phi^2({\bf x}) \Phi_0[\phi(\cdot)].
$$

Since the  term \r{b12} is the only term remaining as $N=\infty$,  one
can say that the theory of $N=\infty$ fields is as follows.  States in
this theory  are  sets \r{b3*} obeying eq.\r{b2*}.  The Hamiltonian of
the model has the form \r{b12},  the evolution equation is $i\dot{f} =
\tilde{H} f$.

\subsection{Representation of the Poincare algebra}

We have  specified  the state space of the theory of $N=\infty$ fields
and evolution operator. However, to construct the {\it relativistic}
quantum
theory, it  is  necessary  to  specify  the  operators $U_{\Lambda,a}$
corresponding to Poincare transformations
$$
x^{'\mu} =   \Lambda^{\mu}_{\nu}    x^{\nu}    +    a^{\mu},    \qquad
\mu,\nu=\overline{0,d},
$$
where the  matrix  $\Lambda$  of  Lorentz transformation satisfies the
property
$$
\Lambda^T g \Lambda = g
$$
($g= diag\{1,-1,-1,...\}$,  $\Lambda^T$ is the  matrix  transposed  to
$\Lambda$). The composition law of the Poincare transformations is
$$
(\Lambda_1,a_1) (\Lambda_2,a_2)     =     (\Lambda_1\Lambda_2,    a_1+
\Lambda_1a_2),
$$
so that   any   Poincare   transformation   can   be   presented    as
$(\Lambda,a)=(0,a)(\Lambda,0)$. Furthermore,   one  can  introduce  the
local coordinates         $\theta_{\lambda\mu}$         ($\lambda,\mu=
\overline{1,d}$, $\theta_{\lambda\mu} = -\theta_{\mu\lambda})$
on the Lorentz group \c{BLOT}, such that
$$
\Lambda = \exp(\frac{1}{2}\theta_{\lambda\mu} l^{\lambda\mu})
$$
with
$$
(l^{\lambda\mu})^{\alpha}_{\beta} =                 -g^{\lambda\alpha}
\delta^{\mu}_{\beta} + g^{\mu\alpha} \delta^{\lambda}_{\beta}.
$$
The operators  $U_{\Lambda,a}$ are required to form the representation
of the Poincare group, so that
$$
U_{\Lambda_1,a_1}U_{\Lambda_2,a_2}
= U_{(\Lambda_1,a_1)(\Lambda_2,a_2)}
$$
Making use  of  the  theory of representations of the Lie groups,  one
finds \c{BLOT}
$$
U_{\Lambda,a} = \exp(i\tilde{P}^{\mu}a_{\mu})
\exp(\frac{i}{2}\tilde{M}^{\lambda\mu} \theta_{\lambda\mu})
$$
for some   operators  $\tilde{P}^{\mu}$  and  $\tilde{M}^{\lambda\mu}$
obeying the commutation relations of the Poincare algebra

$$
[\tilde{P}^{\lambda},\tilde{P}^{\mu}]  = 0, \qquad
[\tilde{M}^{\lambda\mu}, \tilde{P}^{\nu}] =
i(g^{\mu\nu} \tilde{P}^{\lambda} - g^{\lambda\nu}\tilde{P}^{\mu}),
$$
\be
[\tilde{M}^{\lambda\mu}, \tilde{M}^{\rho\sigma} ] = -i (
g^{\lambda\rho} \tilde{M}^{\mu\sigma} -
g^{\mu\rho} \tilde{M}^{\lambda\sigma} +
g^{\mu\sigma} \tilde{M}^{\lambda\rho} -
g^{\lambda\sigma} \tilde{M}^{\mu\rho})
\l{+2}
\ee
Let us    construct    the    operators    $\tilde{P}^{\lambda}$   and
$\tilde{M}^{\lambda\mu}$ for the $N=\infty$-theory.  For the $N$-field
theory, one has \c{BS}:
$$
{\cal P}_N^{\mu} = \int d{\bf x} {\cal T}_N^{\mu 0}({\bf x}),
\quad
{\cal M}_N^{\mu\lambda} = \int d{\bf x}
(x^{\mu} {\cal T}_N^{\lambda 0}({\bf x})
- x^{\lambda} {\cal T}_N^{\mu 0}({\bf x})),
$$
where we integrate over surface $x^0=0$, while
$$
{\cal T}_N^{00}({\bf x}) =
- \frac{1}{2} \frac{\delta^2}{\delta \varphi^a({\bf x})
\delta \varphi^a({\bf x})} + \frac{1}{2} (\nabla \varphi^a)({\bf x})
(\nabla \varphi^a)({\bf x}) + \frac{m^2}{2} \varphi^a({\bf x})
\varphi^a({\bf x})   +   \frac{\lambda}{4N}   (   \varphi^a({\bf   x})
\varphi^a({\bf x}))^2,
$$
$$
{\cal T}_N^{k0}({\bf x}) = \sum_{a=1}^N (\partial^k\varphi^a({\bf x})
\frac{1}{i} \frac{\delta}{\delta \varphi^a({\bf x})})
$$
Let us commute these operators with the multifield canonical operator,
$$
{\cal P}_N^{\mu} K_N = K_N \tilde{\cal P}_N^{\mu},
\qquad
{\cal M}_N^{\mu\nu} K_N = K_N \tilde{\cal M}_N^{\mu\nu},
$$
expand the result in $1/N$:
$$
\tilde{\cal P}_N^{\mu} = N\tilde{\cal P}^{\mu,0} +
N^{1/2} \tilde{\cal P}^{\mu,1}+ \tilde{\cal P}^{\mu,2} + ..., \qquad
\tilde{\cal M}_N^{\mu\nu} = N\tilde{\cal M}^{\mu \nu,0} +
N^{1/2} \tilde{\cal M}^{\mu \nu,1}+ \tilde{\cal M}^{\mu \nu,2} + ...
$$
It will be shown that the operators
$\tilde{\cal P}^{\mu ,0}$,
$\tilde{\cal P}^{\mu ,1}$,
$\tilde{\cal M}^{\mu \nu,0}$,
$\tilde{\cal M}^{\mu   \nu,1}$   vanish,   so   that   the   remaining
nonvanishing at $N=\infty$ parts
$$
\tilde{P}^{\mu} = \tilde{\cal P}^{\mu ,2},
\tilde{M}^{\mu\nu} =
\tilde{\cal M}^{\mu \nu,2}
$$
should be viewed as generators  of  Poincare  transformations  in  the
$N=\infty$-theory.

Remind also that the operator $\tilde{\cal P}^0_N=\tilde{H}$ has been already
constructed in the previous subsection.

Consider the operator
$$
{\cal P}_N^k = \int d {\bf x}  \sum_{a=1}^N  \partial^k\varphi^a({\bf
x}) \frac{1}{i} \frac{\delta}{\delta \varphi^a({\bf x})}.
$$
After commuting with multifield canonical operator, one has
\bea
\tilde{\cal P}^{k,0} = (\Phi_0, \int d{\bf x}
\partial^k \phi({\bf x}) \frac{1}{i} \frac{\delta}{\delta \phi({\bf x})}
\Phi_0),
\nonumber \\
\tilde{\cal P}^{k,1} = \int D\phi (
\tilde{A}^+[\phi(\cdot)] Z^k[\phi(\cdot)]
+ \tilde{A}^-[\phi(\cdot)] Z^{k*}[\phi(\cdot)]),
\l{+1}
\eea
where
$$
Z^k[\phi(\cdot)] =
\int d{\bf x} \partial^k\phi({\bf x}) \frac{1}{i}
\frac{\delta}{\delta \phi({\bf x})}
\Phi_0[\phi(\cdot)].
$$
Since $\Phi_0$ has been chosen to be a vacuum functional for the field
of the mass $\mu$, while the operator
$\int d{\bf x} \partial^k\phi({\bf x}) \frac{1}{i}
\frac{\delta}{\delta \phi({\bf x})}$ is a momentum  operator  for  the
functional Schrodinger representation, one has $Z^k=0$,
$\tilde{\cal P}^{k,0} = 0$, $\tilde{\cal P}^{k,1} = 0$.
Thus, the operator
\be
\tilde{P}^k \equiv \tilde{\cal P}^{k,2} = \int D\phi
\tilde{A}^+[\phi(\cdot)] : \int d{\bf x}
\partial^k\phi({\bf x}) \frac{1}{i}
\frac{\delta}{\delta \phi({\bf x})}:
\tilde{A}^+[\phi(\cdot)]
\l{x1}
\ee
can be viewed as a momentum operator in the $N=\infty$ theory.

Analogously, we find that
\be
\tilde{M}^{ml} =     \tilde{\cal     M}^{ml,2}     =     \int    D\phi
\tilde{A}^+[\phi(\cdot)]
: \int d{\bf x} (x^m \partial^l\phi({\bf x}) - x^l \partial^m\phi({\bf
x})) \frac{1}{i} \frac{\delta}{\delta \phi({\bf x})}:
\tilde{A}^-[\phi(\cdot)].
\l{x2}
\ee
The boost operator presented as
$$
{\cal M}^{k0} = \int d{\bf x} x^k {\cal T}^{00}({\bf x})
$$
after commuting with the multiparticle canonical operator gives us:
$$
\tilde{\cal M}^{k0,0} = \int d{\bf x} x^k (\Phi^0, {\cal E}_0({\bf x})
\Phi_0) + \frac{\lambda}{4} \int d{\bf x} x^k (\Phi_0,  \phi({\bf  x})
\phi({\bf x}) \Phi_0).
$$
Since the integrand is an odd function with respect to $x^m$, it seems
to be   natural   that   $\tilde{\cal   M}^{m0,0}=0$.   The   operator
$\tilde{\cal M}^{k0,1}$ has the structure \r{+1} with
$$
Z^k[\phi(\cdot)] =
\int d{\bf x} x^k {\cal E}({\bf x})
\Phi_0[\phi(\cdot)].
$$
Since the   vacuum   state   $\Phi_0$   is   invariant   under   boost
transformations, while
$$
\int d{\bf x} x^k {\cal E}({\bf x})
$$
is a boost generator,  one has $Z^k=0$ and $\tilde{\cal M}^{k0,1} =0$.
The remaining term is
\bea
\tilde{M}^{k0} =     \tilde{\cal     M}^{k0,2}     =     \int    D\phi
\tilde{A}^+[\phi(\cdot)]
\int d{\bf x} x^k : {\cal E}({\bf x}):
\tilde{A}^-[\phi(\cdot)] +
\frac{\lambda}{4} \int d{\bf x} x^k
Q_0^2({\bf x}).
\l{x3}
\eea
The commutation relations \r{+2} are formally satisfied.
Namely, the operators \r{b12}, \r{x1}, \r{x2}, \r{x3} can be presented
as
\bea
\tilde{H} = \tilde{H}_0 + \lambda \tilde{H}_1, \qquad
\tilde{P}^k = \tilde{P}^k_0 , \nonumber \\
\tilde{M}^{k0} = \tilde{M}^{k0}_0 + \lambda \tilde{M}^{k0}_1, \qquad
\tilde{M}^{kl} = \tilde{M}^{kl}_0.
\eea
For $\lambda=0$ - case,  the check of relations \r{+2} is identical to
the standard check of the Poincare  invariance  of  the  free  quantum
field theory.  For general  case,  it  is sufficiently to justify the
following commutation relations:
\be
[\tilde{H}_1, \tilde{P}_0^k] = 0, \qquad
[\tilde{H}_1, \tilde{M}_0^{kl}] = 0, \qquad
\l{x4}
\ee
\be
[\tilde{M}_1^{k0}, \tilde{P}_0^l] = - ig^{kl} \tilde{H}_1, \qquad
[\tilde{M}_1^{k0}, \tilde{M}_0^{mn}] = - i
(g^{km} \tilde{M}^{0n}_1 - g^{kn} \tilde{M}^{0m}_1).\l{x5}
\ee
\be
[\tilde{M}_1^{k0}, \tilde{M}_1^{l0}] = 0,\qquad
[\tilde{M}_1^{k0}, \tilde{H}_1] = 0,\qquad
\l{x7}
\ee
\be
[\tilde{M}_1^{k0}, \tilde{H}_0] +
[\tilde{M}_0^{k0}, \tilde{H}_1] = 0,
\qquad
[\tilde{M}_1^{k0}, \tilde{M}_0^{l0}] +
[\tilde{M}_0^{k0}, \tilde{M}^{l0}_1] = 0.
\l{x8}
\ee

It is straightforward to check that
$$
[\tilde{P}_0^l, Q_0({\bf x})] = -i \partial^l Q_0({\bf x}),\qquad
[\tilde{M}_0^{mn}, Q_0({\bf  x})] = -i (x^m \partial^n - x^n \partial^m)
Q_0({\bf x}).
$$
We obtain  relations  \r{x4}  and  \r{x5}  then.   Eqs.   \r{x7}   are
corollaries of the property $[Q({\bf x}), Q({\bf y})]=0$. The relation
$[{\cal E}({\bf x}),  Q_0({\bf y})] \sim  \delta({\bf  x}-  {\bf  y})$
imply eq.\r{x8}.  Thus,  the  formal  Poincare  invariance is checked.
However, the divergences and renormalization have not been  considered
yet.

\subsection{Mode decomposition}

We have specified states of the $N=\infty$-theory as sets
\bea
f= \left(
\matrix{
f_0 \nonumber
\\
f_1[\varphi^1(\cdot)]
\nonumber \\
...
\nonumber \\
f_k[\varphi^1(\cdot),...,\varphi^k(\cdot)]
\nonumber \\
...
}
\right)
\l{+3}
\eea
of symmetric   functionals  $f_k[\varphi^1,...,\varphi^k]$  satisfying
relation \r{b2*} such that
$$
||f||^2 =   \sum_{k=0}^{\infty}   \int   D\varphi^1   ...   D\varphi^k
|f[\varphi^1(\cdot),...,\varphi^k(\cdot)]|^2 < \infty.
$$
However, this   definition   is   ill-defined  since  the  measure  of
functional integration is not determined  mathematically.  Instead  of
constructing the   measure,   it   is   convenient   to   use  another
representation for the $k$-field functionals.

Consider the basis functionals
\be
\Phi^{(n)}_{{\bf k}_1...{\bf       k}_n}       [\varphi(\cdot)]      =
\frac{1}{\sqrt{n!}} a^+_{{\bf    k}_1}     ...     a^+_{{\bf     k}_n}
\Phi_0[\varphi(\cdot)], \qquad n=1,2,3,...
\l{b14}
\ee
corresponding to $n$ particles with momenta ${\bf  k}_1$,  ...,  ${\bf
k}_n$. The  operators  $a^+_{\bf  k}$ are usual quantum field creation
operators:
$$
a_{\bf k}^+ = \frac{1}{(2\pi)^{d/2}} \int d{\bf x} e^{i{\bf k}{\bf x}}
\left[
\sqrt{\frac{\omega_{{\bf k}}}{2}} \varphi({\bf x})
- \frac{1}{\sqrt{2\omega_{\bf k}}} \frac{\delta}{\delta \varphi ({\bf x})}
\right]
$$
with $\omega_{\bf k}= \sqrt{{\bf k}^2+\mu^2}$.  Integrating  by  parts
and using  the commutation relations between creation and annihilation
operators, we find that the inner product \r{ba1} for the  functionals
\r{b14} has the form
\bea
(\Phi^{(n)}_{{\bf k}_1...{\bf       k}_n},
\Phi^{(m)}_{{\bf p}_1...{\bf       p}_m}) = 0, \qquad m\ne n;
\nonumber \\
(\Phi^{(n)}_{{\bf k}_1...{\bf       k}_n},
\Phi^{(n)}_{{\bf p}_1...{\bf       p}_n}) =
\frac{1}{n!} \sum_{\sigma}
\delta({\bf k}_1 - {\bf p}_{\sigma_1}) ...
\delta({\bf k}_n - {\bf p}_{\sigma_n}),
\l{b15}
\eea
the sum   is   taken  over  all  transpositions  of  indices  1,...,n.
Eqs.\r{b15} can be viewed as a definition of the  functional  integral
\r{ba1}.

Decompose the functional $f_k$ satisfying eq.\r{b2*} as
$$
f_k(\varphi^1(\cdot),...,\varphi^k(\cdot)) =
\sum_{l_1...l_k=1}^{\infty} \int
d{\bf p}_1^1 ... d{\bf p}_{l_1}^1
...d{\bf p}_1^k ... d{\bf p}_{l_k}^k
f^k_{l_1;{\bf p}_1^1 ... {\bf p}_{l_1}^1;
...l_k,{\bf p}_1^k ... {\bf p}_{l_k}^k}
\Phi^{(l_1)}_{{\bf p}_1^1 ... {\bf p}_{l_1}^1}[\phi^1(\cdot)]
... \Phi^{(l_k)}_{{\bf p}_1^k ... {\bf p}_{l_k}^k}[\phi^k(\cdot)].
$$
One can  uniquely  specify the set \r{+3} of functionals by specifying
the set of functionals
\be
f^k_{l_1;{\bf p}_1^1 ... {\bf p}_{l_1}^1;
...l_k,{\bf p}_1^k ... {\bf p}_{l_k}^k}
\l{+4}
\ee
being symmetric under transpositions of ${\bf p}^m_i$
and ${\bf p}^m_j$, as well as under transpositions of sets
$l_m,{\bf p}_1^m ... {\bf p}_{l_m}^m$ and
$l_s,{\bf p}_1^s ...  {\bf p}_{l_s}^s$. The quantity $||f||^2$ can be
presented as
$$
||f||^2 = \sum_{k=0}^{\infty}
\sum_{l_1...l_k=1}^{\infty} \int
d{\bf p}_1^1 ... d{\bf p}_{l_1}^1
...d{\bf p}_1^k ... d{\bf p}_{l_k}^k
|f^k_{l_1;{\bf p}_1^1 ... {\bf p}_{l_1}^1;
...l_k,{\bf p}_1^k ... {\bf p}_{l_k}^k}|^2.
$$
Creation and annihilation operators can be decomposed as
\bea
A^+[\varphi(\cdot)] =  \sum_{n=0}^{\infty}  \int d{\bf k}_1 ...  d{\bf
k}_n \Phi^{(n)*}_{{\bf     k}_1...{\bf     k}_n}      [\varphi(\cdot)]
A^{+(n)}_{{\bf k}_1 ... {\bf k}_n}.
\nonumber \\
A^-[\varphi(\cdot)] =  \sum_{n=0}^{\infty}  \int d{\bf k}_1 ...  d{\bf
k}_n \Phi^{(n)}_{{\bf     k}_1...{\bf     k}_n}      [\varphi(\cdot)]
A^{-(n)}_{{\bf k}_1 ... {\bf k}_n}.
\l{b16}
\eea
The operators $
A^{\pm(n)}_{{\bf k}_1 ... {\bf k}_n}$ defined as
$$
A^{+(n)}_{{\bf k}_1 ... {\bf k}_n} = \int D\phi
A^+[\phi(\cdot)]
\Phi^{(n)}_{{\bf     k}_1...{\bf     k}_n}[\phi(\cdot)], \qquad
A^{-(n)}_{{\bf k}_1 ... {\bf k}_n} = \int D\phi
A^-[\phi(\cdot)]
\Phi^{*(n)}_{{\bf     k}_1...{\bf     k}_n}[\phi(\cdot)].
$$
create (annihilate) the field in the $n$-particle state  with  momenta
${\bf k}_1$, ..., ${\bf k}_n$. They are invariant under transpositions
of momenta ${\bf  k}_1$,  ...,  ${\bf  k}_n$  and  obey  the  ordinary
canonical commutation relations:
\be
[A^{\pm (m)}_{{\bf k}_1 ... {\bf k}_m}
,A^{\pm (n)}_{{\bf p}_1 ... {\bf p}_n}] = 0, \qquad
[A^{-(m)}_{{\bf k}_1 ... {\bf k}_m}
,A^{+(n)}_{{\bf p}_1 ... {\bf p}_n}] = 0, m\ne n
\ee
\be
[A^{-(m)}_{{\bf k}_1 ... {\bf k}_m },
A^{+(n)}_{{\bf p}_1 ... {\bf p}_n}] =
\frac{1}{n!} \sum_{\sigma}
\delta({\bf k}_1 - {\bf p}_{\sigma_1}) ...
\delta({\bf k}_n - {\bf p}_{\sigma_n}).
\l{b17}
\ee
Any vector $f$ can be written via creation operators and vacuum state
\bea
|0>= \left(
\matrix{
1 \nonumber
\\
0
\nonumber \\
...
\nonumber \\
0
\nonumber \\
...
}
\right)
\eea
as follows:
$$
f = \sum_{k=0}^{\infty} \frac{1}{\sqrt{k!}}
\sum_{l_1...l_k=1}^{\infty} \int
d{\bf p}_1^1 ... d{\bf p}_{l_1}^1
...d{\bf p}_1^k ... d{\bf p}_{l_k}^k
f^k_{l_1;{\bf p}_1^1 ... {\bf p}_{l_1}^1;
...l_k,{\bf p}_1^k ... {\bf p}_{l_k}^k}
A^{+(l_1)}_{{\bf p}_1^1 ... {\bf p}_{l_1}^1}
... A^{+(l_k)}_{{\bf p}_1^k ... {\bf p}_{l_k}^k}|0>.
$$
Making use of the quantum field theory formulas
\bea
:\int d{\bf x}
{\cal E}({\bf x}):
 = \int d{\bf k} \omega_{\bf k} a^+_{\bf k} a^-_{\bf k},
\nonumber \\
\phi^2({\bf x}) \Phi_0 = \frac{1}{(2\pi)^d} \int
\frac{d{\bf k}}{\sqrt{2\omega_{{\bf k}}}}
\frac{d{\bf p}}{\sqrt{2\omega_{{\bf p}}}}
e^{-i ({\bf k}+{\bf p}){\bf x}} a^+_{\bf k} a^+_{\bf p} \Phi_0,
\eea
one transforms expression \r{b12} to the following form:
\bea
\tilde{H} =
\sum_{n=1}^{\infty}   \int   d{\bf   k}_1   ...d   {\bf   k}_n
A^{+(n)}_{{\bf k}_1...{\bf k}_n}
(\omega_{{\bf k}_1} + ... + \omega_{{\bf k}_n} )
A^{-(n)}_{{\bf k}_1...{\bf k}_n}
\nonumber \\
+ \frac{\lambda}{4} \int d {\bf x}
\left(
\frac{\sqrt{2}}{(2\pi)^d} \int
\frac{d{\bf k}_1}{\sqrt{2\omega_{{\bf k}_1}}}
\frac{d{\bf k}_2}{\sqrt{2\omega_{{\bf k}_2}}}
(A^{+(2)}_{{\bf k}_1 {\bf k}_2} e^{-i ({\bf k}_1 + {\bf k}_2) {\bf x}}
+ A^{-(2)}_{{\bf k}_1 {\bf k}_2} e^{i ({\bf k}_1 + {\bf k}_2) {\bf x}})
\right)^2.
\l{b12*}
\eea
Analogously,
\bea
\tilde{P}^l=
\sum_{n=1}^{\infty}   \int   d{\bf   k}_1   ...d   {\bf   k}_n
A^{+(n)}_{{\bf k}_1...{\bf k}_n}
(k_1^l + ... + k_n^l )
A^{-(n)}_{{\bf k}_1...{\bf k}_n}
\nonumber \\
\tilde{M}^{ml}=
\sum_{n=1}^{\infty}   \int   d{\bf   k}_1   ...d   {\bf   k}_n
A^{+(n)}_{{\bf k}_1...{\bf k}_n}
\sum_{s=1}^n (k_s^l i \frac{\partial}{\partial k_s^m} -
k_s^m i \frac{\partial}{\partial k_s^l})
A^{-(n)}_{{\bf k}_1...{\bf k}_n}
\nonumber \\
\tilde{M}^{l0}=
\sum_{n=1}^{\infty}   \int   d{\bf   k}_1   ...d   {\bf   k}_n
A^{+(n)}_{{\bf k}_1...{\bf k}_n}
\sum_{s=1}^n (i \omega_{{\bf k}_s}
\frac{\partial}{\partial k_s^l}  + i
\frac{k_s^l}{2\omega_{{\bf k}_s}})
A^{-(n)}_{{\bf k}_1...{\bf k}_n}
\nonumber \\
+ \frac{\lambda}{4} \int d {\bf x} x^l
\left(
\frac{\sqrt{2}}{(2\pi)^d} \int
\frac{d{\bf k}_1}{\sqrt{2\omega_{{\bf k}_1}}}
\frac{d{\bf k}_2}{\sqrt{2\omega_{{\bf k}_2}}}
(A^{+(2)}_{{\bf k}_1 {\bf k}_2} e^{-i ({\bf k}_1 + {\bf k}_2) {\bf x}}
+ A^{-(2)}_{{\bf k}_1 {\bf k}_2} e^{i ({\bf k}_1 + {\bf k}_2) {\bf x}})
\right)^2.
\l{b13*}
\eea

\subsection{Decomposition of the state space}

We see  that  the  Hilbert  space  of  the  $N=\infty$-theory  can  be
presented as
\be
{\cal F} (\oplus_{n=1}^{\infty} {\cal H}^{\vee n})
\l{b20}
\ee
(the notations of Appendix A are used),  where $\cal H$ is a space  of
complex functions   $f_{\bf   k}$,  ${\bf  k}  \in  {\bf  R}^d$  from
$L^2({\bf R}^d)$.  Analogously to lemma A.8,  the  space  \r{b20}  is
isomorphic to
\be
{\cal F} ({\cal H}^{\vee 2}) \otimes {\cal F}({\cal H} +
\oplus_{n=3}^{\infty} {\cal H}^{\vee n}) \equiv
{\cal F} \otimes \breve{\cal F},
\l{b21}
\ee
while the operators  \r{b12*}  and  \r{b13*}  can  be  viewed  as  the
following operators in the space \r{b21}:
\bea
\tilde{H} = H \otimes 1 + 1 \otimes \breve{H}, \qquad
\tilde{P}^k = P^k \otimes 1 + 1 \otimes \breve{P}^k, \nonumber \\
\tilde{M}^{ml} = M^{ml} \otimes 1 + 1 \otimes \breve{M}^{ml}, \qquad
\tilde{M}^{k0} = M^{k0} \otimes 1 + 1 \otimes \breve{M}^{k0}.
\l{b22}
\eea
The operators   $\breve{H}$,   $\breve{P}^k$,   $\breve{M}^{kl}$   and
$\breve{M}^{k0}$ are the same as in the free theory:
\bea
\breve{H} =
\sum_{n=1,3,4,...}  \int   d{\bf   k}_1   ...d   {\bf   k}_n
A^{+(n)}_{{\bf k}_1...{\bf k}_n}
(\omega_{{\bf k}_1} + ... + \omega_{{\bf k}_n} )
A^{-(n)}_{{\bf k}_1...{\bf k}_n}
\nonumber \\
\breve{P}^l=
\sum_{n=1,3,4...} \int   d{\bf   k}_1   ...d   {\bf   k}_n
A^{+(n)}_{{\bf k}_1...{\bf k}_n}
(k_1^l + ... + k_n^l )
A^{-(n)}_{{\bf k}_1...{\bf k}_n}
\nonumber \\
\breve{M}^{ml}=
\sum_{n=1,3,4,...} \int   d{\bf   k}_1   ...d   {\bf   k}_n
A^{+(n)}_{{\bf k}_1...{\bf k}_n}
\sum_{s=1}^n (k_s^l i \frac{\partial}{\partial k_s^m} -
k_s^m i \frac{\partial}{\partial k_s^l})
A^{-(n)}_{{\bf k}_1...{\bf k}_n}
\nonumber \\
\tilde{M}^{l0}=
\sum_{n=1,3,4,...}   \int   d{\bf   k}_1   ...d   {\bf   k}_n
A^{+(n)}_{{\bf k}_1...{\bf k}_n}
\sum_{s=1}^n (i \omega_{{\bf k}_s}
\frac{\partial}{\partial k_s^l}  + i
\frac{k_s^l}{2\omega_{{\bf k}_s}})
A^{-(n)}_{{\bf k}_1...{\bf k}_n}.
\l{b22*}
\eea
The only nontrivial part of operators \r{b22} correspond to the  space
${\cal F}({\cal H}^{\vee 2})$:
\bea
{H} =
 \int   d{\bf   k}_1  d   {\bf   k}_2
A^{+(2)}_{{\bf k}_1{\bf k}_2}
(\omega_{{\bf k}_1} + \omega_{{\bf k}_2} )
A^{-(2)}_{{\bf k}_1{\bf k}_2}
\nonumber \\
+ \frac{\lambda}{4} \int d {\bf x}
\left(
\frac{\sqrt{2}}{(2\pi)^d} \int
\frac{d{\bf k}_1}{\sqrt{2\omega_{{\bf k}_1}}}
\frac{d{\bf k}_2}{\sqrt{2\omega_{{\bf k}_2}}}
(A^{+(2)}_{{\bf k}_1 {\bf k}_2} e^{-i ({\bf k}_1 + {\bf k}_2) {\bf x}}
+ A^{-(2)}_{{\bf k}_1 {\bf k}_2} e^{i ({\bf k}_1 + {\bf k}_2) {\bf x}})
\right)^2.
\nonumber \\
\tilde{P}^l=
 \int   d{\bf   k}_1 d   {\bf   k}_2
A^{+(2)}_{{\bf k}_1{\bf k}_2}
(k_1^l + k_2^l )
A^{-(2)}_{{\bf k}_1{\bf k}_2}
\nonumber \\
{M}^{ml}=
\int   d{\bf   k}_1 d   {\bf   k}_2
A^{+(2)}_{{\bf k}_1 {\bf k}_2}
\sum_{s=1}^2 (k_s^l i \frac{\partial}{\partial k_s^m} -
k_s^m i \frac{\partial}{\partial k_s^l})
A^{-(2)}_{{\bf k}_1 {\bf k}_2}
\nonumber \\
{M}^{l0}=
\int   d{\bf   k}_1 d   {\bf   k}_2
A^{+(2)}_{{\bf k}_1 {\bf k}_2}
\sum_{s=1}^2 (i \omega_{{\bf k}_s}
\frac{\partial}{\partial k_s^l}  + i
\frac{k_s^l}{2\omega_{{\bf k}_s}})
A^{-(2)}_{{\bf k}_1{\bf k}_2}
\nonumber \\
+ \frac{\lambda}{4} \int d {\bf x} x^l
\left(
\frac{\sqrt{2}}{(2\pi)^d} \int
\frac{d{\bf k}_1}{\sqrt{2\omega_{{\bf k}_1}}}
\frac{d{\bf k}_2}{\sqrt{2\omega_{{\bf k}_2}}}
(A^{+(2)}_{{\bf k}_1 {\bf k}_2} e^{-i ({\bf k}_1 + {\bf k}_2) {\bf x}}
+ A^{-(2)}_{{\bf k}_1 {\bf k}_2} e^{i ({\bf k}_1 + {\bf k}_2) {\bf x}})
\right)^2.
\l{b23}
\eea
The operators \r{b22} correspond to the representation of the Poincare
group in ${\cal F} \otimes \breve{\cal F}$ of the form:
$$
\tilde{U}_{\Lambda,a} = U_{\Lambda,a} \otimes \breve{U}_{\Lambda,a}
$$
with
$$
U_{\Lambda, a} = \exp  (iP_{\mu}a^{\mu})  \exp(\frac{i}{2}  M^{\Lambda
\mu} \theta_{\lambda \mu} ),\qquad
\breve{U}_{\Lambda, a} = \exp  (i \breve{P}_{\mu}a^{\mu})
\exp(\frac{i}{2}  \breve{M}^{\lambda
\mu} \theta_{\lambda \mu})
$$
To express the operators $\breve{U}_{\Lambda,a}$,  it is convenient to
introduce the operators $u_{\Lambda,a}$ of the unitary
representation of the Poincare group in ${\cal H}$:
\be
(u_{\Lambda,a} f)_{\bf k} = \exp(i\omega_{\bf k} a_0 - i {\bf  k}  {\bf
a})
\sqrt{\frac{(\Lambda^{-1})^0_nk^n + (\Lambda^{-1})^0_0\omega_{\bf k}}
{\omega_{\bf k}}} f_{
(\Lambda^{-1})^m_n k^n + (\Lambda^{-1})^m_0 \omega_{\bf k} }
\l{b23x}
\ee
with generators
\bea
p^l = k^l,\qquad p^0 = \omega_{\bf k}\qquad
m^{l0} =     i(\omega_{\bf     k}\frac{\partial}{\partial    k^l}    +
\frac{k^l}{2\omega_{\bf k}}), \nonumber \\
m^{ln} = i (k^n \frac{\partial}{\partial k^l } -
k^l \frac{\partial}{\partial k^n } ).
\eea
By $\tilde{u}_{\Lambda,a}  :
{\cal  H}  \oplus  \oplus_{n=3}^{\infty}
{\cal H}^{\vee n} \to
{\cal  H}  \oplus  \oplus_{n=3}^{\infty}
{\cal H}^{\vee n} $ we denote the operator
$$
\tilde{u}_{\Lambda,a} (f_1,f_3, f_4,...) = (u_{\Lambda,a}f_1,
u_{\Lambda,a}^{\otimes 3} f_3,
u_{\Lambda,a}^{\otimes 4} f_4,...).
$$
We can notice that
$$
\breve{U}_{\Lambda,a} = {\cal U}(\tilde{u}_{\Lambda,a})
$$
(the notations of Appendix A are used).

Thus, the operators $\breve{U}_{\Lambda,a}$ are constructed.  The only
nontrivial problem  is to construct the representation of the Poincare
group corresponding to the generators \r{b23}.

\subsection{Problem of divergences}

\subsubsection{The Haag theorem and volume divergences}

Apply the Hamiltonian \r{b23} to the vacuum
state. The result  will be
$$
H|0> = \frac{\lambda}{4} \frac{2}{(2\pi)^d}
\int
\frac{d{\bf k}_1}{\sqrt{2\omega_{{\bf k}_1}}}
\frac{d{\bf k}_2}{\sqrt{2\omega_{{\bf k}_2}}}
\frac{d{\bf p}_1}{\sqrt{2\omega_{{\bf p}_1}}}
\frac{d{\bf p}_1}{\sqrt{2\omega_{{\bf p}_2}}}
A^+_{{\bf k}_1{\bf k}_2}
A^+_{{\bf p}_1{\bf p}_2}
\delta({\bf k}_1+{\bf k}_2+{\bf p}_1+{\bf p}_2) |0>,
$$
where
$
A^+_{{\bf k}_1{\bf k}_2} \equiv
A^{+(2)}_{{\bf k}_1{\bf k}_2}
$.
Because of  the  $\delta$-function,  the  quantity $||H\Phi^{(0)}||^2$
diverges. This is a volume divergence associated with the Haag theorem
(see, for example,  \c{BLOT}).  An analogous infinite quantity appears
when one  applied  the  perturbation  theory  in  $\lambda$  for   the
evolution operator.

Within the  perturbation theory,  such difficulty can be resolved with
the help of the  Faddeev  transformation  \c{F}.

\subsubsection{The Stueckelberg divergences}

Even after  removing  the  vacuum  divergences,  the  problem  is  not
completely resolved.  If one considers the perturbation theory for the
Schrodinger equation  of motion,  one finds that there are
UV-divergences even in the tree approximation.  Namely,  for the first
order of the perturbation theory one has
\bea
e^{iH_0t} \tilde{H}_1 e^{-iH_0t} \equiv
\tilde{H}_1(t) = \frac{1}{(2\pi)^d}
\int
\frac{d{\bf k}_1}{\sqrt{2\omega_{{\bf k}_1}}}
\frac{d{\bf k}_2}{\sqrt{2\omega_{{\bf k}_2}}}
\frac{d{\bf p}_1}{\sqrt{2\omega_{{\bf p}_1}}}
\frac{d{\bf p}_1}{\sqrt{2\omega_{{\bf p}_2}}}
A^+_{{\bf k}_1{\bf k}_2}
A^-_{{\bf p}_1{\bf p}_2}
\delta({\bf k}_1+ {\bf k}_2 - {\bf p}_1 -{\bf p}_2)
\nonumber \\
\times
e^{-it
(\omega_{{\bf k}_1}
\omega_{{\bf k}_2}+
\omega_{{\bf p}_1}+
\omega_{{\bf p}_2})}.
\eea
Applying the first-order evolution operator
$$
U_t = -i\lambda \int_0^t d\tau \tilde{H}_1(\tau)
$$
to the vector
$$
\Phi^0 =  \int  d{\bf  p}_1  d{\bf  p}_2  A^+_{{\bf   p}_1{\bf   p}_2}
\Phi^0_{{\bf p}_1{\bf p}_2},
$$
we find
$$
U_t\Phi^0 =  \int  d{\bf  p}_1  d{\bf  p}_2  A^+_{{\bf   p}_1{\bf   p}_2}
\Phi^t_{{\bf p}_1{\bf p}_2}
$$
with
$$
\Phi^t_{{\bf k}_1{\bf k}_2} =
\frac{1}{(2\pi)^d} \frac{1}{\sqrt{2\omega_{{\bf k}_1}}}
\frac{1}{\sqrt{2\omega_{{\bf k}_2}}}
\int
\frac{1}{\sqrt{2\omega_{{\bf p}_1}}}
\frac{1}{\sqrt{2\omega_{{\bf p}_2}}}
\delta({\bf k}_1+ {\bf k}_2 - {\bf p}_1 - {\bf p}_2)
\Phi^{(0)}_{{\bf p}_1{\bf p}_2}
\frac{
e^{-it
(\omega_{{\bf k}_1}+
\omega_{{\bf k}_2}-
\omega_{{\bf p}_1}-
\omega_{{\bf p}_2})} - 1
}
{i
(\omega_{{\bf k}_1}+
\omega_{{\bf k}_2}-
\omega_{{\bf p}_1}-
\omega_{{\bf p}_2})
}.
$$
The integral
$$
\int d{\bf k}_1 d{\bf k}_2 |\Phi^t_{{\bf k}_1{\bf k}_2}|^2
$$
diverges for $d\ge 4$.  This is a Stueckelberg  divergence.

\section{Construction of the Hamiltonian}

The purpose  of this section is to define mathematically the operators
in the  Hilbert  space  that  corresponds  to  the  formal  expression
\r{b23}. For  $d+1=4,5$,  it  is  necessary  to  perform  the infinite
renormalization of the coupling constant, for $d+1\ge 6$, the model is
nonrenormalizable.
Subsections A-C deal with heuristic construction of  the  Hamiltonian;
in subsection D representation for space-time translations and spatial
rotations is constructed.

\subsection{Diagonalization of the Hamiltonian}

Since the Hamiltonian \r{b23} is quadratic with  respect  to  creation
and annihilation    operators,   one   can   perform   the   canonical
transformation of creation and annihilation operators in order to take
the Hamiltonian to the canonical form.

It is convenient to introduce new variables,
\bea
Q_{{\bf P}{\bf s}} = \frac{1}{\sqrt{2\Omega_{{\bf P}{\bf s}}}}
(
A^+_{{\bf P}/2-{\bf s},{\bf P}/2+{\bf s}}+
A^-_{-{\bf P}/2-{\bf s},-{\bf P}/2+{\bf s}}
),
\nonumber \\
\Pi_{{\bf P}{\bf s}} = i \sqrt{\frac{\Omega_{{\bf P}{\bf s}}}{2}}
(
A^+_{{\bf P}/2-{\bf s},{\bf P}/2+{\bf s}} -
A^-_{-{\bf P}/2-{\bf s},-{\bf P}/2+{\bf s}}
).
\l{d0a}
\eea
Here
\be
\Omega_{{\bf P}{\bf s}}= \omega_{{\bf P}/2-{\bf s}} +
\omega_{{\bf P}/2-{\bf s}}.
\l{d1a}
\ee
The operators \r{d0a} obey the properties:
$$
Q_{{\bf P}{\bf s}} = Q_{{\bf P},-{\bf s}}= Q^*_{-{\bf P},{\bf s}},
\Pi_{{\bf P}{\bf s}} = \Pi_{{\bf P},-{\bf s}}= \Pi^*_{-{\bf P},{\bf s}}
$$
and canonical commutation relations:
$$
[Q_{{\bf P}{\bf s}}, Q_{{\bf P}'{\bf s}'} ] =0, \qquad
[\Pi_{{\bf P}{\bf s}}, \Pi_{{\bf P}'{\bf s}'} ] =0, \qquad
[Q_{{\bf P}{\bf s}}, \Pi_{{\bf P}'{\bf s}'} ] =
i \delta_{{\bf P}{\bf P}'} \frac{1}{2} (\delta_{{\bf s}-{\bf s}'}
+ \delta_{{\bf s}+{\bf s}'}).
$$
The Hamiltonian takes the following form up to an additive constant:
\be
H = \frac{1}{2} \int d{\bf P} (\Pi_{\bf P}, \Pi_{\bf P}) +
\frac{1}{2} \int d{\bf P} (Q_{\bf P},(M_P)^2 {\bf Q}_{\bf P}).
\l{d1}
\ee
Here $\Pi_{\bf P}$,  $Q_{\bf P}$ are operator-valued even functions of
the variable  ${\bf  s}$.  The  inner  product  $(f,g)$  of  two  even
functions $f_{\bf  s}$  and $g_{\bf s}$ is,  as usual,  $\int d{\bf s}
f_{\bf s}^* g_{\bf s}$.  $(M_{\bf P})^2$
is the following  operator  in  the
space of even functions:
\be
((M_{\bf P})^2 \varphi)_{\bf s} = \Omega_{{\bf P}{\bf s}}^2  \varphi_s
+ \frac{\lambda}{(2\pi)^d} \int d{\bf s}'
\sqrt{\frac{2\Omega_{{\bf P}{\bf s}}}
{2\omega_{{\bf P}/2+{\bf s}}2\omega_{{\bf P}/2-{\bf s}} }}
\sqrt{\frac{2\Omega_{{\bf P}{\bf s'}}}
{2\omega_{{\bf P}/2+{\bf s}'}2\omega_{{\bf P}/2-{\bf s}'} }}
\varphi_{{\bf s}'}
\l{d2}
\ee
The operator \r{d1} can be diagonalized by the following procedure:
\bea
Q_{{\bf P}}  =  \frac{1}{\sqrt{2M_{\bf P}}  } (C^+_{\bf P} + C^-_{-{\bf
P}}),
\nonumber \\
\Pi_{\bf P} = i \sqrt{\frac{M_{\bf P}}{2}} (C^+_{\bf P}  -  C^-_{-{\bf
P}}).
\l{d2*}
\eea
where $C_{\bf P}^{\pm}$ are operator-valued  functions  $C^{\pm}_{{\bf
P}{\bf s}}$  of the variable ${\bf s}$.  They obey the usual canonical
commutation relations:
$$
[C^-_{{\bf P}{\bf s}}, C^+_{{\bf P}{\bf s}} ] =
\delta_{{\bf P}{\bf P}'}
\frac{1}{2} (
\delta_{{\bf s}-{\bf s}'} + \delta_{{\bf s}+{\bf s}'}), \qquad
[C^{\pm}_{{\bf P}{\bf s}}, C^{\pm}_{{\bf P}{\bf s}} ] = 0.
$$
The Hamiltonian takes the form:
\bea
H = \int d{\bf P} d{\bf s} d{\bf s}'
C^+_{{\bf P}{\bf s}}
(M_{\bf P})_{{\bf s}{\bf s}'}
C^-_{{\bf P}{\bf s}'}.
\l{d3}
\eea
The $(M_{\bf P})_{{\bf  s}{\bf  s}'}$  is  a  matrix  element  of  the
operator $M_{\bf P}$.

One should use then another, non-Fock representation for the operators
$A^{\pm}_{{\bf k}_1{\bf k}_2}$,  which is Fock representation for  the
transformed operators  $C^{\pm}_{{\bf  P}{\bf  s}}$.  The  Hamiltonian
\r{d3} is then a self-adjoint operator if $M_{\bf P}$ is self-adjoint.
The evolution   operator   is   expressed  via  the  unitary  operator
$e^{-iM_{\bf P}t}$.  To construct $M_{\bf P}$,  one should first check
that $(M_{\bf  P})^2$  is  a positively definite self-adjoint operator
and define $M_{\bf P}\equiv \sqrt{(M_{\bf P})^2}$,  making use of  the
functional calculus  of self-adjoint operators.  The operator $(M_{\bf
P})^{1/2}$ entering  to  expression  \r{d2*}  can  be  constructed  in
analogous way.

\subsection{Definition of the Hamiltonian and its properties}

Formula \r{d2}  for  the  operator $(M_{\bf P})^2$ is not well-defined
since the vector
$$
\sqrt{\frac{2\Omega_{{\bf P}{\bf s}}}
{2\omega_{{\bf P}/2+{\bf s}'}2\omega_{{\bf P}/2-{\bf s}'} }}
$$
considered as  a  function of ${\bf s}$ does not belong to $L^2$.  The
operator \r{d2} is  therefore  analogous  to  the  quantum  mechanical
Hamiltonian corresponding  to  the  particle  moving  in  the singular
potential like $\delta$-function. The theory of singular potentials is
developed in \c{BF}.

To construct  mathematically  the  (unbounded)  self-adjoint  operator
$(M_{\bf P})^2$  one  may  first  construct  the  (bounded)   operator
$(M_{\bf P})^{-2}$,   prove  that  it  is  invertible  and  positively
definite. Then the operator $(M_{\bf P})^2$  is  defined  as  $(M_{\bf
P})^2 \equiv ((M_{\bf P})^{-2})^{-1}$.

To find the vector
$$
\varphi = (M_{\bf P})^{-2} \psi,
$$
one should  solve  the equation $\psi= (M_{\bf P})^2 \varphi$.  It has
the following form:
\bea
\psi_{\bf s} = \Omega^2_{{\bf P}{\bf s}} \varphi_{\bf s} + c
\sqrt{\frac{2\Omega_{{\bf P}{\bf s}}}
{2\omega_{{\bf P}/2+{\bf s}}2\omega_{{\bf P}/2-{\bf s}} }},
\nonumber \\
c = \frac{\lambda}{(2\pi)^d} \int d{\bf s}
\sqrt{\frac{2\Omega_{{\bf P}{\bf s}}}
{2\omega_{{\bf P}/2+{\bf s}}2\omega_{{\bf P}/2-{\bf s}} }}
\varphi_{\bf s}.
\l{d4}
\eea
Eqs. \r{d4} imply that
\be
c = \frac{\lambda_R^{\bf P}}{(2\pi)^d} \int d{\bf s}
\sqrt{\frac{2\Omega_{{\bf P}{\bf s}}}
{2\omega_{{\bf P}/2+{\bf s}}2\omega_{{\bf P}/2-{\bf s}} }}
\frac{1}{\Omega^2_{{\bf P}{\bf s}}}
\psi_{\bf s}
\l{d5}
\ee
with the   "renormalized"   coupling   constant   $\lambda_R^{\bf  P}$
expressed from the relation
\be
\frac{1}{\lambda_R^{\bf P}}
= \frac{1}{\lambda}
+ \frac{1}{(2\pi)^d}
\int d{\bf s}
{\frac{2\Omega_{{\bf P}{\bf s}}}
{2\omega_{{\bf P}/2+{\bf s}}2\omega_{{\bf P}/2-{\bf s}} }}
\frac{1}{\Omega^2_{{\bf P}{\bf s}}}.
\l{d6}
\ee
The operator $(M_{\bf P})^{-2}$ has then the form
\be
((M_{\bf P})^{-2}\psi)_{\bf s} =
\frac{1}{\Omega^2_{{\bf P}{\bf s}}}\psi_{\bf s}
- \frac{\lambda_R^{\bf P}}{(2\pi)^d}
\chi_{{\bf P}{\bf s}}
\int d{\bf s}' \chi_{{\bf P}{\bf s}'}\psi_{{\bf s}'}
\l{d7}
\ee
with
\be
\chi_{{\bf P}{\bf s}} =
\frac{1}{\Omega^2_{{\bf P}{\bf s}}}.
\sqrt{\frac{2\Omega_{{\bf P}{\bf s}}}
{2\omega_{{\bf P}/2+{\bf s}}2\omega_{{\bf P}/2-{\bf s}} }}.
\l{d8}
\ee
The function \r{d8} treated as a function  of  ${\bf  s}$  belongs  to
$L^2$ for $d+1<6$.  For these values of the space-time dimensionality,
the operator $(M_{\bf P})^{-2}$ is bounded and self-adjoint,  provided
that the  quantity  $\lambda_R^{\bf P}$ is finite.  Since the integral
entering to the right-hand side of eq.\r{d6}  diverges  at  $d+1=4,5$,
for these   values   of  $d$  it  is  necessary  to  perform  infinite
renormalization of the coupling constant.  This means  that  $\lambda$
should be chosen in such a way that $|\lambda_R^{\bf P}|<\infty$.  The
fact that $\lambda$ is ${\bf P}$-independent means that
\be
\frac{1}{\lambda_R^{{\bf P}_1}}
- \frac{1}{\lambda_R^{{\bf P}_2}}
= \frac{1}{(2\pi)^d}
\int d{\bf s}
\left[
\frac{1}{2\Omega_{{\bf P}_1{\bf s}} \omega_{{\bf P}_1/2+{\bf s}}
 \omega_{{\bf P}_1/2-{\bf s}} }
-
\frac{1}{2\Omega_{{\bf P}_2{\bf s}} \omega_{{\bf P}_2/2+{\bf s}}
 \omega_{{\bf P}_2/2-{\bf s}} }
\right]
\l{d9}
\ee
Note that  the  integral  in  the  right-hand  side  of  eq.\r{d9}  is
well-defined at $d+1=4,5$, since
$$
\frac{\partial}{\partial {\bf P}}
\left(
\frac{1}{2\Omega_{{\bf P}_2{\bf s}} \omega_{{\bf P}_2/2+{\bf s}}
 \omega_{{\bf P}_2/2-{\bf s}} }
\right) = O(|{\bf s}|^{-5}), s\to\infty.
$$

The fact  that  the operator \r{d7} is invertible can be understood as
follows. Suppose that $(M_{\bf P})^{-2}\psi=0$ for some  $\psi$.  This
means that
\be
\psi_{\bf s} = c\Omega^2_{{\bf P}{\bf s}} \chi_{{\bf P}{\bf s}}
\l{d10}
\ee
for some multiplier $c$.  But the function \r{d10} does not belong  to
$L^2$. Thus, the operator $(M_{\bf P})^{-2}$ is invertible.

To investigate  the  positive  definiteness  of  the operator $(M_{\bf
P})^{-2}$, calculate the integral
$$
I({\bf P},{{\varepsilon}}) = \frac{1}{(2\pi)^d} \int
\frac{d{\bf s}}
{2 \omega_{{\bf P}_2/2+{\bf s}}
 \omega_{{\bf P}_2/2-{\bf s}} }
\frac{1}{\Omega^2_{{\bf P}_2{\bf s}}+ {\varepsilon}^2},
$$
making use of the dimensional regularization.  First of all, introduce
new variables, ${\bf k}_1 = {\bf P}/2 + {\bf s}$,
${\bf k}_2 = {\bf P}/2 - {\bf s}$, so that $\int d{\bf s} \to
\int d{\bf k}_1 d{\bf k}_2 \delta({\bf k}_1 + {\bf k}_2 - {\bf P})$.
Next, use the identity
$$
\frac{\omega_1 +  \omega_2}{2\omega_1  \omega_2   ({\varepsilon}^2   +
(\omega_1+\omega_2)^2)} = \frac{1}{2\pi} \int \frac{d\xi}
{(\omega_1^2+\xi^2)(\omega_2^2+ (\xi-{\varepsilon})^2)},
$$
so that
$$
I({\bf P},{\varepsilon}) =
\frac{1}{(2\pi)^{d+1}} \int
\frac{d{\bf k}_1 d{\bf k}_2
d{\xi} \delta({\bf k}_1+{\bf k}_2-{\bf P})}
{({\xi}^2+ \omega_{{\bf k}_1}^2)
(({\xi}-{\varepsilon})^2+ \omega_{{\bf k}_2}^2) }.
$$
Introduce, as   usual,   the   $\alpha$-representation:   $a^{-1}   =
\int_0^{\infty} d\alpha e^{-\alpha a}$.
We get
\be
I({\bf P},{\varepsilon}) =
\frac{1}{(2\pi)^{d+1}}
\int_0^{\infty}
d\alpha
\int_0^{\infty} d\beta
e^{-\mu^2(\alpha+\beta)} \left(
\frac{\pi}{\alpha+\beta}
\right)^{\frac{d+1}{2}}
e^{-\frac{\alpha\beta}{\alpha+\beta}({\bf P}^2+ {\varepsilon}^2)}.
\l{d11*}
\ee
Therefore,
\be
\frac{1}{\lambda_R^{\bf P}} -
\frac{1}{\lambda_R^{0}} =
\frac{1}{(2\pi)^{d+1}}
\int_0^{\infty}
d\alpha
\int_0^{\infty} d\beta
e^{-\mu^2(\alpha+\beta)} \left(
\frac{\pi}{\alpha+\beta}
\right)^{\frac{d+1}{2}}
(e^{-\frac{\alpha\beta}{\alpha+\beta}{\bf P}^2}-1)<0.
\l{d12}
\ee
The requirement  $(M_{\bf  P})^{-2}  \ge  0$  is  a  corollary  of the
condition $\lambda_R^{\bf P} <0$.  Inequality \r{d12} implies that  it
is sufficient   to  require  $\lambda_R  <0$.  This  is  a  well-known
condition of absence of tachyons \c{T} in the large-$N$ theory.

Thus, we have constructed the Hamiltonian.

Another way to define the Hamiltonian is the following \c{BF}. One can
use the   theory  of  self-adjoint  extensions  \c{AG}.  Consider  the
operator $\Omega^2_{{\bf P}{\bf s}}$ defined on the domain  consisting
of such $\varphi$ that
$$
\int d{\bf s} \varphi_{\bf s}
\sqrt{
\frac{2\Omega_{{\bf P}{\bf s}}}{2\omega_{{\bf P}/2+{\bf s}}
2\omega_{{\bf P}/2-{\bf s}}}
}
=0.
$$
If $d+1  \ge  6$,  the  operator  is  essentially  self-adjoint.  This
corresponds to  the  "triviality"  (or  nonrenormalizability)  of  the
model. If $d+1<6$,  there is a one-parametric family  of  self-adjoint
extensions specified  by  the parameter $\lambda_R^{\bf P}$.  However,
the condition \r{d9} cannot be obtained by the self-adjoint  extension
method. One should use another argumentation like Poincare invariance.

\subsection{Momentum and angular momentum}

Let us express the momentum and angular  momentum  operators  via  new
creation and  annihilation  operators  $C^{\pm}_{{\bf P}{\bf s}}$.  It
follows from eqs.\r{d0a} that operators \r{b23} can be written as
\bea
P^l = \int d{\bf P} d{\bf s}  Q_{{\bf  P}{\bf  s}}  P^l  i  \Pi_{-{\bf
P}{\bf s}}, \nonumber \\
M^{ml} =  \int  d{\bf  P}  d{\bf   s}   Q_{{\bf   P}{\bf   s}}
(iP^l
\frac{\partial}{\partial P^m}  +  is^l \frac{\partial}{\partial s^m} -
iP^m \frac{\partial}{\partial P^l} -  i  s^m  \frac{\partial}{\partial
s^l} )
i\Pi_{-{\bf P}{\bf s}}.
\eea
Since the   kernel  of  the  operator  $M_{{\bf  P}}^{-2}$  \r{d7}  is
invariant under spatial rotations
$$
{\bf P} \to O{\bf P},\qquad
{\bf s} \to O{\bf s},\qquad
{\bf s}' \to O{\bf s}'
$$
with orthogonal matrix $O$,  it commutes with the rotation operator of
the form   $Of_{{\bf   P}{\bf   s}}   =  f_{{O}{\bf  P},O{\bf  s}  }$.
Analogously, any function of $M$ obey this property. Since the operator
$$
iP^l
\frac{\partial}{\partial P^m}  +  is^l \frac{\partial}{\partial s^m} -
iP^m \frac{\partial}{\partial P^l} -  i  s^m  \frac{\partial}{\partial
s^l}
$$
is a  generator  of a rotation,  it commutes with any function of $M$.
Making use of this property, we find
$$
M^{ml} = \int d{\bf P} d{\bf s} C^+_{{\bf P}{\bf s}} (iP^l
\frac{\partial}{\partial P^m}  +  is^l \frac{\partial}{\partial s^m} -
iP^m \frac{\partial}{\partial P^l} -  i  s^m  \frac{\partial}{\partial
s^l}) C^-_{{\bf P}{\bf s}}.
$$
Analogously,
$$
P^l = \int d{\bf P} d{\bf s} C^+_{{\bf P}{\bf s}} P^l C^-_{{\bf P}{\bf
s}}.
$$

\subsection{Representation for space-time translations and space rotations}

The problem  of  divergences  made us to change the representation for
the operators $A^{\pm(2)}_{{\bf k}_1 {\bf k}_2}$. We have considered the
space ${\cal  H}_2  \subset  L^2({\bf  R}^{2d})$ of functions $f_{{\bf
P}{\bf s}}$ which obey the property $f_{{\bf P},-{\bf  s}}  =  f_{{\bf
P},{\bf s}}$.  The  space  ${\cal  F}({\cal H}_2)$ has been considered
instead of ${\cal F}({\cal H}^{\vee 2})$. Therefore, the space \r{b21}
is substituted  by ${\cal F}({\cal H}_2) \otimes \breve{\cal F}$.  One
should define  then  operators  $H$,  $P^l$,  $M^{ml}$,  $M^{m0}$  and
$U_{\Lambda,a}$ in  ${\cal  F}({\cal H}_2)$ that corresponds to formal
expressions \r{b23}.

The operators $H$, $P^l$, $M^{ml}$ have been considered above.

Let $\lambda_R^0$ be a fixed negative  quantity.  Set  ${\bf  P}_2=0$,
${\bf P}_1   =   {\bf   P}$  in  eq.\r{d9}  and  define  the  quantity
$\lambda_R^{\bf P}$.  Consider the operator $M^{-2}:  {\cal  H}_2  \to
{\cal H}_2$ of the form
$$
(M^{-2}\psi)_{{\bf P}{\bf   s}}   =   \Omega_{{\bf   P}{\bf   s}}^{-2}
\psi_{{\bf P}{\bf s}} - \frac{\lambda_R^{\bf P}}{(2\pi)^d}  \chi_{{\bf
P}{\bf s}}  \int  d{\bf  s}'  \chi_{{\bf P}{\bf s}'} \psi_{{\bf P}{\bf
s}'},
$$
where $\chi_{{\bf P}{\bf s}}$ has the form \r{d8}.  Since the operator
$M^{-2}$ is  positively  definite  and self-adjoint,  the self-adjoint
positive operator $M \equiv (M^{-2})^{-1/2}$ is uniquely defined.  The
Hamiltonian operator is
$$
H = {\cal F} (M)
$$
(the notations from appendix A are used), while
$$
e^{-iHt} = {\cal U}(e^{-iMt})
$$
Analogously,
$$
P^l = {\cal F}(P^l),
$$
$$
M^{ml} = {\cal F}(iP^l
\frac{\partial}{\partial P^m}  +  is^l \frac{\partial}{\partial s^m} -
iP^m \frac{\partial}{\partial P^l} -  i  s^m  \frac{\partial}{\partial
s^l}).
$$
The space rotations being Lorentz transformations with
\be
\Lambda_i^0 =0, \qquad
\Lambda_0^i =0,\qquad
\Lambda_0^0 =1
\l{xx1}
\ee
are represented   by   the   operators    $U_{\Lambda,0}    =    {\cal
U}(u_{\Lambda,0})$ with
\be
(u_{\Lambda,0}f)_{{\bf P}{\bf    s}}    =    f_{\Lambda^{-1}{\bf   P},
\Lambda^{-1}{\bf s}}.
\l{xx1*}
\ee
For space-time translations, one has
\be
U_{1,a} =  e^{iHt}  e^{-iP^la^l}  =  {\cal U} (e^{iMt}e^{-i{\bf P}{\bf
a}}).
\l{xxx}
\ee
Thus, we  have constructed the operators $U_{\Lambda,a}$ corresponding
to the Poincare transformations obeying eq.\r{xx1}:  $U_{\Lambda,a}  =
U_{0,a} U_{\Lambda,0}$.

{\bf Lemma 3.1.} {\it The group property
$$
U_{\Lambda_1,a_1} U_{\Lambda_2,a_2}
=
U_{\Lambda_1\Lambda_2,a_1 + \Lambda_1 a_2}
$$
is satisfied for Poincare transformations obeying eq.\r{xx1}.}

{\bf Proof.} It is sufficient to show that
\bea
U_{\Lambda_1\Lambda_2, 0 } = U_{\Lambda_1,0} U_{\Lambda_2,0}
\l{xx2} \\
U_{1,a_1} U_{1,a_2} = U_{1,a_1+a_2},\l{xx3} \\
U_{\Lambda,0} U_{1,a} U^{-1}_{\Lambda,0} = U_{1,\Lambda a}.
\l{xx4}
\eea
The property  \r{xx2}  is  an  obvious  corollary  of  the  definition
\r{xx1*}. Relation \r{xx3} is a corollary of the Stone theorem and the
property
\be
[e^{iMt}, e^{-i{\bf P}{\bf a}}] =0.
\l{xx5}
\ee
Definition \r{xx1*} and commutation relation
\be
[u_{\Lambda,0}, e^{iMt}] = 0
\ee
imply property \r{xx4}. Lemma 3.1 is proved.

Let us check now some axioms of quantum field theory.

{\bf Lemma 3.2.} (Existence and uniqueness of vacuum).
{\it
For  vector  $\Phi  \in  {\cal  F}({\cal  H}_2)$  the
following statements are equivalent:\\
(i) invariance under space-time translations: for all $a$
$$
U_{0,a} \Phi =\Phi;
$$
(ii) $\Phi = c|0>$ for some multiplier $c\in {\bf C}$.
}

The proof is obvious.

Investigate now spectral properties.

{\bf Lemma  3.3.}  {\it  The  spectrum  of the operator $P^{\mu}$ is a
subset of   a   set   $\{0\}   \cup   \{({\varepsilon},{\bf   p})    |
{\varepsilon}^2- {\bf p}^2 > 0$}.

{\bf Proof.}  It  is  sufficient  to  prove  that $\sigma(M_{\bf P}^2)
\subset ({\bf P}^2,\infty)$.  The property ${\varepsilon}^2 \in \sigma
(M_{\bf P}^2)$ means that the operator ${\varepsilon}^2 - M_{\bf P}^2$
is not boundedly invertible. Since
$$
(({\varepsilon}^2 - M_{\bf P}^2)^{-1} \psi)_{\bf s} = ({\varepsilon}^2
- \Omega_{{\bf P}{\bf s}}^2)^{-1} \psi_{\bf s} +
\frac{(\Omega_{{\bf P}{\bf s}}^2 - {\varepsilon}^2)^{-1}
\Omega_{{\bf P}{\bf s}}^2
\chi_{{\bf P}{\bf s}} \int d{\bf  s}'
\chi_{{\bf P}{\bf s}'} \Omega^2_{{\bf P}{\bf s}'} (\Omega_{{\bf P}{\bf
s}'}^2 - {\varepsilon}^2)^{-1}  }
{ \frac{(2\pi)^d}{\lambda_R^{\bf P}} - \int d{\bf s}
\chi_{{\bf P}{\bf s}}^2 \Omega_{{\bf P}{\bf s}}^4
(\Omega_{{\bf P}{\bf s}}^{-2} -
(\Omega_{{\bf P}{\bf s}}^{2}
-{\varepsilon}^2)^{-1})
},
$$
${\varepsilon}^2\in \sigma(M_{\bf ^2})$ if and only if
\be
{\varepsilon} = \omega_{{\bf P}/2-{\bf s}} + \omega_{{\bf P}/2 +  {\bf
s}}
\l{xx7}
\ee
for some ${\bf s}$ or
\be
\frac{(2\pi)^d}{\lambda_R^{\bf P}} - \int d{\bf s}
\chi_{{\bf P}{\bf s}}^2 \Omega_{{\bf P}{\bf s}}^4
(\Omega_{{\bf P}{\bf s}}^{-2} -
(\Omega_{{\bf P}{\bf s}}^{2}
-{\varepsilon}^2)^{-1}) = 0.
\l{xx8}
\ee
Since $\omega_{{\bf  P}/2+{\bf  s}} + \omega_{{\bf P}/2 - {\bf s}} \ge
2\omega_{{\bf P}/2} = \sqrt{{\bf P}^2+ 4\mu^2}  >  |{\bf  P}|$,  these
values of ${\varepsilon}$ obey the property ${\varepsilon}^2 \in ({\bf
P}^2, \infty)$.  It follows from eq.\r{d11*} that  eq.\r{xx8}  can  be
transformed to the form
\be
\frac{1}{(2\pi)^{d+1}} \int_0^{\infty}  d\alpha \int_0^{\infty} d\beta
e^{-\mu^2(\alpha+\beta)} \left(          \frac{\pi}{\alpha           +
\beta}\right)^{\frac{d+1}{2}} (e^{-\frac{\alpha\beta}{\alpha   +\beta}
({\bf P}^2 - {\varepsilon}^2)} -1) = - \frac{1}{\lambda_R^0}.
\l{xx9}
\ee
Since $\lambda_R^0<0$,  the  left-hand  side  of  eq.\r{xx9} should be
positive. This means ${\varepsilon}^2 > {\bf P}^2$. Lemma is proved.

\section{Composed field operators}

In the  previous  section  we  have constructed the Hamiltonian of the
theory of ''infinite number of  fields''  which  was  shown  to  be  a
self-adjoint operator  in  the  Hilbert  space.  However,  it  is also
necessary to check the property of the Poincare invariance.

To simplify the investigation, it is convenient to introduce an analog
of the notion of a field which is very useful in traditional QFT:  the
Wightman axiomatic approach  allows  us  to  reduce  the  problem  of
Poincare invariance   of   the  theory  to  the  problem  of  Poincare
invariance of Wightman functions.

However, it is not easy to introduce the field  $\varphi^a(x)$  since
we have not considered the nonsymmetric $N$-field states yet. However,
one can investigate the properties of the ''multifield operators'':
\be
{\cal W}_{N,k} (x_1,...,x_k) = \frac{1}{N} \sum_{a=1}^N \varphi^a(x_1)
... \varphi^a(x_k).
\l{g0}
\ee
Consider this operator at $x_1^0=...=x_k^0=0$.  The results of section
II imply that
\be
{\cal W}_{N,k}   (x_1,...,x_k)   K_N  f  =  K_N  \tilde{\cal  W}_{N,k}
(x_1,...,x_k) f
\l{g0a}
\ee
with
\be
\tilde{\cal W}_{N,k} ({\bf x}_1,...,{\bf x}_k)
=  \int  D\phi |\Phi_0[\phi(\cdot)]|^2 \phi({\bf
x}_1)... \phi({\bf x}_k) + N^{-1/2} \tilde{W}_k({\bf x}_1,...{\bf x}_k)
+ O(N^{-1}).
\l{g1}
\ee
The operator $\tilde{W}_k$ can be presented as
\bea
\tilde{W}_k({\bf x}_1,...,{\bf       x}_k)      =      \int      D\phi
(\tilde{A}^+[\phi(\cdot)] + \tilde{A}^-[\phi(\cdot)]) \phi({\bf  x}_1)
... \phi({\bf x}_k) \Phi_0[\phi(\cdot)] = \nonumber \\
\sum_{n=1}^{\infty} \int d{\bf p}_1 ... d{\bf p}_n
[A^{(n)+}_{{\bf p}_1 ... {\bf p}_n}
(\Phi^{(n)}_{{\bf p}_1 ...  {\bf p}_n}, \phi({\bf x}_1) ... \phi({\bf
x}_k) \Phi_0)
+ A^{(n)-}_{{\bf p}_1 ... {\bf p}_n}
(\Phi^{(n)}_{{\bf p}_1 ...  {\bf p}_n}, \phi({\bf x}_1) ... \phi({\bf
x}_k) \Phi_0)^*].
\l{g1a}
\eea
One can expect that Heisenberg operator will also obey the relation of
the type \r{g1}:
\be
\tilde{\cal W}_{N,k} ({x}_1,...,{x}_k)
=  (\Phi_0[\phi(\cdot)],
\phi({x}_1)... \phi({x}_k)
\Phi_0[\phi(\cdot)] )
+ N^{-1/2} \tilde{W}_k({x}_1,...{x}_k)
+ O(N^{-1}).
\l{g2}
\ee
Here $\phi(x)$ is a Heisenberg operator of the free field of the  mass
$\mu$. The property \r{g2} is to be checked in appendix B.

The multifield  operators  $\tilde{W}_k(x_1,...,x_k)$ being analogs of
fields are to be investigated.

\subsection{Multifield operators}

In this subsection we compute the explicit form of the operators
$\tilde{W}_k(x_1,...,x_k)$.

The "$k$-field" \r{g0} satisfies the Heisenberg equation
$$
\left(
\frac{\partial}{\partial x^{\mu}_A}
\frac{\partial}{\partial x_{A\mu}}
+ m^2   +    \frac{\lambda}{N}    \sum_{b=1}^N    \varphi^b    (x_A)
\varphi^b(x_A) \right)
{\cal W}_{N,k} (x_1,...,x_k) =0.
$$
The property \r{g0a} implies that
$$
\left(
\frac{\partial}{\partial x^{\mu}_A}
\frac{\partial}{\partial x_{A\mu}}
+ m^2   +   {\lambda}
\tilde{\cal W}_{N,2} (x_A,x_A) \right)
\tilde{\cal W}_{N,k} (x_1,...,x_k) =0.
$$
Use now  the  property  \r{g2}.  One  can  notice  that $m^2 + \lambda
(\Phi_0, \phi(x_A)\phi(x_A) \Phi_0) = \mu^2$. Therefore,
\bea
\left(
\frac{\partial}{\partial x^{\mu}_A}
\frac{\partial}{\partial x_{A\mu}}
+ \mu^2   +   \frac{\lambda}{\sqrt{N}}
\tilde{W}_{2} (x_A,x_A) + O(N^{-1}) \right)
\nonumber \\
( (\Phi_0, \phi(x_1)...\phi(x_k) \Phi_0) + N^{-1/2}
\tilde{W}_{k} (x_1,...,x_k)+ O(N^{-1})) =0.
\l{g2*}
\eea

The terms of order $O(1)$ give us an equation on  the  vacuum  average
value. It has the form
$$
\left(
\frac{\partial}{\partial x^{\mu}_A}
\frac{\partial}{\partial x_{A\mu}}
+ \mu^2
\right)
(\Phi_0, \phi(x_1)...\phi(x_k) \Phi_0)  = 0.
$$
This equation  is  automatically  satisfied.  The   terms   of   order
$O(N^{1/2})$ lead to the nontrivial equation:
\be
\left(
\frac{\partial}{\partial x^{\mu}_A}
\frac{\partial}{\partial x_{A\mu}}
+ \mu^2
\right)
\tilde{W}_{k} (x_1,...,x_k) + Q(x_A)
(\Phi_0, \phi(x_1)...\phi(x_k) \Phi_0)  = 0.
\l{g3}
\ee
with
\be
Q(x_A) = \lambda \tilde{W}_2(x_A,x_A).
\l{g3*}
\ee

To perform investigation of eq.\r{g3},  it is convenient to  introduce
the linear  combinations  of the multifields $\tilde{W}_k$.  It follows
from the Wick theorem that the operators \r{g1a} can be presented as
\be
\tilde{W}_k({\bf x}_1,...,  {\bf  x}_k)  =  \sum
(\Phi_0, \phi({\bf x}_{l_1^1}) \phi({\bf x}_{l_1^2}) \Phi_0)
...
(\Phi_0, \phi({\bf x}_{l_{\nu}^1}) \phi({\bf x}_{l_{\nu}^2}) \Phi_0)
\hat{W}_{k-2\nu} ({\bf x}_{m_1},..., {\bf x}_{m_{k-2\nu}}).
\l{g4}
\ee
Here the summation is performed over all decompositions of the set
$$
\{ 1,2,...,k \} =
\{ l_1^1, l_1^2 \}
\cup
...
\cup
\{ l_{\nu}^1, l_{\nu}^2 \}
\cup
\{ m_1, ..., m_{k-2\nu} \},
$$
while
$$
l_1^1 < l_1^2,
\qquad
l_{\nu}^1 < l_{\nu}^2,
\qquad
m_1< ... < m_{k-2\nu}, \qquad k-2\nu >0.
$$
The operator $\hat{W}_s({\bf  x}_1,...,{\bf  x}_s)$  entering  to  the
formula \r{g4} has the form
\bea
\hat{W}_s({\bf x}_1,...,{\bf       x}_s) =
\int d{\bf p}_1 ... d{\bf p}_s
[A^{(s)+}_{{\bf p}_1 ... {\bf p}_s}
(\Phi^{(s)}_{{\bf p}_1 ...  {\bf p}_s},:\phi({\bf x}_1) ... \phi({\bf
x}_s): \Phi_0) \nonumber \\
+ A^{(s)-}_{{\bf p}_1 ... {\bf p}_s}
(\Phi^{(s)}_{{\bf p}_1 ...  {\bf p}_s}, :\phi({\bf x}_1) ... \phi({\bf
x}_s): \Phi_0)^*].
\l{g5}
\eea
The notation  $:  :$ is used for the Wick ordering of  combinations
of fields
$$
\phi({\bf x})     =     \frac{1}{(2\pi)^{d/2}}     \int    \frac{d{\bf
k}}{\sqrt{2\omega_{\bf k}}} [a_{\bf k}^+e^{-i{\bf k}{\bf x}} +
a_{\bf k}^-e^{i{\bf k}{\bf x}}]
$$
Analogously, define the Heisenberg operators
$\hat{W}_s(x_1,...,x_s)$ from the recursive relations
\be
\tilde{W}_k({x}_1,...,  {x}_k)  =  \sum
(\Phi_0, \phi({x}_{l_1^1}) \phi({x}_{l_1^2}) \Phi_0)
...
(\Phi_0, \phi({x}_{l_{\nu}^1}) \phi({x}_{l_{\nu}^2}) \Phi_0)
\hat{W}_{k-2\nu} ({x}_{m_1},..., {x}_{m_{k-2\nu}}).
\l{g6}
\ee
Applying the Wick theorem to the combinations of the
field and momenta operators
$$
\pi({\bf x}) = \frac{1}{i}
\frac{\delta}{\delta \phi({\bf x})}
= \frac{1}{(2\pi)^{d/2}}     \int    \frac{d{\bf
k}}{\sqrt{2}}  i\sqrt{\omega_{\bf k}}[a_{\bf k}^
+e^{-i{\bf k}{\bf x}} +
a_{\bf k}^-e^{i{\bf k}{\bf x}}]
$$
we obtain in an analogous way that
\bea
\frac{\partial}{\partial x_{i_1}^0}...
\frac{\partial}{\partial x_{i_l}^0}
\hat{W}_s({x}_1,...,{       x}_s) =
\int d{\bf p}_1 ... d{\bf p}_s
[A^{(s)+}_{{\bf p}_1 ... {\bf p}_s}
(\Phi^{(s)}_{{\bf p}_1 ...  {\bf p}_s},:\phi({\bf x}_1) ...
\pi_{i_1}({\bf x}_{i_1}) ...
\pi_{i_l}({\bf x}_{i_1})...
\phi({\bf
x}_s): \Phi_0) \nonumber \\
+ A^{(s)-}_{{\bf p}_1 ... {\bf p}_s}
(\Phi^{(s)}_{{\bf p}_1 ...  {\bf p}_s}, :\phi({\bf x}_1) ...
\pi_{i_1}({\bf x}_{i_1}) ...
\pi_{i_l}({\bf x}_{i_1})
\phi({\bf
x}_s): \Phi_0)^*]
\l{g6*}
\eea
for $i_1<...<i_l$, $x_1^0=...=x_s^0=0$.

Let us find an  equation  on  the  operator  $\hat{W}_s$.  For  $k=2$,
$\tilde{W}_2 = \hat{W}_2$, so that
\be
\left(
\frac{\partial}{\partial x^{\mu}_A}
\frac{\partial}{\partial x_{A\mu}}
+ \mu^2
\right)
\hat{W}_{2} (x_1,x_2) + Q(x_A)
(\Phi_0, \phi(x_1)\phi(x_2) \Phi_0)  = 0.
\l{g7}
\ee
For odd values of $k$,  one has $(\Phi_0, \phi(x_1)...\phi(x_k)\Phi_0)
= 0$, so that
$$
\left(
\frac{\partial}{\partial x^{\mu}_A}
\frac{\partial}{\partial x_{A\mu}}
+ \mu^2
\right)
\tilde{W}_k(x_1,...,x_k) = 0.
$$
It follows from the recursive relations that
\be
\left(
\frac{\partial}{\partial x^{\mu}_A}
\frac{\partial}{\partial x_{A\mu}}
+ \mu^2
\right)
\hat{W}_k(x_1,...,x_k) = 0.
\l{g8}
\ee

Let us show that eq.\r{g8} is also satisfied for even
values of $k \ne 2$.  For definiteness,  consider the case $A=1$.  The
general case can be investigated analogously. The quantity
$$
\left(
\frac{\partial}{\partial x^{\mu}_1}
\frac{\partial}{\partial x_{1\mu}}
+ \mu^2
\right)
\tilde{W}_{k} (x_1,...,x_k)
$$
entering to the left-hand side of eq.\r{g3} can be decomposed into two
parts. One  of  them corresponds to the case $k-2\nu=2$,  another - to
$k-2\mu >2$. The first part is
\bea
\left(
\frac{\partial}{\partial x^{\mu}_1}
\frac{\partial}{\partial x_{1\mu}}
+ \mu^2
\right)
\sum
(\Phi_0, \phi({x}_{l_1^1}) \phi({x}_{l_1^2}) \Phi_0)
...
(\Phi_0, \phi({x}_{l_{\nu}^1}) \phi({x}_{l_{\nu}^2}) \Phi_0)
\hat{W}_{2} ({x}_{1}, {x}_{m_{2}})
=  \nonumber \\ - Q(x_1)
\sum
(\Phi_0, \phi({x}_{1}) \phi({x}_{m_2}) \Phi_0)
(\Phi_0, \phi({x}_{l_1^1}) \phi({x}_{l_1^2}) \Phi_0)
...
(\Phi_0, \phi({x}_{l_{\nu}^1}) \phi({x}_{l_{\nu}^2}) \Phi_0)
\nonumber \\
= - Q(x_1) (\Phi_0, \phi(x_1) ... \phi(x_k) \Phi_0).
\eea
The second part reads
\be
\left(
\frac{\partial}{\partial x^{\mu}_1}
\frac{\partial}{\partial x_{1\mu}}
+ \mu^2
\right)
\sum_{k-2\nu>2}
(\Phi_0, \phi({x}_{l_1^1}) \phi({x}_{l_1^2}) \Phi_0)
...
(\Phi_0, \phi({x}_{l_{\nu}^1}) \phi({x}_{l_{\nu}^2}) \Phi_0)
\hat{W}_{k-2\nu} ({x}_{m_1},..., {x}_{m_{k-2\nu}}).
\l{g11}
\ee
It follows from  eq.\r{g3}  then  that  the  quantity  \r{g11}  should
vanish. We   obtain  by  induction  that  the  functions  $\hat{W}_4$,
$\hat{W}_6$ etc. obey eq.\r{g8}.

To find  an  explicit  form  of  $\hat{W}_k$,  prove   the   following
proposition.

{\bf Proposition 4.1.} {\it Let
\be
\left(
\frac{\partial}{\partial x^{\mu}_A}
\frac{\partial}{\partial x_{A\mu}}
+ \mu^2
\right)
f_k(x_1,...,x_k) = 0, \qquad A=\overline{1,k}
\l{g8*}
\ee
and
$$
\frac{\partial}{\partial x_{i_1}^0}...
\frac{\partial}{\partial x_{i_l}^0}
f_k(x_1,...,x_k) = 0,  \qquad i_1 <...<i_l, \qquad x_1^0 = ... = x_k^0
=0
$$
Then $f_k=0$.
}

{\bf Proof.}    Consider    the    spatial    Fourier   transformation
$\tilde{f}_k({\bf p}_1,t_1;...;{\bf p}_k,t_k)$ of the function  $f_k$.
It obeys the set of equations
$$
\left(
\frac{\partial}{\partial t_{A}}
\frac{\partial}{\partial t_{A}}
+ \omega_{{\bf p}_A^2}
\right)
\tilde{f}_k({\bf p}_1,t_1;...;{\bf    p}_k,t_k)    =     0,     \qquad
A=\overline{1,k}
$$
with $\omega_{\bf p} = \sqrt{{\bf p}^2+\mu^2}$. This implies that
$$
\tilde{f}_k ({\bf         p}_1,t_1;...;{\bf         p}_k,t_k)        =
\sum_{\sigma_1,...,\sigma_k \in \{-1,1\}} \alpha_{\sigma_1...\sigma_k}
({\bf p}_1,...,{\bf p}_k) e^{i\sigma_1\omega_{{\bf p}_1t_1}
+ ... + i\sigma_k\omega_{{\bf p}_kt_k} }.
$$
One can express the coefficients $\alpha$ as
$$
\alpha_{\sigma_1...\sigma_k}({\bf p}_1,...,{\bf p}_k) =
\left(
\frac{1}{2} - \frac{i\sigma_1}{2\omega_{{\bf p}_1}}
\right)
...
\left(
\frac{1}{2} - \frac{i\sigma_k}{2\omega_{{\bf p}_k}}
\right)
\tilde{f}_k({\bf p}_1,t_1,...,{\bf p}_k,t_k)
$$
Therefore,
$\alpha_{\sigma_1...\sigma_k}({\bf p}_1,...,{\bf   p}_k)  =  0$.  This
implies $f_k=0$. Proposition is proved.

Denote
\bea
\hat{W}_s^0(x_1,..,x_s) =
\int d{\bf p}_1 ... d{\bf p}_s
[A^{(s)+}_{{\bf p}_1 ... {\bf p}_s}
(\Phi^{(s)}_{{\bf p}_1 ...  {\bf p}_s},:\phi({x}_1) ... \phi({
x}_s): \Phi_0)
\nonumber \\
+ A^{(s)-}_{{\bf p}_1 ... {\bf p}_s}
(\Phi^{(s)}_{{\bf p}_1 ...  {\bf p}_s}, :\phi({x}_1) ... \phi({
x}_s): \Phi_0)^*].
\l{g9}
\eea
Consider the operator distribution
$
f_s =
\hat{W}_s - \hat{W}^0_s$
obeying the condition of proposition 4.1. This implies $f_k=0$, so that
$\hat{W}_s=\hat{W}_s^0$. The explicit form of the operator $\hat{W}_s$
is
\bea
\hat{W}_s({\bf x}_1,t_1,...,{\bf x}_s,t_s) = \frac{1}{(2\pi)^{sd/2}}
\int \frac{d{\bf p}_1}{\sqrt{2\omega_{{\bf p}_1}}}
... \frac{d{\bf p}_s}{\sqrt{2\omega_{{\bf p}_s}}} \sqrt{k!}
(A^{+(s)}_{{\bf p}_1... {\bf p}_s}
e^{i\omega_{{\bf p}_1}t_1  +  ...  +  i\omega_{{\bf  p}_s}t_s  - i{\bf
p}_1{\bf x}_1 - ... - i{\bf p}_s{\bf x}_s }
\nonumber \\
+
A^{-(s)}_{{\bf p}_1... {\bf p}_s}
e^{-i\omega_{{\bf p}_1}t_1  -  ...  -  i\omega_{{\bf  p}_s}t_s  + i{\bf
p}_1{\bf x}_1 + ... + i{\bf p}_s{\bf x}_s })
\l{g9x}
\eea

Let $f\in {\cal S}({\bf R}^{ds})$. Consider the operators
$$
\hat{W}_s[f] = \int dx_1...dx_s \hat{W}_s(x_1,...,x_s) f(x_1,...,x_s).
$$
They are defined on the set  $\breve{D}$  of  all  finite  vectors  of
$\breve{\cal F}$. The set $\breve{D}$ is invariant under the operator
$\hat{W}_s[f]$.

{\bf Proposition 4.2.}
{\it
1. $\hat{W}_s(x_1,...,x_s)$ ($s\ne 2$)is an operator distribution.
2. The set
$$
\hat{W}_{s_1}[f_1] ...  \hat{W}_{s_k}[f_k]  |0>,  \qquad f_j \in {\cal
S}({\bf R}^{dj})
$$
is a total set in $\breve{\cal F}$.
}
The first statement is obvious. The second statement is a corollary of
lemma A.14.

Thus, we see that the "k-field" $\tilde{W}_k$
has the form \r{g6},  the operators $\hat{W}_s$ having the form \r{g9}
are already found.  The remaining problem is to find the explicit form
of the bifield $\hat{W}_2(x_1,x_2)$. Since an equation for the bifield
contains the operator $Q(x)$ being an analog of the composed  $\lambda
\varphi^a\varphi^a$ field, let us investigate first its properties.

\subsection{The $\lambda \varphi^a\varphi^a$ composed field}

The operator $Q(x) = Q({\bf x},t)$ can be presented as
\bea
Q({\bf x},t) =
e^{iHt} \lambda \hat{W}_2({\bf x},0,{\bf x},0) e^{-iHt} =
\nonumber \\
\lambda \frac{\sqrt{2}}{(2\pi)^d}
\int
\frac{d{\bf k}_1}{\sqrt{2\omega_{{\bf k}_1}}}
\frac{d{\bf k}_2}{\sqrt{2\omega_{{\bf k}_2}}}
(A^+_{{\bf k}_1{\bf k}_2}(t) e^{-i({\bf k}_1+{\bf k}_2) {\bf x}}
+ A^-_{{\bf k}_1{\bf k}_2}(t) e^{i({\bf k}_1+{\bf k}_2) {\bf x}}),
\l{d16}
\eea
where
$$
A^{\pm}_{{\bf k}_1{\bf k}_2} (t) = e^{iHt}
A^{\pm (2)}_{{\bf k}_1{\bf k}_2} e^{-iHt}.
$$
After transformations \r{d0a} and \r{d2*} expression \r{d16} takes the
form
$$
Q({\bf x},t) = \lambda \frac{\sqrt{2}}{(2\pi)^d} \int d{\bf P} d{\bf s}
\sqrt{ \frac{2\Omega_{{\bf P}{\bf s}}}{2\omega_{{\bf P}/2+{\bf s}}
2\omega_{{\bf P}/2-{\bf s}}}} e^{-i{\bf P}{\bf x}} Q_{{\bf P}{\bf s}}(t)
=
$$
$$
=
\int d{\bf  P}d{\bf  s}  (C^+_{{\bf  P}{\bf  s}}  \gamma_{{\bf  P}{\bf
s}}({\bf x},t)
+ C^-_{{\bf P}{\bf s}} \gamma^*_{{\bf P}{\bf s}} ({\bf x},t))
= C^+[\gamma({\bf x},t)] + C^-[\gamma({\bf x},t)],
$$
where
\be
\gamma({\bf     x},t)     =
\frac{\sqrt{2}}{(2\pi)^d}
e^{-i{\bf P}{\bf x}} \frac{1}{\sqrt{2M}}  e^{iMt}
\lambda  \Omega \chi.
\l{y1}
\ee
Our purpose is to prove  that  $Q(x)$  is  an  operator  distribution.
Therefore, we  should  show  that $(\chi({\bf x},t))_{{\bf P}{\bf s}}$
can be viewed as a vector distribution.  However, an infinite quantity
$\lambda^{-1}$ and   function   $\Omega^2\chi  \notin  L^2$  enter  to
eq.\r{y1}. It is remarkable that these divergences can be  eliminated:
one can use the property:
$$
M^{-2} \lambda \Omega^2\chi = \lambda_R \chi.
$$
Here $\lambda_R$  is  an operator of multiplication by $\lambda_R^{\bf
P}$. Thus, the vector function \r{y1} can be written as
\be
\gamma({\bf x},t) = (2\pi)^{-d} e^{-i{\bf P}{\bf x}}  e^{iMt}  M^{3/2}
\lambda_R \chi.
\l{y2}
\ee
One can present it as
$$
\gamma({\bf x},t) = (2\pi)^{-d} (-\Delta +1)^m \left(
- \frac{\partial^2}{\partial t^2} \right)
e^{-i{\bf P}{\bf x}} e^{iMt} M^{-1/2} ({\bf P}^2 +  1)^{-m}  \lambda_R
\chi.
$$
Since $M^{-1/2}  ({\bf P}^2+1)^{-m} \lambda_R \chi \in {\cal H}_2$ for
sufficiently large m, the function
$e^{-i{\bf P}{\bf x}} e^{iMt} M^{-1/2} ({\bf P}^2 +  1)^{-m}  \lambda_R
\chi$ is a bounded continuous vector function,  we obtain  from  lemmas
A.12 and  A,13  that  $\gamma$  is  a vector distribution.  Lemma A.14
implies that $Q({\bf x},t)$ is an operator distribution.

\subsection{Canonical variables as operator distributions}

The purpose of this subsection is to investigate the properties of the
operators $Q_{{\bf  P}{\bf  s}}$  and  $\Pi_{{\bf  P}{\bf s}}$.  These
properties will be essentially used.

First of all, notice that
$$
Q_{{\bf P}{\bf s}}  = C^+[\xi_{{\bf P}{\bf s}}] +
C^-[\xi_{-{\bf P}{\bf s}}], \qquad
\Pi_{{\bf P}{\bf s}}  = C^+[\pi_{{\bf P}{\bf s}}] +
C^-[\pi_{-{\bf P}{\bf s}}],
$$
where $\xi_{{\bf P}{\bf s}}$ and $\pi_{{\bf P}{\bf s}}$ have the form
$$
(\xi_{{\bf P}{\bf  s}})_{{\bf  P}'{\bf s}'} = \delta_{{\bf P}{\bf P}'}
((2M_{\bf P}^{-1/2})_{{\bf s}{\bf s}'},
\qquad
(\pi_{{\bf P}{\bf  s}})_{{\bf  P}'{\bf s}'} = i\delta_{{\bf P}{\bf P}'}
((M_{\bf P}/2)^{-1/2})_{{\bf s}{\bf s}'},
$$
Consider the integrals
\bea
\int d{\bf   P}d{\bf   s}  \varphi_{{\bf P}{\bf s}}
\xi_{{\bf   P}{\bf   s}}   =   (2M)^{-1/2}
\overline{\varphi},
\l{m1}\\
\int d{\bf   P}d{\bf   s} \varphi_{{\bf P}{\bf s}}
\pi_{{\bf   P}{\bf   s}}   =  i (M/2)^{1/2}
\overline{\varphi}
\l{m2}
\eea
where $\varphi_{{\bf P}{\bf s}}$ are  complex  functions  from  ${\cal
S}({\bf R}^{2d})$,
\be
\overline{\varphi}_{{\bf P}{\bf   s}}   =  \frac{1}{2}  (\varphi_{{\bf
P}{\bf s}} + \varphi_{{\bf P},-{\bf s}})
\l{m3}
\ee
Since $(2M)^{-1/2}$  is  a  bounded  operator,  the integral \r{m1} is
always defined.  However,  $(M/2)^{1/2}$ is not a bounded operator, so
that quantity  \r{m2}  may be not defined.  We see that the expression
for $\xi_{{\bf  P}{\bf  s}}$  defines  a  vector  distribution,  while
$\pi_{{\bf P}{\bf s}}$ is not a vector distribution.

To consider   objects   like   $\pi_{{\bf   P}{\bf   s}}$   as  vector
distributions, it  is  necessary  to   perform   the   renormalization
procedure. Let
$$
Q_{\bf P}(t)  =  \frac{1}{(2\pi)^d}  \int d{\bf x} e^{i{\bf P}{\bf x}}
Q({\bf x},t) = - \ddot{R}_{\bf P}(t),
$$
$$
R_{\bf P}(t)  =  \frac{1}{(2\pi)^d}  \int d{\bf x} e^{i{\bf P}{\bf x}}
R({\bf x},t) = C^+[r_{\bf P}(t)] + C^-[r_{-\bf P}(t)],
$$
where
$$
(r_{\bf P}(t))_{{\bf P}'{\bf s}'} =  (2\pi)^{-d}  \delta_{{\bf  P}{\bf
P}'} (e^{iMt} M^{-1/2} \lambda_R \chi)_{{\bf P}'{\bf s}'}
$$
is a vector distribution.

Denote
\be
\pi_{{\bf P}{\bf s}}^{ren} = \pi_{{\bf P}{\bf s}} - \frac{1}{\sqrt{2}}
\Omega_{{\bf P}{\bf s}}^2 \chi_{{\bf P}{\bf s}} \dot{r}_{\bf P}(0),
\l{m4}
\ee
where $\chi_{{\bf P}{\bf s}}$ has the form \r{d8}. One has
$$
\int d{\bf P}d{\bf s} \pi_{{\bf P}{\bf s}}^{ren} \varphi_{{\bf P}{\bf
s}} = i(M/2)^{1/2}  [\overline{\varphi}  -  \frac{\lambda^R}{(2\pi)^d}
\chi \int    d{\bf    s}'    \overline{\varphi}_{{\bf    P}{\bf   s}'}
\Omega^2_{{\bf P}{\bf s}'}  \chi_{{\bf  P}{\bf  s}'}]  =  i(M/2)^{1/2}
M^{-2} \Omega^2\overline{\varphi}.
$$
Since $\Omega^2  \overline{\varphi}  \in  L^2$,  while $M^{-3/2}$ is a
bounded operator,   $\pi^{ren}_{{\bf   P}{\bf   s}}$   is   a   vector
distribution. We obtain from lemma A.14 the following proposition.

{\bf Proposition 4.5.} $Q_{{\bf P}{\bf s}}$ {\it and
\be
\Pi_{{\bf P}{\bf s}}^{ren} = \Pi_{{\bf P}{\bf s}} - \frac{1}{\sqrt{2}}
\Omega^2_{{\bf P}{\bf s}} \chi_{{\bf P}{\bf s}} \dot{R}_{\bf P}(0)
\l{m5}
\ee
are operator distributions.
}

Investigate now the transformation properties of these  distributions.
Analogously to the previous subsection, we obtain

{\bf Proposition 4.6.} {\it  For spatial rotations,
the following properties are satisfied:
$$
u_{\Lambda,0} \xi_{{\bf P},{\bf s}} = \xi_{\Lambda{\bf P},\Lambda{\bf s}},
\qquad
u_{\Lambda,0} \pi^{ren}_{{\bf P},{\bf s}} =
\pi^{ren}_{\Lambda{\bf P},\Lambda{\bf s}},
\qquad
u_{\Lambda,0} r_{{\bf P}}(t) = r_{\Lambda{\bf P}}(t).
$$
}

{\bf Corollary}. {\it Under conditions \r{xx1} the operators
$Q_{{\bf P}{\bf s}}$, $\Pi^{ren}_{{\bf P}{\bf s}}$ obey
the following transformation properties
\bea
U_{\Lambda,0} Q_{{\bf  P}{\bf s}} U_{\Lambda,0}^{-1} = Q_{\Lambda {\bf
P}, \Lambda {\bf s}},
\qquad
U_{\Lambda,0} \Pi^{ren}_{{\bf  P}{\bf s}}
U_{\Lambda,0}^{-1} = \Pi^{ren}_{\Lambda {\bf P}, \Lambda {\bf s}},
\qquad
U_{\Lambda,0} R_{\bf  P} (t)
U_{\Lambda,0}^{-1} = R_{\Lambda {\bf P}}(t),
\nonumber \\
U_{1,{\bf a}} Q_{{\bf  P}{\bf s}} U_{1,{\bf a}}^{-1} =
e^{-i{\bf P}{\bf a}} Q_{{\bf P},{\bf s}},
\qquad
U_{1,{\bf a}} \Pi^{ren}_{{\bf  P}{\bf s}}
U_{1,{\bf a}}^{-1} =
e^{-i{\bf P}{\bf a}} \Pi^{ren}_{{\bf P}, {\bf s}},
\qquad
U_{1,a} R_{\bf  P} (t)
U_{1,a}^{-1} = R_{{\bf P}}(t+a_0).
\eea
}

\subsection{The bifield operator}

In this section we construct the bifield operator $\hat{W}_2(x_1,x_2)$
which obey eq.\r{g7} and initial conditions \r{g6*}.  We show it to be
an operator distribution of ${\bf x}_1$,  ${\bf x}_2$ at fixed  values
of $x_1^0$, $x_2^0$. It can be also viewed as an operator distribution
of $x_1,x_2$.

Firs of all, consider the spatial Fourier transformation
$$
\hat{W}_2({\bf x},t_x;{\bf y},t_y) = \frac{1}{(2\pi)^d} \int d{\bf  k}
d{\bf p}  w_2({\bf  k},t_x;{\bf  p},t_y)  e^{-i{\bf  k}{\bf x} - i{\bf
p}{\bf y}}.
$$
Initial conditions \r{g6*} can be presented in the following form
\bea
w_2({\bf k},0;{\bf p},0) = \sqrt{\frac{\omega_{\bf k}+ \omega_{\bf p}}
{\omega_{\bf k}\omega_{\bf  p}}}  Q_{{\bf  k}+{\bf p},  \frac{{\bf k}-
{\bf p}}{2}}, \nonumber\\
\frac{\partial}{\partial t_x}
w_2({\bf k},0;{\bf p},0) = \sqrt{\frac{\omega_{\bf k}
(\omega_{\bf k}+ \omega_{\bf p})}
{\omega_{\bf  p}}}  \Pi_{{\bf  k}+{\bf p},  \frac{{\bf k}-
{\bf p}}{2}}, \nonumber \\
\frac{\partial}{\partial t_y}
w_2({\bf k},0;{\bf p},0) = \sqrt{\frac{\omega_{\bf p}
(\omega_{\bf k}+ \omega_{\bf p})}
{\omega_{\bf  k}}}  \Pi_{{\bf  k}+{\bf p},  \frac{{\bf k}-
{\bf p}}{2}}, \nonumber \\
\frac{\partial}{\partial t_x}
\frac{\partial}{\partial t_y}
w_2({\bf k},0;{\bf p},0) = \sqrt{({\omega_{\bf p}
(\omega_{\bf k}+ \omega_{\bf p})})
{\omega_{\bf  k}}}  Q_{{\bf  k}+{\bf p},  \frac{{\bf k}-
{\bf p}}{2}}.
\eea
Eqs.\r{g7} can be written as
\bea
\left( \frac{\partial^2}{\partial t_x \partial t_x} + \omega_{\bf k}^2
\right) w_2({\bf  k},t_x;  {\bf  p},t_y)  +  \frac{1}{2\omega_{\bf p}}
e^{-i\omega_{\bf p}(t_x-t_y)} Q_{{\bf k}+{\bf p}}(t_x) =0.
\nonumber \\
\left( \frac{\partial^2}{\partial t_y \partial t_y} + \omega_{\bf p}^2
\right) w_2({\bf  k},t_x;  {\bf  p},t_y)  +  \frac{1}{2\omega_{\bf k}}
e^{-i\omega_{\bf k}(t_x-t_y)} Q_{{\bf k}+{\bf p}}(t_y) =0.
\eea
These equations and initial conditions lead to  the  following  formal
solution
\bea
w_2({\bf k},t_x;    {\bf   p},t_y)   =   \sqrt{\frac{\omega_{\bf   k}+
\omega_{\bf p}}{\omega_{\bf k}\omega_{\bf p}}}
Q_{{\bf k}+ {\bf p},
\frac{{\bf k}-{\bf p}}{2} } cos (\omega_{\bf k}t_x + \omega_{\bf p}t_y)
\nonumber \\
+ \frac{1}{\sqrt{\omega_{\bf   k}\omega_{\bf   p}   (\omega_{\bf   k}+
\omega_{\bf p}) }} \Pi_{{\bf k}+ {\bf p},
\frac{{\bf k}-{\bf p}}{2} } sin(\omega_{\bf k}t_x + \omega_{\bf p}t_y)
\nonumber \\
- \int_0^{t_x} d\tau \frac{sin(\omega_{\bf k}(t_x-\tau))}{\omega_{\bf k}}
\frac{1}{2\omega_{\bf p}}  e^{-i\omega_{\bf p} (\tau-t_y)} Q_{{\bf k}+
{\bf p}}(\tau)
\nonumber \\
- \int_0^{t_y} d\tau \frac{sin(\omega_{\bf p}(t_y-\tau))}{\omega_{\bf p}}
\frac{1}{2\omega_{\bf k}}  e^{-i\omega_{\bf k} (t_x - \tau)} Q_{{\bf k}+
{\bf p}}(\tau).
\eea
This form is not suitable for investigation  since  $\Pi_{{\bf  P}{\bf
s}}$ has  been  discovered  to  be not a distribution,  while $Q_{{\bf
P}{\bf s}}$ is a distribution rather that ordinary function.  However,
we can  use  the relation $Q = - \ddot{R}$ and integrate by parts.  We
obtain:
\bea
w_2({\bf k},t_x;    {\bf   p},t_y)   =   \sqrt{\frac{\omega_{\bf   k}+
\omega_{\bf p}}{\omega_{\bf k}\omega_{\bf p}}}
Q_{{\bf k}+ {\bf p},
\frac{{\bf k}-{\bf p}}{2} } cos (\omega_{\bf k}t_x + \omega_{\bf p}t_y)
\nonumber \\
+ \frac{1}{\sqrt{\omega_{\bf   k}\omega_{\bf   p}   (\omega_{\bf   k}+
\omega_{\bf p}) }} \Pi^{ren}_{{\bf k}+ {\bf p},
\frac{{\bf k}-{\bf p}}{2} } sin(\omega_{\bf k}t_x + \omega_{\bf p}t_y)
\nonumber \\
- \int_0^{t_x} d\tau
\frac{\partial}{\partial\tau}
\left(
\frac{sin(\omega_{\bf k}(t_x-\tau))}{\omega_{\bf k}}
\frac{1}{2\omega_{\bf p}}  e^{-i\omega_{\bf p} (\tau-t_y)}
\right)
\frac{\partial}{\partial\tau}
{R}_{{\bf k}+
{\bf p}}(\tau)
\nonumber \\
- \int_0^{t_y} d\tau
\frac{\partial}{\partial\tau}
\left(
\frac{sin(\omega_{\bf p}(t_y-\tau))}{\omega_{\bf p}}
\frac{1}{2\omega_{\bf k}}  e^{-i\omega_{\bf k} (t_x-\tau)}
\right)
\frac{\partial}{\partial\tau}
{R}_{{\bf k}+
{\bf p}}(\tau).
\l{mm1}
\eea
Since for any smooth function $\varphi$ the integral
$$
\int_0^{t} d\tau  \dot{R}_{\bf  P}  (\tau)  \varphi(\tau)   =   R_{\bf
P}(\tau) \varphi(\tau)  |_0^t  -  \int_0^t  d\tau  \dot{\varphi}(\tau)
R_{\bf P}(\tau)
$$
is defined as an operator distribution of ${\bf P}$ ($R_{\bf P}(\tau)$
is a distribution of ${\bf P}$ at fixed $\tau$), while $Q_{{\bf P}{\bf
s}}$ and $\Pi^{ren}_{{\bf  P}{\bf  s}}$  are  operator  distributions,
expression \r{mm1} gives us an operator distribution.  Thus, we obtain
the following proposition.

{\bf Proposition 4.7.} {\it  $w_2({\bf k},t_x; {\bf p},t_y)$ is:
\\
(1) an  operator distribution of ${\bf k}$,  ${\bf p}$ at fixed $t_x$,
$t_y$;\\
(2) an operator distribution of ${\bf k},t_x,{\bf p},t_y$.
}

Corollary of proposition 4.6 implies the following statement.

{\bf Proposition  4.8.}  {\it  The  transformation properties of $w_2$
under spatial rotations and translations are:
\bea
U_{\Lambda,0} w_2({\bf  k},t_x,  {\bf  p},t_y)  U_{\Lambda,0}^{-1}   =
w_2(\Lambda {\bf k}, t_x, \Lambda {\bf p}, t_y)
\nonumber \\
U_{1,a} w_2({\bf  k},t_x,  {\bf  p},t_y)  U_{1,a}^{-1}   =
e^{-i({\bf k}+{\bf p}){\bf a}} w_2({\bf k},t_x;{\bf p},t_y).
\eea
}

{\bf Corollary.}  {\it $\hat{W}({\bf   x},t_x;   {\bf   y},t_y)$   is   an
$t_x,t_y$-dependent operator distribution of ${\bf x}$, ${\bf y}$ with
the following transformation properties under spatial rotations
and translations:
\bea
U_{\Lambda,0} \hat{W}_2({\bf  x},t_x,  {\bf  y},t_y)  U_{\Lambda,0}^{-1}   =
\hat{W}_2(\Lambda {\bf x}, t_x, \Lambda {\bf y}, t_y)
\nonumber \\
U_{1,a} \hat{W}_2({\bf  x},t_x,  {\bf  y},t_y)  U_{1,a}^{-1}   =
\hat{W}_2({\bf x} + {\bf a},t_x;{\bf y}+ {\bf a},t_y).
\eea
$\hat{W}(x,y)$ is also an operator distribution of $x$, $y$.}

Consider the operators
$$
W_2[f,t_x,t_y] =  \int  d{\bf x} d{\bf y} \hat{W}_2({\bf x},t_x;  {\bf
y},t_y) f({\bf x},{\bf y}).
$$

{\bf Proposition 4.9.} {\it
The set of all finite linear combinations
\be
\sum_n
W_2[f_{n,1},t_{x,1}^n,t_{y,1}^n] ...
W_2[f_{n,s_n},t_{x,s_n}^n,t_{y,s_n}^n]|0>
\l{mm2}
\ee
is dense in ${\cal F}({\cal H}_2)$.
}

To prove this proposition,  it is  sufficient  to  consider  the  case
$t^n_{x,i} = t^n_{y,i} =0$ only and use lemma A.15.

One can also consider the operators
$$
W_2[g] =  \int dt_x
dt_y d{\bf x} d{\bf y} W_2({\bf x},t_x;  {\bf y},t_y) g({\bf
x},t_x,{\bf y},t_y).
$$

{\bf Proposition 4.10.} {\it The set of all finite linear combinations
\be
\sum_n W_2[g_{n,1}]...W_2[g_{n,s_n}]|0>
\l{mm3}
\ee
is dense in ${\cal F}({\cal H}_2)$.
}

To prove this proposition,  it is sufficient to approximate the vector
\r{mm2} by the vector \r{mm3} by choosing
$$
g_{n,k}({\bf x},t_x;{\bf  y},t_y)  =   f_{n,k}   ({\bf   x},{\bf   y})
\frac{1}{{\varepsilon}^2} \varphi(t_x/{\varepsilon}, t_y/{\varepsilon})
$$
for any  smooth function $\varphi(\tau_x,\tau_y)$ with compact support
such that $\int d\tau_x d\tau_y \varphi(\tau_x,\tau_y)=1$.

Thus, the cyclic property of the vacuum state is checked.

\subsection{Invariance under time translations}

The purpose of this subsection is to check the invariance property  of
the bifield under time translations:
\be
U_{1,a} \hat{W}_2({\bf x},t_x,{\bf y},t_y) U^{-1}_{1,a} =
\hat{W}_2({\bf x},t_x+t, {\bf y}, t_y+t)
\l{n0}
\ee
if $a^0=t$,   ${\bf  a}  =  0$.  Let  us  prove  first  the  following
proposition. Denote
$$
Q_{{\bf P}{\bf s}}(T) = e^{iHT} Q_{{\bf P}{\bf s}} e^{-iHT},
\Pi^{ren}_{{\bf P}{\bf s}}(T) = e^{iHT}
\Pi^{ren}_{{\bf P}{\bf s}} e^{-iHT}.
$$
For smooth function $f(\tau)$ let
$$
\int_0^T d\tau f(\tau) \frac{\partial}{\partial \tau} R_{\bf  P}(\tau)
\equiv f(\tau) R_{\bf P}(\tau)|_0^T - \int_0^T d\tau \frac{\partial f}
{\partial \tau} R_{\bf P}(\tau).
$$

{\bf Proposition 4.11.} {\it The following properties are satisfied:
\bea
Q_{{\bf P}{\bf  s}}  (T)  = Q_{{\bf P}{\bf s}} cos (\Omega_{{\bf P}{\bf
s}}T) + \Pi_{{\bf P}{\bf s}}^{ren} \frac{sin(\Omega_{{\bf P}{\bf s}}) T}
{\Omega_{{\bf P}{\bf s}}} -
\nonumber \\
\int_0^T \frac{d\tau}{2\sqrt{\omega_{{\bf P}/2-{\bf  s}}  \omega_{{\bf
P}/2+{\bf s}} \Omega_{{\bf P}{\bf s}} }} \frac{\partial}{\partial \tau}
[sin(\Omega_{{\bf P}{\bf s}} (T-\tau))] \frac{\partial}{\partial \tau}
R_{{\bf P}}(\tau), \nonumber \\
\Pi_{{\bf P}{\bf  s}}^{ren}(T)  =  \Pi_{{\bf  P}{\bf   s}}^{ren}   cos
(\Omega_{{\bf P}{\bf s}} T) - Q_{{\bf P}{\bf s}}
\Omega_{{\bf P}{\bf s}} sin(\Omega_{{\bf P}{\bf s}}T)
- \nonumber \\
\int_0^T \frac{d\tau}{2} \sqrt{\frac{\Omega_{{\bf P}{\bf s}} }
{\omega_{{\bf P}/2-{\bf s}} \omega_{{\bf P}/2+{\bf s}}  }}
\frac{\partial}{\partial\tau}
[cos(\Omega_{{\bf P}{\bf s}} (T-\tau))] \frac{\partial}{\partial \tau}
R_{\bf P}(\tau)
\l{n1}
\eea
}

{\bf Proof.} First of all, notice that
\bea
Q_{{\bf P}{\bf s}}(T) =
C^+[\xi_{{\bf P}{\bf s}}(T)]  +  C^-[\xi_{- {\bf P}{\bf s}}(T)],
\nonumber \\
\Pi_{{\bf P}{\bf s}}^{ren} (T) =
C^+[\pi^{ren}_{{\bf P}{\bf s}}(T)]  +
C^-[\pi^{ren}_{- {\bf P}{\bf s}}(T)],
\eea
where $\xi_{{\bf  P}{\bf  s}}(T)$  and $\pi^{ren}_{{\bf P}{\bf s}}(T)$
are the following vector distributions:
\bea
(\xi_{{\bf P}{\bf s}}(T))_{{\bf P}' {\bf s}'}  =  \delta_{{\bf  P}{\bf
P}'} (e^{iM_{\bf P}T} (2M_{\bf P})^{-1/2})_{{\bf s}{\bf s}'},
\nonumber \\
(\pi_{{\bf P}{\bf  s}}^{ren}(T))_{{\bf  P}'{\bf  s}'}  =  \delta_{{\bf
P}{\bf P}'} (i\Omega^2 M_{\bf P}^{-2} (M_{\bf  P}/2)^{1/2}  e^{iM_{\bf
P}T})_{{\bf s}{\bf s}'}.
\eea
Formulas \r{n1} mean that
\bea
\xi_{{\bf P}{\bf  s}}  (T)  = \xi_{{\bf P}{\bf s}} cos (\Omega_{{\bf P}{\bf
s}}T) + \pi_{{\bf P}{\bf s}}^{ren} \frac{sin(\Omega_{{\bf P}{\bf s}}) T}
{\Omega_{{\bf P}{\bf s}}} -
\nonumber \\
\int_0^T \frac{d\tau}{2\sqrt{\omega_{{\bf P}/2-{\bf  s}}  \omega_{{\bf
P}/2+s} \Omega_{{\bf P}{\bf s}} }} \frac{\partial}{\partial \tau}
[sin(\Omega_{{\bf P}{\bf s}} (T-\tau))] \frac{\partial}{\partial \tau}
r_{{\bf P}}(\tau), \nonumber \\
\pi_{{\bf P}{\bf  s}}^{ren}(T)  =  \pi_{{\bf  P}{\bf   s}}^{ren}   cos
(\Omega_{{\bf P}{\bf s}} T) - \xi_{{\bf P}{\bf s}}
\Omega_{{\bf P}{\bf s}} sin(\Omega_{{\bf P}{\bf s}}T)
\nonumber \\
-
\int_0^T \frac{d\tau}{2} \sqrt{\frac{\Omega_{{\bf P}{\bf s}}}
{\omega_{{\bf P}/2-{\bf s}} \omega_{{\bf P}/2+{\bf s}}  }}
\frac{\partial}{\partial\tau}
[cos(\Omega_{{\bf P}{\bf s}} (T-\tau))] \frac{\partial}{\partial \tau}
r_{\bf P}(\tau)
\l{n2}
\eea
Integrating relations \r{n2} with  the  function  $\varphi  \in  {\cal
S}({\bf R}^{2d})$   and   applying  the  operator  $(2M_P)^{1/2}$,  we
transform them to the form
\bea
e^{iMT} \overline{\varphi}  =  cos  (\Omega  T)  \overline{\varphi}  +
iM^{-1}\Omega sin  (\Omega  T)  \overline{\varphi}
\nonumber \\+  \int_0^T  d\tau
(e^{iM\tau})^{\dot{}} (\Omega^{-2}   -   M^{-2})   cos(\Omega(T-\tau))
\Omega^2 \overline{\varphi},
\l{n3} \\
ie^{iMT} M^{-1}  \Omega^2  \overline{\varphi}  = iM^{-1} cos(\Omega T)
\overline{\varphi} -  \Omega  sin(\Omega   T)   \overline{\varphi}   -
\nonumber \\
\int_0^T d\tau  (e^{iMT})^{\dot{}}  (\Omega^{-2}  -  M^{-2})  \Omega^3
sin(\Omega (T-\tau)) \overline{\varphi}.
\l{n4}
\eea
Here $\Omega$  is  the  operator  of  multiplication  by $\Omega_{{\bf
P}{\bf s}}$. We have used the definition of the operator $M^{-2}$.

Relation  \r{n4}  is  a  corollary  of  the  relation  \r{n3}   is
sufficient to consider the time derivatives of eq.\r{n3}.
The simplest way to check eq.\r{n3} is to consider the
Laplace transformations of the left-hand side
\be
\int_0^{\infty} e^{iMT} e^{-\omega T} dT = \frac{1}{\omega-iM}
\l{n5}
\ee
and of the right-hand side:
\be
\frac{\omega}{\omega^2 + \Omega^2} + iM^{-1}  \frac{\Omega^2}{\omega^2
+ \Omega^2}  +  \frac{iM}{\omega - iM} (\Omega^{-2} - M^{-2}) \Omega^2
\frac{\omega}{\omega^2 + \Omega^2}.
\l{n6}
\ee
Formulas \r{n5} and \r{n6} coincide.  Thus,  eq.\r{n3} is satisfied at
$T>0$, the check procedure for $T<0$ is analogous. Proposition is proved.

{\bf Proposition 4.12}. {\it Relation \r{n0} is satisfied.}

{\bf Proof.} One has
\bea
e^{iHT} w_2({\bf k},t_x; {\bf p},t_y) e^{-iHT} =
\sqrt{\frac{\omega_{\bf k}+    \omega_{\bf    p}}{    \omega_{\bf   k}
\omega_{\bf p}}}  Q_{{\bf k}+{\bf p}; \frac{{\bf k}-{\bf p}}{2}} (T)
cos (\omega_{\bf k}t_x + \omega_{\bf p}t_y) +
\nonumber \\
\frac{1}{\sqrt{\omega_{\bf k}\omega_{\bf    p}    (\omega_{\bf     k}+
\omega_{\bf p})}} \Pi_{{\bf k}+{\bf p}; \frac{{\bf p}-{\bf k}}{2}} (T)
sin(\omega_{\bf k}t_x + \omega_{\bf p} t_y)
\nonumber \\
- \int_0^{t_x} d\tau \frac{sin(\omega_{\bf k}(t_x-\tau))}
{2\omega_{\bf k}\omega_{\bf p}} e^{-i\omega_{\bf p}(\tau-t_y)}
Q_{{\bf k}+{\bf p}}(\tau +T)
\nonumber \\
- \int_0^{t_y} d\tau \frac{sin(\omega_{\bf p}(t_y-\tau))}
{2\omega_{\bf k}\omega_{\bf p}} e^{-i\omega_{\bf k}(t_x - \tau)}
Q_{{\bf k}+{\bf p}}(\tau +T)
\nonumber \\
w_2({\bf k},t_x+T; {\bf p},t_y+T) =
\sqrt{\frac{\omega_{\bf k}+    \omega_{\bf    p}}{    \omega_{\bf   k}
\omega_{\bf p}}}  Q_{{\bf k}+{\bf p}; \frac{{\bf k}-{\bf p}}{2}}
cos (\omega_{\bf k}(t_x+T) + \omega_{\bf p}(t_y+T)) +
\nonumber \\
\frac{1}{\sqrt{\omega_{\bf k}\omega_{\bf    p}    (\omega_{\bf     k}+
\omega_{\bf p})}} \Pi_{{\bf k}+{\bf p}; \frac{{\bf p}-{\bf k}}{2}}
sin(\omega_{\bf k}(t_x+T) + \omega_{\bf p} (t_y+T))
\nonumber \\
- \int_0^{t_x+T} d\tau \frac{sin(\omega_{\bf k}(t_x+T-\tau))}
{2\omega_{\bf k}\omega_{\bf p}} e^{-i\omega_{\bf p}(\tau-t_y-T)}
Q_{{\bf k}+{\bf p}}(\tau)
\nonumber \\
- \int_0^{t_y+T} d\tau \frac{sin(\omega_{\bf p}(t_y+T-\tau))}
{2\omega_{\bf k}\omega_{\bf p}} e^{-i\omega_{\bf k}(t_x +T- \tau)}
Q_{{\bf k}+{\bf p}}(\tau)
\eea
It follows from proposition 4.11 that
$$
e^{iHT} w_2({\bf k},t_x; {\bf p},t_y) e^{-iHT} =
w_2({\bf k},t_x+T; {\bf p},t_y+T).
$$
We obtain relation \r{n0}. Proposition is proved.

\section{Poincare invariance of the theory}

The purpose  of  this section is to check the property of relativistic
invariance of the theory which mean that:

(a) the unitary representation  of  the  Poincare  group  $(\Lambda,a)
\mapsto \tilde{U}_{\Lambda,a}$ is constructed:
\be
\tilde{U}_{\Lambda_1,a_1}
\tilde{U}_{\Lambda_2,a_2}
= \tilde{U}_{(\Lambda_1,a_1)(\Lambda_2,a_2)};
\l{t1-}
\ee

(b) the  $k$-field  operators  $\tilde{W}_k(x_1,...,x_k)$ are Poincare
invariant:
\be
\tilde{U}_{\Lambda,a}
\tilde{W}_k(x_1,...,x_k)
\tilde{U}_{\Lambda,a}^{-1}
= \tilde{W}_k(\Lambda x_1+a,...,\Lambda x_k+a);
\l{t1}
\ee

(c) the vacuum state is Poincare-invariant:
\be
\tilde{U}_{\Lambda,a}|0> = |0>
\l{t1+}
\ee

To simplify  the  problem,  remind  that  the  state  space  has  been
decomposed according to eq.\r{b21}, while the operators
$\tilde{U}_{\Lambda,a}$ are looked for in the form
${U}_{\Lambda,a} \otimes \breve{U}_{\Lambda,a}$. The operators
$\breve{U}_{\Lambda,a}$ have  been  already  constructed in subsection
II.F.

First of all, investigate the property of invariance of the operators
$\hat{W}_k(x_1,...x_k)$ of  the  form  \r{g9}  (at  $k\ne  2$).  These
operators are of the form $1 \otimes \hat{W}_k(x_1,...x_k)$.

{\bf Lemma 5.1.}
{\it
1. The  vacuum  state  is  invariant   under   action   of   operators
$\breve{U}_{\Lambda,a}$.\\
2. For $k\ne  2$,  the  operators  $\hat{W}_k(x_1,...,x_k)$  obey  the
property
\be
\breve{U}_{\Lambda,a} \hat{W}_k(x_1,...,x_k)
\breve{U}_{\Lambda,a}^{-1}
= \hat{W}_k(\Lambda x_1 + a,...,\Lambda x_k + a).
\l{t2}
\ee
}

{\bf Proof.}
The first property is obvious. Prove the second property.
It follows from eq.\r{g9x}, lemma A.11 and formula \r{b23x} imply that
\bea
\breve{U}_{\Lambda,a} \hat{W}_s(x_1,...,x_s) \breve{U}_{\Lambda,a}^{-1}
= \frac{\sqrt{s!}}{(2\pi)^{sd/2}}
\int \frac{d{\bf p}_1}{\sqrt{2\omega_{{\bf p}_1}}}
... \frac{d{\bf p}_s}{\sqrt{2\omega_{{\bf p}_s}}}
(A^{+(s)}_{{\bf p}_1... {\bf p}_s}
e^{i(p_1+...+p_s)a} e^{i(\Lambda^{-1}p_1 \cdot x_1
+ ... + \Lambda^{-1}p_s \cdot x_s)}
\nonumber \\
+ A^{-(s)}_{{\bf p}_1... {\bf p}_s}
e^{-i(p_1+...+p_s)a} e^{-i(\Lambda^{-1}p_1 \cdot x_1
+ ... + \Lambda^{-1}p_s \cdot x_s)}.
\eea
Property $\Lambda^{-1}p\cdot  x = p\cdot \Lambda x$.  imply eq.\r{t2}.
Lemma is proved.

{\bf Lemma 5.2.}
{\it Let $U_{\Lambda,a}$ be unitary operators in $\cal F$ such that \\
(a) the group property
\be
{U}_{\Lambda_1,a_1}
{U}_{\Lambda_2,a_2}
= {U}_{(\Lambda_1,a_1)(\Lambda_2,a_2)}
\l{t3-}
\ee
is satisfied;\\
(b) the bifield operator is invariant
\be
U_{\Lambda,a} \hat{W}_2(x,y)  U^{-1}_{\Lambda,a}  =  \hat{W}_2(\Lambda
x+a, \Lambda y + a);
\l{t3}
\ee
(c) the vacuum state is invariant:
\be
U_{\Lambda,a} |0> = |0>.
\l{t3+}
\ee
Then the  operators  $\tilde{U}_{\Lambda,a}  =  U_{\Lambda,a}  \otimes
\breve{U}_{\Lambda,a}$ obey properties \r{t1-} - \r{t1+}.
}

This lemma  is  a direct corollary of lemma 5.1 and formula \r{g4} for
the operators $\tilde{W}_k(x_1,...,x_k)$.

The remaining  problem  is  to  construct  operators   $U_{\Lambda,a}$
satisfying relations  \r{t3-}  - \r{t3+}.  The possible way may be the
following. The operators $P^{\mu}$ and $M^{mn}$ in $\cal F$ have  been
already constructed.  One  should  then  try to construct the operator
$M^{0k}$, check the commutation relations of the Poincare algebra.

However, the following problems arise in the approach.  It is not easy
to check  the self-adjointness of the operator $M^{0k}$ since it is an
unbounded operator. Further,  to construct the  group  representation
from  the algebra representation,  one should check the conditions of
the Nelson theorem \c{N}  or  investigate  the  properties  of
analytic  vectors \c{FS3}.

Therefore, another   approach   will  be  used  for  constructing  the
operators $U_{\Lambda,a}$.  First of all, we will check the invariance
of the Wightman function
\be
<0| \hat{W}_2(x,y)  \hat{W}_2(x',y')|0>  =  <0|\hat{W}_2(\Lambda  x+a,
\Lambda y + a) \hat{W}_2(\Lambda x' + a, \Lambda y' + a)|0>
\l{t4}
\ee
Then we will define the operator $U_{\Lambda,a}$ from the property
\be
U_{\Lambda,a} \hat{W}_2(x_1,y_1) ... \hat{W}_2(x_k,y_k) |0>
= \hat{W}_2(\Lambda x_1 + a,  \Lambda y_1 + a)  ...  \hat{W}_2(\Lambda
x_k + a, \Lambda y_k + a)|0>.
\l{t5}
\ee
This definition will be shown  to  be  correct  if  and  only  if  the
property \r{t4} is satisfied. Let us investigate the properties of the
Wightman functions.

\subsection{The $QQ$-propagator}

First of all,  investigate the vacuum average value  $<0|Q(x)Q(y)|0>$.
It has the form
\be
<0|Q(x)Q(y)|0> =   \frac{1}{(2\pi)^{2d}}   \int   d{\bf   P}  (\lambda
\Omega^2\chi, e^{i{\bf  P}({\bf  x}-{\bf  y})  -  iM(x^0-y^0)}  M^{-1}
\lambda \Omega^2\chi).
\l{t6}
\ee
The vector $\lambda \Omega^2\chi$ is viewed as $\lambda_R M^2\chi$. To
check the Poincare invariance of the average \r{t6}, present it as
$$
<0|Q(x)Q(y)|0> = \frac{1}{(2\pi)^{d+1}} \int dP V(P) e^{-iP(x-y)}
$$
with
\be
V(P^0, {\bf   P})  =  \frac{1}{(2\pi)^{d-1}}  \int  (\lambda  \Omega^2
\chi)_{{\bf P}{\bf s}} (\delta(M_{\bf P}-P^0) M_{\bf  P}^{-1}  \lambda
\Omega^2\chi)_{{\bf P}{\bf s}} d{\bf s}
\l{t7}
\ee
Making use of the relations
$$
\delta(M_{\bf P}-P^0) M_{\bf P}^{-1}  =  2\theta(P^0)  \delta  (M_{\bf
P}^2 - (P^0)^2),\qquad
2\pi i \delta(x) = \frac{1}{x-i0} - \frac{1}{x+i0}
$$
and
$$
\int d{\bf s} (\lambda \Omega^2 \chi)_{{\bf P}{\bf s}}
\left( \frac{1}{M_{\bf P}^2+ {\varepsilon}^2} \lambda \Omega^2\chi
\right)_{{\bf P}{\bf s}} =
\lambda (2\pi)^d [1 - \frac{1}{1+ \lambda I({\bf P},{\varepsilon})}]
$$
we take the formula \r{t7} to the form
\be
V(P^0,{\bf P}) = 2i\theta(P^0) \left[
\frac{\lambda}{1+\lambda I({\bf P},i(P^0+i0))}
- \frac{\lambda}{1+\lambda I({\bf P},i(P^0-i0))}
\right]
\l{t7a}
\ee
where $I$   is  of  the  form  \r{d11*}.  We  see  that  the  function
$<0|Q(x)Q(y)|0>$ is Poincare invariant.

It will be also necessary to calculate the  propagator  of  the  field
$Q$. Formally, one has
\be
<0|TQ(x)Q(y)|0> =  \theta(x^0-y^0)  <0|Q(x)Q(y)|0>  -  \theta(y^0-x^0)
<0|Q(y)Q(x)|0>.
\l{t8}
\ee
Eq.\r{t6} implies
$$
<0|TQ(x)Q(y)|0> =      \frac{1}{(2\pi)^{2d}}     \int     d{\bf     P}
(\lambda\Omega^2\chi, e^{i{\bf  P}({\bf  x}-{\bf  y})  -  iM|x^0-y^0|}
M^{-1} \lambda \Omega^2\chi).
$$
The Fourier transformation of the propagator which is defined from the
relation
$$
<0|TQ(x)Q(y)|0> = \frac{1}{(2\pi)^{d+1}} \int dP G_Q(P) e^{-iPx}
$$
can be presented as
$$
G_Q(P^0,{\bf P}) = -2i  \int  d{\bf  s}  (\lambda  \Omega^2\chi)_{{\bf
P}{\bf s}}  ((M_{\bf  P}^2-P_0^2-i0)^{-1}  \lambda\Omega^2 \chi)_{{\bf
P}{\bf s}} = -2i\lambda +
\frac{2i\lambda}{1+\lambda
I({\bf P},i(P^0+i0))}
$$
We see  that  formally  calculated propagator consists of the singular
part
$$
<0|TQ(x)Q(y)|0>^{sing} = -2i\lambda \delta(x-y)
$$
and of the regular (renormalized) part with the Fourier transformation
\be
G_Q^{ren}(P^0,{\bf P}) =
\frac{2i\lambda}{1+\lambda
I({\bf P},i(P^0+i0))}
\l{t9}
\ee
However, this difficulty is usual:  the $T$-product is defined up to a
quasilocal quantity  being  proportional  to $\delta(x-y)$.  Note also
that the result \r{t9} is in agreement with the approach based on  the
summation of Feynman graphs \c{Wilson}.

Thus, we have obtained the following result.

{\bf Proposition  5.3.}
{\it The average value $<0|Q(x)Q(y)|0>$ is Poincare
invariant. Its Fourier transformation has the form \r{t7a}.}

\subsection{The $W_2Q$-average}

The purpose of this subsection is to compute the average values
\be
F_1(x,y,z) = <0|\hat{W}_2(x,y) Q(z)|0>,\qquad
F_2(x,y,z) = <0|Q(z)\hat{W}_2(x,y)|0>.
\l{t9*}
\ee
However, explicit  formulas  for   the   operators   $\hat{W}_2$   are
complicated, so that direct calculations are too difficult. Therefore,
the indirect method will be used. First of all, these averages will be
calculated at $x^0=y^0>z^0$ and $z^0>x^0=y^0$ correspondingly. Then we
will investigate the properties of the Fourier transformation  of  the
averages. The  equations  on  the averages will be obtained.  Then the
solution of the equations will be found.

\subsubsection{The $x^0=y^0$-case}

Consider the  average  value  $<0|Tw_2({\bf  p},0;{\bf  k},0)   Q_{\bf
P}(t)|0>$. According to subsection IV,
\be
w_2({\bf p},0;{\bf k},0) = \sqrt{\frac{\omega_{\bf k}+\omega_{\bf p}}
{\omega_{\bf k}\omega_{\bf p}}} Q_{{\bf k}+{\bf p},\frac{{\bf  k}-{\bf
p}}{2}}.
\l{t10}
\ee
Therefore,
\bea
<0|Tw_2({\bf p},0;{\bf k},0)Q_{\bf P}(t)|0>
\nonumber \\
= \frac{1}{(2\pi)^{2d}}
\sqrt{\frac{\omega_{\bf k}+\omega_{\bf p}}{2\omega_{\bf k}\omega_{\bf p}}}
\delta_{{\bf k}+{\bf p}+{\bf P},0}
[
\theta(-t) (e^{iMt}  M^{-1} \lambda \Omega^2\chi)_{{\bf P},\frac{{\bf k}
-{\bf p}}{2}} +
\theta(t) (e^{-iMt}  M^{-1} \lambda \Omega^2\chi)_{{\bf P},\frac{{\bf k}
-{\bf p}}{2}}]
\eea
Consider the Fourier transformation of the average \r{t10} defined as
\be
G({\bf k},{\bf p},{\varepsilon}) \delta_{{\bf k}+{\bf p},{\bf P}}
= \int dt e^{-i{\varepsilon}t}  <0|Tw_2({\bf  p},0;{\bf  k},0)  Q_{\bf
P}(t)|0>.
\l{t11}
\ee
One has
$$
G({\bf k},{\bf      p},{\varepsilon})       =       \frac{1}{(2\pi)^d}
\sqrt{\frac{2(\omega_{\bf k}+        \omega_{\bf       p}}{\omega_{\bf
k}\omega_{\bf p}}} \frac{1}{i} \left(
\frac{1}{M^2_{{\bf k}+{\bf p}} -{\varepsilon}^2-i0} \lambda\Omega^2\chi
\right)_{{\bf k}+{\bf p},\frac{{\bf k}-{\bf p}}{2}}.
$$
Making use of the definition of the operator $M_{\bf P}^2$, we obtain
$$
G({\bf k},{\bf    p},{\varepsilon})     =     -     \frac{1}{(2\pi)^d}
\frac{2(\omega_{\bf k}+\omega_{\bf  p})}{ 2\omega_{\bf k} 2\omega_{\bf
p} } \frac{1}{(\omega_{\bf k}+\omega_{\bf p})^2- {\varepsilon}^2 -i0}
G_Q^{ren} ({\varepsilon},{\bf k}+{\bf p}).
$$
Applying the Fourier transformation to eq.\r{t11}, we obtain that
\be
<0|T \hat{W}_2(x,y) Q(z)|0> = -i \int d\xi <0|TQ(\xi)Q(z)|0>^{ren}
<0|T\phi(x)\phi(y)|0> <0|T\phi(y)\phi(\xi)|0>,
\l{t12}
\ee
provided that $x=({\bf x},0)$, $y=({\bf y},0)$, $z=({\bf z},t)$, while
$<0|T\phi(x)\phi(y)|0>$ is  the  usual  propagator  of the free scalar
field
$$
<0|T\phi(x)\phi(y)|0> =    \frac{1}{(2\pi)^d}     \int     \frac{d{\bf
k}}{2\omega_{\bf k}}  e^{-i{\bf  k}({\bf x}-{\bf y})} e^{-i\omega_{\bf
k}|x^0-y^0|}.
$$
Formula \r{t12}  is  valid  not  only  at  $x^0=y^0=0$  but  also   at
$x^0=y^0\ne 0$  because  of translation invariance properties (section
IV). Note also that formula \r{t12} is in agreement with the  approach
based on the summation of Feynman graphs \c{Wilson}.

Thus, the following result is obtained.

{\bf Proposition 5.4.}
{\it
 The Green function $<0|T\hat{W}_2(x,y) Q(z)|0>$ has the form \r{t12},
provided that $x^0=y^0$.
}

\subsubsection{Properties of the $W_2$-field}

Consider the  state  $\hat{W}_2(x,y)|0>$.  Our purpose is to prove the
following property.

{\bf Lemma 5.5.} {\it The Fourier transformation
\be
\int dy \hat{W}_2(x,y) e^{-ipy} |0>
\l{t13}
\ee
vanish at $p^0<0$.}

{\bf Proof.} Consider the vector
\be
\int dt_y e^{i{\varepsilon}t_y} w_2({\bf k},t_x;{\bf p},t_y)|0>
\l{t14}
\ee
provided that ${\varepsilon}>0$.  It is sufficient  to  show  that  it
vanishes. The vector \r{t14} can be presented as
$$
C^+[\alpha({\bf k},t_x;{\bf p},{\varepsilon})]|0>
$$
with the following vector $\alpha$:
\bea
(\alpha({\bf k},t_x;{\bf p},{\varepsilon}))_{{\bf P}'{\bf s}'} =
\frac{1}{2} e^{-i\omega_{\bf k}t_x} \delta_{{\bf k}+{\bf p},{\bf P}'}
\left[
\sqrt{\frac{\omega_{\bf k}+\omega_{\bf p}}
{\omega_{\bf k}\omega_{\bf p}}}
(2M_{{\bf k}+{\bf p}})^{-1/2}_{\frac{{\bf k}-{\bf p}}{2} s'} -
\frac{1}{\sqrt{\omega_{\bf k}\omega_{\bf    p}(\omega_{\bf    k}     +
\omega_{\bf p})}} (M_{{\bf k}+{\bf p}}/2)^{1/2}_{\frac{{\bf k}
-{\bf p}}{2}   s'}    +
\right. \nonumber \\ \left.
\frac{1}{(2\pi)^d}    \frac{1}{2\omega_{\bf
k}\omega_{\bf p}}
\left(
\frac{1}{\omega_{\bf k}+\omega_{\bf     p}+M_{{\bf     k}+{\bf    p}}}
M^{-1/2}_{{\bf k}+{\bf p}} \lambda\Omega^2\chi
\right)_{P's'}
\right]
\l{t15}
\eea
It follows  from  the  definition of the operator $M_{\bf P}$ that the
quantity \r{t15} vanish. Lemma is proved.

{\bf Corollary.} {\it  The Fourier transformations
\be
\int F_1(x,y,z) e^{ipx} dp, \qquad \int dy e^{-ipy} F_2(x,y,z)
\l{t15*}
\ee
vanish at $p^0<0$.
}

\subsubsection{Equations for average values}

Let us obtain equations for vacuum averages \r{t9*}.  It follows  from
definition \r{mm1} that
$$
\left(
\frac{\partial}{\partial x^{\mu}}
\frac{\partial}{\partial x^{\mu}}
+ \mu^2
\right)
\hat{W}_2(x,y) + Q(x) <0|\phi(x)\phi(y)|0> = 0.
$$
$$
\left(
\frac{\partial}{\partial y^{\mu}}
\frac{\partial}{\partial y^{\mu}}
+ \mu^2
\right)
\hat{W}_2(x,y) + Q(y) <0|\phi(x)\phi(y)|0> = 0.
$$
Here $<0|\phi(x)\phi(y)|0>$  is the vacuum average for the free scalar
field.

Thus, we obtain the following statement.

{\bf Proposition  5.6.}  {\it The functions \r{t9*} obey the following
equations:
\bea
\left(
\frac{\partial}{\partial x^{\mu}}
\frac{\partial}{\partial x^{\mu}}
+ \mu^2
\right) F_1(x,y,z)
+ <0| Q(x)Q(z)|0> <0|\phi(x)\phi(y)|0> = 0.
\nonumber \\
\left(
\frac{\partial}{\partial y^{\mu}}
\frac{\partial}{\partial y^{\mu}}
+ \mu^2
\right) F_1(x,y,z)
+ <0| Q(y)Q(z)|0> <0|\phi(x)\phi(y)|0> = 0.
\nonumber \\
\left(
\frac{\partial}{\partial x^{\mu}}
\frac{\partial}{\partial x^{\mu}}
+ \mu^2
\right) F_2(x,y,z)
+ <0| Q(z)Q(x)|0> <0|\phi(x)\phi(y)|0> = 0.
\nonumber \\
\left(
\frac{\partial}{\partial y^{\mu}}
\frac{\partial}{\partial y^{\mu}}
+ \mu^2
\right) F_2(x,y,z)
+ <0| Q(z)Q(y)|0> <0|\phi(x)\phi(y)|0> = 0.
\l{t16}
\eea
}

Let us prove the following lemma.

{\bf Lemma   5.7.}   {\it   Let   $F_1(x,y,z)$   and  $F_2(x,y,x)$  be
distributions obeying the following properties.\\
(a) For some distributions $\Phi_1$ and $\Phi_2$
$$
F_1(x,y,z) = \Phi_1(x-z,y-z),
\qquad
F_2(x,y,z) = \Phi_2(x-z,y-z),
$$
(b) The functions $F_1$ and $F_2$ obey eqs.\r{t16}.\\
(c) The Fourier transformations \r{t15*} vanish at $p^0<0$. \\
(d)
$F_1(x,y,z)=<0|T\hat{W}_2(x,y)Q(z)|0>$ at $x^0=y^0>z^0$;\\
$F_2(x,y,z)=<0|T\hat{W}_2(x,y)Q(z)|0>$ at $x^0=y^0<z^0$.\\
Then
$$
F_1(x,y,z) = <0|\hat{W}_2(x,y) Q(z)|0>,
\qquad
F_2(x,y,z) = <0|Q(z) \hat{W}_2(x,y)|0>.
$$
}

{\bf Proof.} Consider the functions
$$
\tilde{F}_1(x,y,z) =
F_1(x,y,z) - <0|\hat{W}_2(x,y) Q(z)|0>,
\qquad
\tilde{F}_2(x,y,z) =
F_2(x,y,z) - <0|Q(z) \hat{W}_2(x,y)|0>.
$$
One has   $\tilde{F}_{1,2}(x,y,z)=\tilde{\Phi}_{1,2}(x-z,y-z)$.    The
functions $\tilde{\Phi}_{1,2}$ obey the following properties:\\
(a)
$$
( \frac{\partial}{\partial x^{\mu}}
\frac{\partial}{\partial x^{\mu}} + \mu^2) \tilde{\Phi}_{1,2}(x,y) = 0,
\qquad
( \frac{\partial}{\partial y^{\mu}}
\frac{\partial}{\partial y^{\mu}} + \mu^2) \tilde{\Phi}_{1,2}(x,y) = 0,
$$
(b) Fourier transformations
$\int dx \tilde{\Phi}_1(x,y) e^{ipx}$
and $\int dy \tilde{\Phi}_1(x,y) e^{-ipy}$
vanish if $p^0<0$;\\
(c)
$\tilde{\Phi}_1(x,y) = 0$ at $x^0=y^0>0$,
\\ $\tilde{\Phi}_2(x,y) = 0$ at $x^0=y^0<0$.

Properties (a) and (b) mean that
$$
\tilde{\Phi}_1(x,y) = \int d{\bf k} d{\bf p}
[\alpha^+_{{\bf k}{\bf p}}
e^{-i\omega_{\bf k}x^0 + i\omega_{\bf p}y^0} +
\alpha^-_{{\bf k}{\bf p}}
e^{-i\omega_{\bf k}x^0 - i\omega_{\bf p}y^0}]
e^{-i{\bf k}{\bf x}-i{\bf p}{\bf y}};
$$
$$
\tilde{\Phi}_2(x,y) = \int d{\bf k} d{\bf p}
[\beta^+_{{\bf k}{\bf p}}
e^{i\omega_{\bf k}x^0 + i\omega_{\bf p}y^0} +
\beta^-_{{\bf k}{\bf p}}
e^{-i\omega_{\bf k}x^0 + i\omega_{\bf p}y^0}]
e^{-i{\bf k}{\bf x}-i{\bf p}{\bf y}}
$$
for some
$\alpha^{\pm}_{{\bf k}{\bf p}}$, $\beta^{\pm}_{{\bf k}{\bf p}}$.
Property (c) means that
$$
\alpha^+_{{\bf k}{\bf p}} e^{i\omega_{\bf p}x^0}
+ \alpha^-_{{\bf k}{\bf p}} e^{-i\omega_{\bf p}x^0} = 0,
\qquad x^0>0;
$$
$$
\beta^+_{{\bf k}{\bf p}} e^{i\omega_{\bf k}x^0}
+ \beta^-_{{\bf k}{\bf p}} e^{-i\omega_{\bf k}x^0} = 0,
\qquad x^0<0.
$$
We obtain that $\alpha^{\pm}_{{\bf k}{\bf p}} =0$,
$\beta^{\pm}_{{\bf k}{\bf   p}}=0$.   Therefore,   $\tilde{\Phi}_1=0$,
$\tilde{\Phi}_2=0$. Lemma is proved.

\subsubsection{Explicit form of the averages}

{\bf Lemma 5.8.} {\it The average values have the form
\bea
F_1(x,y,z) = <0|\hat{W}_2(x,y)Q(z)|0> =
\nonumber \\ \frac{1}{i} \int d\xi
[(<0|T\phi(x)\phi(\xi)|0> - <0|\phi(x)\phi(\xi)|0>)  <0|Q(\xi)Q(z)|0>
<0|\phi(\xi)\phi(y)|0> +
\nonumber \\
<0|T\phi(y)\phi(\xi)|0> - <0|\phi(y)\phi(\xi)|0>)  <0|Q(\xi)Q(z)|0>
<0|\phi(x)\phi(\xi)|0> +
\nonumber \\
<0|\phi(x)\phi(\xi)|0><0|\phi(y)\phi(\xi)|0> <0|TQ(\xi)Q(z)|0>^{ren}]
\nonumber \\
F_2(x,y,z) = <0|Q(z) \hat{W}_2(x,y)|0> =
\nonumber \\
\frac{1}{i} \int d\xi
[(<0|T\phi(x)\phi(\xi)|0> - <0|\phi(\xi)\phi(x)|0>)  <0|Q(z)Q(\xi)|0>
<0|\phi(\xi)\phi(y)|0> +
\nonumber \\
(<0|T\phi(y)\phi(\xi)|0> - <0|\phi(\xi)\phi(y)|0>)  <0|Q(z)Q(\xi)|0>
<0|\phi(x)\phi(\xi)|0> +
\nonumber \\
<0|\phi(\xi)\phi(x)|0><0|\phi(\xi)\phi(y)|0> <0|TQ(z)Q(\xi)|0>^{ren}]
\eea
}

{\bf Proof.} It is sufficient to check the conditions  of  lemma  5.7.
Condition (a) is obvious. Eqs.\r{t16} are corollaries of the property
\be
\left(
\frac{\partial}{\partial x^{\mu}}
\frac{\partial}{\partial x^{\mu}}
+ \mu^2 \right)
<0|T\phi(x)\phi(y)|0> = \frac{1}{i} \delta(x-y).
\l{t16a}
\ee
Let us check the condition (c). Note that the Fourier transformations
$$
\int dx e^{ipx} <0|\phi(x)\phi(\xi)|0>,
\qquad
\int dx e^{ipx} <0|Q(x)Q(\xi)|0>
$$
vanish if $p^0<0$. The same property is also valid for the function
$$
\int dx e^{ipx} <0|\phi(x)\phi(\xi)|0> <0|Q(x)Q(\xi)|0>.
$$
Since the integral operator with the kernel  $<0|T\phi(x)\phi(\xi)|0>$
multiplies the  Fourier  transformation by $(\mu^2-p^2-i0)^{-1}$,  the
quantity
$$
\int dx e^{ipx} F_1(x,y,z)
$$
vanishes if $p^0<0$.  The analogous property for the function $F_2$ is
checked in the same way. Property (c) is checked.

To check  property  (d),  note  that under condition $x^0=y^0>z^0$ the
function $F_1$ can be presented as
\bea
\frac{1}{i} \left[
\int d\xi                    (<0|T\phi(x)\phi(\xi)-\phi(x)\phi(\xi)|0>
<0|TQ(\xi)Q(z)|0>^{ren} <0|T\phi(\xi)\phi(y)|0> +
\right.
\nonumber \\
\int d\xi                    (<0|T\phi(y)\phi(\xi)-\phi(y)\phi(\xi)|0>
<0|TQ(\xi)Q(z)|0>^{ren} <0|T\phi(x)\phi(\xi)|0>
\nonumber \\
\left.
+ \int      d\xi     <0|\phi(x)\phi(\xi)|0>     <0|\phi(y)\phi(\xi)|0>
<0|TQ(\xi)Q(z)|0>^{ren}
\right]
\eea
since the  first  and second integrands may be nonzero only at $\xi^0>
x^0=y^0>z^0$. The obtained expression coincides  with  \r{t12}.  Thus,
property (d)  is  checked  for  the function $F_1$.  The check for the
function $F_2$ is analogous. Conditions of lemma 5.7 are checked. This
implies lemma 5.8.

\subsection{The $W_2W_2$-averages}

The purpose of this subsection  is  to  find  explicit  forms  of  the
average value
\be
F(x,y:x',y') = <0| \hat{W}_2(x,y) \hat{W}_2(x'.y')|0>
\l{t17a}
\ee
We consider first the $x^0=y^0>x^0{}'=y^0{}'$ case.  Then equations on the
function $F$  will be obtained.  The solution of this equation will be
found.

\subsubsection{The equal-time case}

Consider the Green function
$$
<0|Tw_2({\bf p},t;{\bf k},t) w_2({\bf p}',\tau;{\bf k}',\tau)|0> =
\sqrt{\frac{\omega_{\bf k}+\omega_{\bf p}}
{\omega_{\bf k}\omega_{\bf p}}}
\sqrt{\frac{\omega_{{\bf k}'}+\omega_{{\bf p}'}}
{\omega_{{\bf k}'}\omega_{{\bf p}'}}}
\delta_{{\bf k}+{\bf p}; {\bf k}'+{\bf p}'}
\left(
\frac{1}{2M_{{\bf k}+{\bf p}}} e^{-iM_{{\bf k}+{\bf p}} |t-\tau|}
\right)_{\frac{{\bf k}-{\bf p}}{2};
\frac{{\bf k}'-{\bf p}'}{2}}
$$
and its Fourier transformation
\be
\delta_{{\bf k}+{\bf p}; {\bf k}'+{\bf p}'}
G_2({\bf p},{\bf k},{\bf p}',{\bf k}',{\varepsilon})
= \int  dt  e^{i{\varepsilon}t}  <0|Tw_2({\bf p},t;{\bf k},t) w_2({\bf
p}',0;{\bf k}',0)|0>
\l{t17}
\ee
One has
$$
G_2({\bf p},{\bf k},{\bf p}',{\bf k}',{\varepsilon}) =
\sqrt{\frac{\omega_{\bf k}+\omega_{\bf p}}
{\omega_{\bf k}\omega_{\bf p}}}
\sqrt{\frac{\omega_{{\bf k}'}+\omega_{{\bf p}'}}
{\omega_{{\bf k}'}\omega_{{\bf p}'}}}
\frac{1}{i}
\left( \frac{1}{M_{{\bf k}+{\bf p}}^2 - {\varepsilon}^2 -i0}
\right)_{\frac{{\bf k}-{\bf p}}{2};
\frac{{\bf k}'-{\bf p}'}{2}}.
$$
Making use of the definition of the operator $M_{\bf P}^2$, we obtain:
\bea
G_2({\bf p},{\bf k},{\bf p}',{\bf k}',{\varepsilon}) =
\sqrt{\frac{\omega_{\bf k}+\omega_{\bf p}}
{\omega_{\bf k}\omega_{\bf p}}}
\sqrt{\frac{\omega_{{\bf k}'}+\omega_{{\bf p}'}}
{\omega_{{\bf k}'}\omega_{{\bf p}'}}}
\frac{1}{2i}
\frac{1}{(\omega_{\bf k}+\omega_{\bf p})^2 - {\varepsilon}^2 -i0}
(\delta_{\frac{{\bf k}-{\bf p}}{2},\frac{{\bf k}'-{\bf p}'}{2}} +
\delta_{\frac{{\bf k}-{\bf p}}{2},\frac{{\bf p}'-{\bf k}'}{2}})
\nonumber \\
+ \frac{1}{(2\pi)^d}
\frac{\omega_{\bf k}+\omega_{\bf p}}{2\omega_{\bf k}\omega_{\bf p}}
\frac{\omega_{{\bf k}'}+\omega_{{\bf p}'}}{2\omega_{{\bf  k}'}
\omega_{{\bf p}'}}
\frac{1}{(\omega_{\bf k}+\omega_{\bf p})^2 - {\varepsilon}^2 -i0}
\frac{1}{(\omega_{{\bf k}'}+\omega_{{\bf p}'})^2 - {\varepsilon}^2 -i0}
G^{ren}_Q({\varepsilon},{\bf k}+{\bf p}).
\eea
Applying the Fourier transformation to expression \r{t17},  we  obtain
that
\bea
<0|T\hat{W}_2(x,y)\hat{W}_2(x',y')|0> =
\nonumber \\
<0|T\phi(x)\phi(x')|0> <0|T\phi(y)\phi(y')|0> +
<0|T\phi(x)\phi(y')|0> <0|T\phi(y)\phi(x')|0> -
\nonumber \\ \int d\xi d\xi' <0|TQ(\xi)Q(\xi)|0>^{ren}
<0|T\phi(x)\phi(\xi)|0>
<0|T\phi(y)\phi(\xi)|0>
\nonumber \\
<0|T\phi(\xi')\phi(x')|0>
<0|T\phi(\xi')\phi(y')|0>
\nonumber \\
\l{t18}
\eea
provided that  $x^0=y^0=t$,  $x^0{}'=y^0{}'=\tau$.  Eq.\r{t18}  is  in
agreement with the approach based on summation of Feynman graphs.

We have obtained the following statement.

{\bf Proposition 5.9.}
{\it The  Green  function  $F$  has  the  form  \r{t18}  at $x^0=y^0$,
$x^0{}'=y^0{}'$.}

\subsubsection{Equations for average values}

The following statements are analogs of proposition 5.6 and  corollary
of lemma 5.5.

{\bf Proposition  5.10.}  {\it  The  function  $F$  \r{t17a} obeys the
following equations
\bea
\left(
\frac{\partial}{\partial x^{\mu}}
\frac{\partial}{\partial x^{\mu}} +\mu^2
\right)
F(x,y,x',y') = - <0|\phi(x)\phi(y)|0><0|Q(x)\hat{W}_2(x',y')|0>,
\nonumber \\
\left(
\frac{\partial}{\partial y^{\mu}}
\frac{\partial}{\partial y^{\mu}} +\mu^2
\right)
F(x,y,x',y') = - <0|\phi(x)\phi(y)|0><0|Q(y)\hat{W}_2(x',y')|0>,
\nonumber \\
\left(
\frac{\partial}{\partial x^{\prime \mu}}
\frac{\partial}{\partial x^{\prime \mu}} +\mu^2
\right)
F(x,y,x',y') = - <0|\phi(x')\phi(y')|0><0|\hat{W}_2(x',y')Q(x')|0>,
\nonumber \\
\left(
\frac{\partial}{\partial y^{\prime \mu}}
\frac{\partial}{\partial y^{\prime \mu}} +\mu^2
\right)
F(x,y,x',y') = - <0|\phi(x')\phi(y')|0><0|\hat{W}_2(x',y')Q(y')|0>,
\l{t19}
\eea
}

{\bf Proposition 5.11.}
{\it The Fourier transformations
\be
\int dx e^{ipx} F(x,y,x',y'),
\qquad
\int dy' e^{-ipy'} F(x,y,x',y')
\l{t20}
\ee
vanish at $p^0<0$.
}

{\bf Lemma  5.12.}  {\it Let  $F(x,y,x',y')$  be a distribution obeying the
following properties:\\
(a) the function $F$ obey eq.\r{t19};\\
(b) the Fourier transformations \r{t20} vanish at $p^0<0$;\\
(c) at $x^0=y^0>x^{0\prime} = y^0{}'$
$$
F(x,y;x',y') = <0|T\hat{W}_2(x,y) \hat{W}_2(x',y')|0>.
$$
Then
$$
F(x,y;x',y') = <0|\hat{W}_2(x,y) \hat{W}_2(x',y')|0>.
$$
}

{\bf Proof.}
Consider the function
$$
\tilde{F}(x,y;x',y') =
F(x,y;x',y') - <0|\hat{W}_2(x,y) \hat{W}_2(x',y')|0>
$$
obeying the properties: \\
(a)
\bea
\left(
\frac{\partial}{\partial x^{\mu}}
\frac{\partial}{\partial x^{\mu}} +\mu^2
\right)
\tilde{F}(x,y,x',y') = 0,
\qquad
\left(
\frac{\partial}{\partial y^{\mu}}
\frac{\partial}{\partial y^{\mu}} +\mu^2
\right)
\tilde{F}(x,y,x',y') = 0,
\nonumber \\
\left(
\frac{\partial}{\partial x^{\mu}{}'}
\frac{\partial}{\partial x^{\mu}{}'} +\mu^2
\right)
\tilde{F}(x,y,x',y') = 0,
\qquad
\left(
\frac{\partial}{\partial y^{\mu}{}'}
\frac{\partial}{\partial y {\mu}{}'} +\mu^2
\right)
\tilde{F}(x,y,x',y') = 0,
\eea
(b) the Fourier transformations
$$
\int dx e^{ipx} \tilde{F}(x,y,x',y'),
\qquad
\int dy' e^{-ipy'} \tilde{F}(x,y,x',y')
$$
vanish at $p^0<0$.\\
(c) $\tilde{F}(x,y;x',y')=0$ at $x^0=y^0>x^0{}'=y^0{}'$.

Therefore,
\bea
\tilde{F}(x,y;x',y') = \int d{\bf k} d{\bf p} d{\bf k}' d{\bf p}'
[
\alpha^{++}_{{\bf k}{\bf p}{\bf k}'{\bf p}'}
e^{i\omega_{\bf k}x^0 + i\omega_{\bf p}y^0 + i\omega_{{\bf k}'}x^0{}'
- i\omega_{{\bf p}'}y^0{}'}
+
\alpha^{+-}_{{\bf k}{\bf p}{\bf k}'{\bf p}'}
e^{i\omega_{\bf k}x^0 + i\omega_{\bf p}y^0 - i\omega_{{\bf k}'}x^0{}'
- i\omega_{{\bf p}'}y^0{}'}
\nonumber \\
+
\alpha^{-+}_{{\bf k}{\bf p}{\bf k}'{\bf p}'}
e^{i\omega_{\bf k}x^0 - i\omega_{\bf p}y^0 + i\omega_{{\bf k}'}x^0{}'
- i\omega_{{\bf p}'}y^0{}'}
+
\alpha^{--}_{{\bf k}{\bf p}{\bf k}'{\bf p}'}
e^{i\omega_{\bf k}x^0 - i\omega_{\bf p}y^0 - i\omega_{{\bf k}'}x^0{}'
- i\omega_{{\bf p}'}y^0{}'}
]
e^{-i({\bf k}{\bf  x}  +  {\bf  p}{\bf  y}  +  {\bf k}'{\bf x}' + {\bf
p}'{\bf y}')}
\eea
Relation (c) implies that
$$
\alpha^{++}_{{\bf k}{\bf p}{\bf k}'{\bf p}'}
e^{i\omega_{\bf p}y^0 + i\omega_{{\bf k}'}x^0{}'}
+
\alpha^{+-}_{{\bf k}{\bf p}{\bf k}'{\bf p}'}
e^{i\omega_{\bf p}y^0 - i\omega_{{\bf k}'}x^0{}'}
+
\alpha^{-+}_{{\bf k}{\bf p}{\bf k}'{\bf p}'}
e^{-i\omega_{\bf p}y^0 + i\omega_{{\bf k}'}x^0{}'}
+
\alpha^{--}_{{\bf k}{\bf p}{\bf k}'{\bf p}'}
e^{-i\omega_{\bf p}y^0 - i\omega_{{\bf k}'}x^0{}'} = 0
$$
at $x^0=y^0>x^0{}'=y^0{}'$.  Therefore,  all $\alpha$ vanish,  so that
$\tilde{F}=0$. Lemma 5.12 is proved.

\subsubsection{Explicit form of average values}

{\bf Lemma 5.13.}
{\it  The function $F$ has the form
\bea
F(x,y;x',y') =
<0|\phi(x)\phi(x')|0> <0|\phi(y)\phi(y')|0> +
<0|\phi(x)\phi(y')|0> <0|\phi(y)\phi(x')|0>
\nonumber \\
- i\int d\xi <0|\phi(x)\phi(\xi)|0> <0|\phi(y)\phi(\xi)|0>
<0|TQ(\xi)\hat{W}_2(x',y')|0> -
\nonumber \\
i\int d\xi [
<0|T\phi(x)\phi(\xi) - \phi(x) \phi(\xi)|0> <0|\phi(\xi)\phi(y)|0>
\nonumber \\
+ <0|T\phi(y)\phi(\xi) - \phi(y) \phi(\xi)|0> <0|\phi(x)\phi(\xi)|0>
] <0|Q(\xi)\hat{W}_2(x',y')|0>
\nonumber \\
+ \int d\xi d\xi'
<0|T\phi(x')\phi(\xi') - \phi(x') \phi(\xi')|0>
<0|T\phi(\xi')\phi(y') - \phi(\xi') \phi(y')|0>
\nonumber \\
\times
<0|\phi(x)\phi(\xi)|0> <0|\phi(y)\phi(\xi)|0> <0|TQ(\xi)Q(\xi')|0>,
\nonumber \\
\eea
where $<0|TQ(\xi)\hat{W}_2(x',y')|0>$ is  the  function  of  the  form
\r{t12}.
}

Since the straightforward check of the conditions  of  lemma  5.12  is
analogous to proof of lemma 5.8, proof of lemma 5.13 is obvious.

{\bf Corollary.}
{\it The function $F$ is Poincare-invariant:
$$
F(x,y;x',y')= F(\Lambda x+ a, \Lambda y+a, \Lambda x'+a, \Lambda y'+a).
$$
}

\subsection{Check of Poincare invariance}

First of all, note that all Wightman functions are Poincare invariant.

{\bf Lemma 5.14.}
{\it The following property is satisfied:
$$
<0|\hat{W}_2(x_1,y_1) ... \hat{W}_2(x_n,y_n)|0> =
<0|\hat{W}_2(\Lambda x_1+ a,\Lambda y_1+a) ...
\hat{W}_2(\Lambda x_n + a,\Lambda y_n+ a)|0>.
$$
}

To prove  this  lemma,  it  is  sufficient  to  notice  that operators
$\hat{W}_2(x,y)$ are linear combinations of creation and  annihilation
operators, so that the Wick theorem is applicable.

{\bf Lemma 5.15.} \\
{\it 1.There exists a unique unitary operator $U_{\Lambda,a}$  obeying
the properties: $U_{\Lambda,a}|0>=|0>$,
\be
U_{\Lambda,a} \hat{W}_2(x_1,y_1) ... \hat{W}_2(x_n,y_n)|0>
=
\hat{W}_2(\Lambda x_1 + a,\Lambda y_1+a)
... \hat{W}_2(\Lambda x_n+a,\Lambda y_n+a)|0>
\l{t21a}
\ee
2. The group property \r{t3-} is satisfied.
\\
3. The invariance property \r{t3} is satisfied.
}

{\bf Proof.} Let $W_2[f] = \int dx dy \hat{W}_2(x,y) f(x,y)$,
\be
\Phi = c|0> + \sum_{n=1}^N W_2[f_{n,1}]... W_2[f_{n,i_n}]|0>
\l{t21}
\ee
Set
$$
U_{\Lambda,a}\Phi =
c|0> + \sum_{n=1}^N W_2[u_{\Lambda,a}f_{n,1}]...
W_2[u_{\Lambda,a}f_{n,i_n}]|0>
$$
with
$$
(u_{\Lambda,a}f)(x,y) = f(\Lambda^{-1}(x-a),\Lambda^{-1}(y-a)).
$$
It follows from lemma 5.14 that $(U_{\Lambda,a} \Phi,
U_{\Lambda,a} \Phi) = (\Phi,\Phi)$.  This  means  that  $U_{\Lambda,a}
\Phi=0$, provided  that  $\Phi=0$.  Thus,  the mapping $U_{\Lambda,a}:
{\cal D} \to {\cal D}$ is defined (here ${\cal D}$ is  a  set  of  all
vectors of  the  form  \r{t21a}).  This  mapping is a linear isometric
(and therefore bounded) operator,  while ${\cal D}$ is a dense  subset
of $\cal F$.  Therefore,  the operator $U_{\Lambda,a}$ can be uniquely
extended to the space $\cal F$.  Thus, there exists a unique isometric
operator $U_{\Lambda,a}$ obeying the property \r{t21a}.

Check the group property. Consider the operator
$$
V = U_{\Lambda_1,a_1} U_{\Lambda_2,a_2}
U_{((\Lambda_1,a_1)(\Lambda_2a_2))^{-1}}.
$$
It satisfies the property:
$$
V \hat{W}_2(x_1,y_1) ...\hat{W}_2(x_n,y_n)|0>
= \hat{W}_2(x_1,y_1) ...\hat{W}_2(x_n,y_n)|0>
$$
Thus, $V=1$. The group property is checked. One analogously proves that
$U_{\Lambda,a}^{-1} = U_{(\Lambda,a)^{-1}}$,  so  that  the  isometric
operator $U_{\Lambda,a}$ is unitary.

One also has
$$
U_{\Lambda,a} \hat{W}_2(x,y) U^{-1}_{\Lambda,a}
\hat{W}_2(x_1,y_1) ...\hat{W}_2(x_n,y_n)|0>
= \hat{W}(\Lambda x + a,\Lambda y +a)
\hat{W}_2(x_1,y_1) ...\hat{W}_2(x_n,y_n)|0>.
$$
Thus, the property \r{t3} is  satisfied  on  the  subspace  ${\cal  D}
\subset {\cal F}$. Lemma 5.15 is proved.

Thus, we  have  checked  the  property  of  Poincare invariance of the
theory.

\section{Conclusions}

An old problem of  axiomatic  and  constructive  field  theory  is  to
construct a  nontrivial  model of relativistic QFT which obey Wightman
axioms. The known models successfully  constructed  \c{GJ}  in  2  and
3-dimensional space-time   do   not   contain   such  difficulties  as
Stueckelberg divergences and  infinite  renormalization  of  the  wave
function.

Large-$N$ conception   enables   us  to  construct  a  wide  class  of
relativistic quantum theory models. One the one hand, these models are
trivial since  the  Hamiltonian  is quadratic with respect to creation
and annihilation operators. On the other hand, such phenomena as bound
states, quasistationary    states   and   scattering   processes   are
successfully prescribed by the quadratic Hamiltonian of the model.

A suitable language to describe the  states  and  observables  of  the
large-$N$ theory in the leading order of $1/N$-expansion is the notion
of third quantization introduced  in  quantum  cosmology  \c{GS,C}  in
order to describe processes with variable number of universes.

The third-quantized model considered in this paper may be viewed as a
large-$N$ limit of the ordinary field theory.  However, it can be also
interpreted as an independent model of relativistic quantum theory. We
have seen that such properties as renormalizability are  satisfied  in
higher dimensions    with   respect   to   ordinary   field   theories
(cf.\c{Parisi}): the model \r{b23} is renormalized  at  $d+1  \le  5$,
while the  $(\varphi^a\varphi^a)^2$-model  is  renormalized  at $d+1\le 4$
only. Thus,  usage of third-quantized models leads  to  new  types  of
renormalizable theories in higher dimensions.

For the simplicity, we have considered the large-$N$ approximation for
the $(\varphi^a\varphi^a)^2$-model only.  One can  also  consider  the
$\varphi^a\varphi^a\Phi$-model. For   this  case,  the  phenomenon  of
infinite renormalization of the wave function should be  investigated:
it happens  that  indefinite inner product should be introduced in the
state space \c{Shv}.
Investigation of the large-$N$ QED in the third-quantized
formulation gives  us  a  good  example of renormalizable gauge theory
beyond perturbation theory.

\section*{Acknowledgements}

This work  was supported by the Russian foundation for Basic Research,
project 99-01-01198.

\appendix

\section{Some properties of the Fock space}

Let ${\cal H}$ be a Hilbert space.  Denote by ${\cal H}^{\otimes n}  =
{\cal H}  \otimes  ...  \otimes  {\cal H}$ the $n$-th tensor degree of
space ${\cal H}$.  Let $\pi$ be a  transposition  $(\pi_1,...,\pi_n)$,
$1\le \pi_1\ne ...\ne \pi_n\le n$ of numbers $(1,...,n)$. Consider the
operator $\hat{\pi}$ in  ${\cal  H}^{\otimes  n}$  which  is  uniquely
defined from the relation
$$
\hat{\pi} (e_1  \otimes  ...  \otimes  e_n)  =  e_{\pi_1}  \otimes ...
\otimes e_{\pi_n}, \qquad e_1,...,e_n \in {\cal H}.
$$
By $Sym$ we denote the symmetrization operator
$$
Sym \Phi_n = \frac{1}{n!} \sum_{\pi} \hat{\pi} \Phi_n,  \qquad  \Phi_n
\in {\cal H}^{\otimes n}
$$
which is a projector. Introduce the notation ${\cal H}^{\vee n} = Sym
{\cal H}^{\otimes n}$ for the  symmetrized  $n$-th  tensor  degree  of
${\cal H}$. Denote also ${\cal H}^{\vee 0} = {\bf C}$.

{\bf Lemma  A.1.}  {\it  The set $\{ f^{\otimes n} \equiv f\otimes ...
\otimes f | f\in {\cal H} \}$ is a total set in ${\cal H}^{\vee n}$.
}

{\bf Proof.}  Let  $\Phi_n  \in  {\cal  H}^{\vee n}$,  $\Phi_n \perp f
\otimes ...  \otimes f$ for all $f\in {\cal H}$.  It is  necessary  to
prove that $\Phi_n=0$. For
$$
f = \alpha_1 e_1 + ... + \alpha_ne_n, \qquad \alpha_1,...,\alpha_n \in
{\bf C}, \qquad e_1,...,e_n \in {\cal H}
$$
one has
$$
0= \sum_{i_1 ...  i_n =1}^n  \alpha_{i_1}  ...  \alpha_{i_n}  (\Phi_n,
e_{i_1} \otimes ... \otimes e_{i_n}).
$$
The right-hand side of this relation is a polynomial in $\alpha_1,...,
\alpha_n$. The coefficient of $\alpha_1...\alpha_n$ should be equal to
zero:
\be
n! (\Phi_n, Sym e_1 \otimes ... \otimes e_n) =0.
\l{c1}
\ee
Relation \r{c1} is satisfied for all $e_1,...,e_n \in {\cal H}$.

Let $f_1,f_2,...$  be  an  orthonormal  basis  in  ${\cal  H}$.   Then
$\{f_{i_1} \otimes     ...     \otimes    f_{i_n},    i_1,...,i_n    =
\overline{1,\infty} \}$ is an orthonormal basis in ${\cal  H}^{\otimes
n}$. The vector $\Phi_n$ can be presented as
$$
\Phi_n =    \sum_{i_1...i_n=1}^{\infty}   \Phi^n_{i_1...i_n}   f_{i_1}
\otimes ... \otimes f_{i_n}.
$$
Since
$$
Sym \Phi_n =
\sum_{i_1...i_n=1}^{\infty}
\frac{1}{n!} \sum_{\pi}
\Phi^n_{i_{\pi_1}...i_{\pi_n}}   f_{i_1}
\otimes ... \otimes f_{i_n},
$$
$\Phi_n\in {\cal H}^{\vee n}$ if and only if  $\Phi^n_{i_1...i_n}$  is
symmetric with respect to transpositions of $i_1,...,i_n$.

For symmetric $\Phi^n$ one has
$$
(Sym f_{j_1} \otimes ... f_{j_n}, \Phi_n) = \Phi^n_{j_1...j_n}.
$$
Thus, eq.\r{c1}  implies  that  $\Phi^n_{i_1...i_n}=0$ and $\Phi_n=0$.
Lemma is proved.

{\bf Definition A.1.} {\it The space
$$
{\cal F}({\cal H}) = \oplus_{n=0}^{\infty} {\cal H}^{\vee n}
$$
is a Fock space.
}

Let $f\in {\cal H}$. The creation and annihilation operators
$A_n^+(f): {\cal H}^{\vee n-1} \to {\cal H}^{\vee n}$,
$A_n^-(f): {\cal H}^{\vee n} \to {\cal H}^{\vee n-1}$
are defined from the relations:
\bea
A_n^+ (f)  e^{\otimes  n-1}  =   \frac{1}{\sqrt{n}}   \sum_{k=0}^{n-1}
e^{\otimes k} \otimes f \otimes e^{\otimes (n-k-1)},
\nonumber \\
A_n^-(f) e^{\otimes n} = \sqrt{n} (f,e) e^{\otimes n-1}.
\l{c2}
\eea

{\bf Lemma A.2.} {\it The definition \r{c2} is correct. $A^{\pm}_n(f)$
are bounded operators and $||A_n^{\pm}(f)|| \le \sqrt{n} ||f||$}.

{\bf Proof.} One has
$$
||A_n^+(f) \sum_i   e_i^{\otimes   n-1}||    \le    \frac{1}{\sqrt{n}}
||\sum_i e_i^{\otimes k} \otimes f \otimes e_i^{\otimes (n-k-1)}|| =
\frac{1}{\sqrt{n}} \sum_{k=0}^{n-1} ||f|| ||\sum_i e_i^{\otimes n-1} ||
= \sqrt{n}||f|| ||\sum_i e_i^{\otimes n-1}||.
$$
$$
||A_n^-(f) \sum_i  e_i^{\otimes   n}||^2   =   ||A_{n+1}^+(f)   \sum_i
e_i^{\otimes n}  ||^2  -  ||f||^2  ||\sum_i  e_i^{\otimes n}||^2 \le n
||f||^2 ||\sum_i e_i^{\otimes n}||^2.
$$
Lemma 2 is proved.

{\bf Definition A.2.} {\it The operators
$$
A^+(f) (\Phi_0, \Phi_1, \Phi_2,... ) =
(0, A_1^+(f) \Phi_0, A_2^+(f) \Phi_1,...)
$$
$$
A^-(f) (\Phi_0, \Phi_1, ... ) =
(A_1^-(f) \Phi_1, A_2^-(f) \Phi_2,...)
$$
are called creation and annihilation operators in the Fock space.  The
finite vectors of the form $(\Phi_0,...,  \Phi_n,0,0,...)$  belong  to
the domains of $A^{\pm}(f)$.
}

{\bf Definition A.3.} {\it The vector $|0> = (1,0,0,...)$ is a  vacuum
vector.}

{\bf Lemma  A.3.}  {\it  The vector $f= (0,...,0,  Sym f_1 \otimes ...
\otimes f_n,0,...)$ can be presented as
$$
f = \frac{1}{\sqrt{n!}} A^+(f_1) ... A^+(f_n) |0>
$$}

The proof is straightforward.

{\bf Lemma A.4.} {\it The following commutation relations take place
$$
[A^{\pm}(f_1), A^{\pm}(f_2)]   =   0,   \qquad  [A^-(f_1),A^+(f_2)]  =
(f_1,f_2).
$$
The operators $A^+(f)$ and $A^-(f)$ are conjugated.
}

{\bf Definition  A.4.}  {\it  A  coherent  state  $C(f)$  is  a vector
$\Phi\in {\cal    F}$    of    the    form    $C(f)    =    \Phi     =
(\Phi_0,\Phi_1,...,\Phi_n,...)$ with   $\Phi_n  =  \frac{1}{\sqrt{n!}}
f^{\otimes n}$.
}

{\bf Lemma A.5.} {\it The following relations take place:
$$
(C(f),C(f)) =\exp (f,f), \qquad A^-(f) C(g) = (g,f) C(g).
$$
}

{\bf Lemma  A.6.}  {\it Let $g\in {\cal H}$ and $g_n$,  $n=1,2,...$ be
such a  sequence  of  elements   of   ${\cal   H}$   that   $||g_n-g||
\to_{n\to\infty} 0$. Then $||C(g_n)-C(g)|| \to_{n\to\infty} 0$.}

{\bf Proof.} Let $\xi_n = g_n-g$. Then
$$
||C(g+\xi_n) -   C(g)||^2   =   e^{(g,g)}  [e^{(g,\xi_n)}e^{(\xi_n,g)}
e^{(\xi_n,\xi_n)} - e^{(\xi_n,g)} - e^{(g,\xi_n)} +1] \to_{n\to\infty}
0.
$$

{\bf Lemma A.7.} {\it The set $\{C(f) | f\in {\cal H}\}$ is a total set
in ${\cal F}({\cal H})$.}

{\bf Proof.}   Let   $\Phi  =  (\Phi_0,\Phi_1,...,  \Phi_n,...)  \perp
C(\alpha f)$ for all  $\alpha  \in  {\bf  C}$  and  $f\in  {\cal  H}$.
Therefore,
$$
\sum_{n=0}^{\infty} \frac{\alpha^n}{\sqrt{n!}} (f^{\otimes n}, \Phi_n)
= 0.
$$
The series absolutely converges for all $\alpha$
because of the Cauchy-Bunyakovski inequality.
Therefore, $(f^{\otimes  n},  \Phi_n)  =0$  for  all  $f$,   so   that
$\Phi_n=0$. Lemma is proved.

{\bf Lemma A.8.} {\it  Let ${\cal H} = {\cal H}_1 \oplus {\cal H}_2$.
Then there exists a unique isomorphism
$I: {\cal F}({\cal H}_1) \otimes {\cal F}({\cal H}_2) \to
{\cal F}({\cal H})$ such that
\be
I(C(f_1) \otimes C(f_2)) = C(f_1 \oplus f_2), \qquad f_1 \in {\cal H}_1,
f_2 \in {\cal H}_2.
\l{c3}
\ee
}

{\bf Proof.}  The  mapping  \r{c3}  conserves   the   inner   product.
Therefore, formula  \r{c3}  uniquely  defines  an  isometric operator.
Lemma A.7 applied for the space ${\cal F}({\cal H})$ implies that  the
set $I(C(f_1)\otimes  C(f_2))$  is a total set in ${\cal F}({\cal H})$,
so that $I({\cal F}({\cal H}_1) \otimes {\cal F}({\cal H}_2)) =
{\cal F}({\cal H})$. Thus, $I$ is an isomorphism. Lemma is proved.

{\bf Lemma A.9.} {\it Let $f=f_1+f_2$, $f_1 \in {\cal H}_1$, $ f_2 \in
{\cal H}_2$. Then
\be
I^{-1} A^{\pm}(f) I = A^{\pm}(f_1) \otimes 1 + 1 \otimes A^{\pm}(f_2).
\l{c4}
\ee
}

{\bf Proof.}  It is sufficient to note that the matrix elements of the
left-hand and right-hand sides of eq.\r{c4}  between
$C(f_1)  \otimes C(f_2)$ and $C(\tilde{f}_1)  \otimes C(\tilde{f}_2)$
coincide.

Let $U$ be a bounded operator in ${\cal H}$. By ${\cal U}(U)$ we denote
the operator ${\cal U}(U):  {\cal F}({\cal H}) \to {\cal F}({\cal H})$
of the form
$$
{\cal U}(U) (f_0,f_1,...,f_n,...) = (f_0,  U f_1,  U\otimes U  f_2,...,
U^{\otimes n} f_n,...).
$$
Let $\cal  H$  be  a  self-adjoint operator in $\cal H$.  Consider the
one-parametric group of unitary operators $e^{-iHt}$.  The
operator-valued mapping  $t \mapsto {\cal  U}(e^{-iHt})$
can be also viewed as a one-parametric group.  According to the  Stone
theorem, it has the form
$$
{\cal U}(e^{-iHt}) = e^{-i{\cal F}(H)t}
$$
for some  self-adjoint operator ${\cal F}(H)$ in ${\cal F}({\cal H})$.
The explicit form of this operator is
$$
({\cal F}(H)f)_n = \sum_{k=0}^{n-1} 1^{\otimes k}  \otimes  H  \otimes
1^{\otimes (n-k-1)} f_n.
$$
Let $\varphi_1,\varphi_2,...  $ be an orthonormal basis in ${\cal H}$.
Let
$$
Hf = \sum_{ij=1}^{\infty} H_{ij} \varphi_i (\varphi_j, f).
$$

{\bf Proposition A.10.}
$$
{\cal F}(H)= \sum_{ij=1}^{\infty} H_{ij} A^+[\varphi_i] A^-[\varphi_j].
$$

The proof is straightforward.

{\bf Lemma A.11.} {\it Let $U$ be an unitary  operator  in  $\cal  H$,
$f\in {\cal H}$. Then
\be
{\cal U}(U) A^{\pm}(f) {\cal U}(U)^{-1} = A^{\pm} (Uf).
\l{c77} \ee
}

To prove  the lemma,  it is sufficient to consider the matrix elements
of the sides of eq.\r{c77} between coherent states.

Formulate now   some   results   concerning   vector   and    operator
distributions. Let  ${\cal S}({\bf R}^n)$ be a space of complex smooth
functions $u:{\bf R}^n \to {\bf C}$ such that
$$
||u||_{l,m} = \max_{\alpha_1 +...  + \alpha_n \le l}  \sup_{x\in  {\bf
R}^n} (1+|x|)^m   |\frac{\partial^{\alpha_1+...\alpha_n}u(x)}{\partial
x_1^{\alpha_1} ... \partial x_n^{\alpha_n}}| \to_{k\to\infty} 0.
$$
We say  that  the  sequence  $\{u_k\}  \in  {\cal  S}({\bf  R}^n),  k=
\overline{1,\infty}$ tends to zero  if
$$
||u_k||_{l,m} \to_{k\to\infty} 0
$$
for all $l,m$.

{\bf Definition  A.5.} {\it Let $\cal H$ be a Hilbert space.  A vector
distribution $f$ on ${\bf R}^n$
is a linear mapping $f:{\cal S}({\bf R}^n) \to  {\cal
H}$ such that $||f(u_k)|| \to_{k\to\infty} 0$ if $u_k \to_{k\to\infty}
0$. }

{\bf Remark.} We will write $f(\varphi) = \int dx f(x)\varphi(x)$  and
say that  $f(x)$  is  a vector distribution of the argument $x\in {\bf
R}^n$.

{\bf Lemma A.12.} {\it Let  $f:{\bf  R}^n  \to  {\cal  H}$  be  a  strongly
continuous bounded  vector  function.  Then $f(\varphi) = \int dx f(x)
\varphi(x)$ is a vector distribution.}

{\bf Lemma  A.13.}  {\it  Let  $f$  be  a  vector  distribution.  Then
$\frac{\partial f}{\partial x^{\alpha}}$ is a vector distribution.}

The proof is straightforward.

{\bf Definition  A.6.} {\it Let ${\cal D} \subset {\cal H}$ be a dense
subset of ${\cal H}$. An operator distribution $A$ is a linear mapping
$$
\varphi \in {\cal S}({\bf R}^n) \mapsto  A(\varphi)  :  {\cal  D}  \to
{\cal D}
$$
such that  for  all  $\Phi\in  {\cal  D}$ the mapping $\varphi \mapsto
A(\varphi) \Phi$ is a vector distribution.
}

Let $\cal  D$ be a subset of the Fock space ${\cal F}({\cal H})$ which
consists of all finite vectors $(f_0,f_1,...,f_k,0,...)$.

{\bf Lemma  A.14.}  {\it  Let  $f$  be  a  vector  distribution.  Then
$A^{\pm}(f)$ is an operator distribution.}

Investigate now the cyclic property of the vacuum vector.

Let ${\cal G} \in {\cal H} \oplus {\cal H}$. Consider the operators
$$
B(f,g) = A^+(f) + A^-(g), \qquad (f,g) \in {\cal G}.
$$
By $I_1:{\cal  H}  \oplus {\cal H}\to {\cal H}$ we denote the operator
$I_1(f,g) =f$.

{\bf Lemma A.15.} {\it Let $I_1{\cal G}$ be a dense subset  of  ${\cal
H}$. Then the set of all linear combinations
\be
\sum_n c_n B(f_{n,1},g_{n,1}) ... B(f_{n,{k_n}},g_{n,k_n})|0>,
\qquad (f_{i,k_i}, g_{i,k_i}) \in {\cal G}
\l{c5}
\ee
is dense in ${\cal F}({\cal H})$.
}

{\bf Proof.} Let $\Phi \in {\cal F}({\cal H})$.  One should prove that
it can  be  approximated  by the linear combination \r{c5}.  Lemma A.7
implies that it is sufficient to prove this statement for the coherent
states $C(\varphi)$. Choose such a sequence $f_n \in I_1{\cal G}$ that
$f_n \to  \varphi$.  Lemma  A.6  implies  that  $C(\varphi)$  can   be
approximated by $C(f_n)$. Furthermore, the coherent state $C(f_n)$ can
be approximated  a   by   finite   linear   combination   of   vectors
$(A^+(f_n))^m|0>$. For  some  $g_n$  one has $(f_n,g_n) \in {\cal G}$.
The vector $(A^+(f_n))^m|0>$ can be presented as a linear  combination
of vectors $(B(f_n,g_n))^k|0>$, Lemma A.15 is proved.

\section{What is field?}

In section  IV  we  have  investigated  the  commutation  rule between
operators $\frac{1}{N}  \sum_{a=1}^N  \varphi^a(x_1)...\varphi^a(x_k)$
and multifield canonical operator
$$
\frac{1}{N} \sum_{a=1}^N  \varphi^a(x_1) ...  \varphi^a(x_k) K_N = K_N
[(\Phi_0, \phi(x_1)    ...    \phi(x_k)     \Phi_0)     +     N^{-1/2}
\tilde{W}_k(x_1,...,x_k) + O(N^{-1})]
$$
The operators  $\tilde{W}_k(x_1,...,x_k)$  acting in the space \r{b20}
of the theory of  infinite  number  of  fields  were  interpreted  as
multifield operators.

The purpose  of this appendix is to construct analogs of the operators
$\varphi^a(x)$ in the $N=\infty$-theory.

One can notice that conception of symmetric states only  is  not  valid
for this purpose.  If the state $\Psi_N[\varphi^1,...,\varphi^N]$ were
symmetric with respect to transpositions of  the  fields  $\varphi^1$,
... ,$\varphi^N$,       the       state       $\varphi^1({\bf       x})
\Psi_N[\varphi^1,...,\varphi^N]$ is  not  symmetric.   Thus,   it   is
necessary to consider the nonsymmetric solutions of eq.\r{ba1}.

Consider the  states  of  large-$N$  theory  which  are symmetric with
respect to $N-s$ fields  $\varphi^{s+1},...,\varphi^{N}$  only,  where
$s$ is a finite quantity. Analogously to eq.\r{b3}, let $\Psi_N$ be of
the form
\be
(K_N^sf)[\varphi^1,...,\varphi^N] =                   \sum_{k=0}^{N-s}
\frac{\sqrt{k!}}{N^{k/2}} \sum_{s+1   \le   a_1  <  ...  <a_k  \le  N}
f_k[\varphi^1,...,          \varphi^s,\varphi^{a_1},...,\varphi^{a_k}]
\prod_{a>s,a\ne a_1...a_k} \Phi_0[\varphi^a],
\l{v1}
\ee
where $f_k[\varphi^1,...,\varphi^s,      \phi^1,...,\phi^k]$       are
functionals being    symmetric    under   transpositions   of   fields
$\phi^1,..., \phi^k$ and obeying the condition
\be
\int D\phi_1                                          \Phi_0^*[\phi_1]
f_k[\varphi^1,...,\varphi^s,\phi^1,...,\phi^k] = 0.
\l{v2}
\ee
We see that states under consideration are specified by infinite sets
\bea
\left( \matrix{
f_0[\varphi^1,...,\varphi^s]
\nonumber \\
f_1[\varphi^1,...,\varphi^s,\phi^1]
\nonumber \\
...
\nonumber \\
f_k[\varphi^1,...,\varphi^s,\phi^1,...,\phi^k]
\nonumber \\
...}
\right)
\eea
The state space is then isomorphic to
\be
\tilde{\cal F}_s   =    {\cal    H}^{\otimes    s}    \otimes    {\cal
F}(\oplus_{n=1}^{\infty} {\cal H}^{\vee n}).
\l{v3}
\ee
Since the symmetric state can be viewed as a state of the form \r{v1},
there should   be   exist  an  operator  $I_s:  \tilde{\cal  F}_0  \to
\tilde{\cal F}_s$ such that
$$
K_N^s I_s f = K_N f.
$$
Let us present the explicit form of the operator $I_1$. One has
\bea
(K_Nf)[\varphi^1,...,\varphi^N] =                         \sum_{k=0}^N
\frac{\sqrt{k!}}{N^{k/2}} \sum_{1\le   a_1   <   ...   <  a_k  \le  N}
f_k[\varphi^{a_1},...,\varphi^{a_k}] \prod_{a\ne            a_1...a_k}
\Phi_0[\varphi^a] =
\nonumber \\
\sum_{k=0}^N \frac{\sqrt{k!}}{N^{k/2}} \sum_{2\le a_2 < ...  < a_k \le
N} f_k[\varphi^1,     \varphi^{a_2},...,\varphi^{a_k}]     \prod_{a\ne
1,a_2,...,a_k} \Phi_0[\varphi^a] +
\nonumber \\
\sum_{k=0}^N \frac{\sqrt{k!}}{N^{k/2}} \Phi_0[\varphi^1]
\sum_{2\le a_1 < ...  < a_k \le
N} f_k[ \varphi^{a_1},...,\varphi^{a_k}]     \prod_{a\ne
1,a_2,...,a_k} \Phi_0[\varphi^a].
\eea
We see that
\be
(I_1f)_k[\varphi^1,\phi^1,...,\phi^k] =              \Phi_0[\varphi^1]
f_k[\phi^1,...,\phi^k] +    N^{-1/2}    (\tilde{A}^-[\varphi^1]   f)_k
[\phi^1,...,\phi^k].
\l{v7}
\ee
One can  also  perform  the  symmetrization  procedure  for the vector
\r{v1} and obtain the symmetric  state.  Therefore,  there should  exist
an operator $S_s: \tilde{\cal F}_s \to \tilde{\cal F}_0$ such that
$$
Sym K_N^sf = K_NS_s f.
$$
Construct the operator $S_1$. If
\be
\int D\varphi^1  f_k[\varphi^1,\phi^1,...,\phi^k]  \Phi_0^*[\varphi^1]
=0,
\l{v4}
\ee
one obtains from direct calculation that
$$
(S_1f)_k[\phi^1,...,\phi^k] =       N^{-1/2}       \int       D\varphi
(\tilde{A}^+[\varphi] f)_k[\varphi,\phi^1,...,\phi^k].
$$
If
\be
f_k[\varphi^1,\phi^1,...,\phi^k] = \Phi_0[\varphi^1] g_k[\phi^1,  ...,
\phi^k],
\l{v5}
\ee
then
$$
(S_1f)_k[\phi^1,...,\phi^k] = \frac{N-k}{N} g_k[\phi^1,...,\phi^k].
$$
Generally, $f_k$ can be viewed as a superposition of  vectors  obeying
conditions \r{v4} and \r{v5} correspondingly, so that
\be
(S_1f)_k[\phi^1,...,\phi^k] =   [1-\frac{\hat{n}}{N}]   \int  D\varphi
\Phi_0^*[\varphi] f_k[\varphi,\phi^1,...,\phi^k]   +   N^{-1/2}   \int
D\varphi (\tilde{A}^+[\varphi]f)_k [\varphi, \phi^1,...,\phi^k].
\l{v8}
\ee
The Schrodinger  field  $\varphi^1({\bf  x})$ in $\tilde{\cal F}_1$
may  be  viewed  as  an
operator of multiplication by $\varphi^1({\bf x})$, since
$$
\varphi^1({\bf x}) K_N^1 = K_N^1 \varphi^1({\bf x}).
$$
The multifield can be constructed from the field $\varphi^1$ as
\be
\tilde{\cal W}_{N,k}   (x_1,...,x_k)   =   S_1   \varphi^1(x_1)    ...
\varphi^1(x_k) I_1
\l{v6}
\ee
since
$$
\frac{1}{N} \sum_{a=1}^N  \varphi^a(x_1)  ...  \varphi^a(x_k)  =   Sym
\varphi^1(x_1) ... \varphi^1(x_k).
$$
One can  notice  from  eqs.\r{v7} and \r{v8} that formula \r{v6} is in
agreement with the results of section IV.

To construct the Heisenberg field operator $\varphi^1(x):
\tilde{\cal F}_1 \to \tilde{\cal F}_1$, it is necessary to commute the
Hamiltonian operator ${\cal H}_N$ with the operator $K_N^1$. We obtain:
\be
{\cal H}_N K_N^1 = K_N^1 \tilde{\cal H}_N^1,
\l{v8*}
\ee
where
$$
\tilde{\cal H}_N^1 =
\tilde{\cal H}_N + \int d{\bf x} \left[
- \frac{1}{2}   \frac{\delta^2   }{\delta  \varphi^1({\bf  x})  \delta
\varphi^1({\bf x})} +  \frac{1}{2}  (\nabla  \varphi^1)^2({\bf  x})  +
\frac{\mu^2}{2} (\varphi^1({\bf x}))^2 \right] + O(N^{-1/2}).
$$
Therefore, the Heisenberg operator
$$
\varphi^1({\bf x},t) = e^{i\tilde{\cal H}_N^1t} \varphi^1({\bf x})
 e^{-i\tilde{\cal H}_N^1t}
$$
coincides with   the   operator   of  the  free  scalar  field  up  to
$O(N^{-1/2})$. Therefore, for operator $\tilde{\cal W}_{N,k}$ one has
\be
\tilde{\cal W}_{N,k}(x_1,...,x_k) = \int D\varphi^1 \Phi^*_0[\varphi^1]
\varphi^1(x_1) ... \varphi^1(x_k) \Phi_0[\varphi^1] + O(N^{-1/2}).
\l{v9} \ee
This result confirms the hypothesis of section IV.

In order  to  obtain  the  explicit  form of the $O(N^{-1/2})$-term of
formula \r{v9},  it is necessary to compute the $O(N^{-1/2})$-term  in
eq.\r{v8*}. The result is in agreement with section IV.

\end{document}